\documentclass{article}
\usepackage{amsmath,amsfonts,amsthm}
\usepackage{amstext}
\usepackage{amsgen}
\usepackage{amsbsy}
\usepackage{amsopn}
\usepackage{amssymb}
\usepackage{graphicx}
\usepackage{epsfig}
\newtheorem{thm}{Theorem}[section]

\newtheorem{prop}[thm]{Proposition}
\theoremstyle{definition}
\newtheorem{defn}[thm]{Definition}
\theoremstyle{remark}
\newtheorem{rem}[thm]{Remark}
\numberwithin{equation}{section}
\newtheorem{theorem}{Theorem}

\newtheorem{algorithm}[theorem]{Algorithm}

\newtheorem{definition}[theorem]{Definition}

\newcommand{\Real}{\mathbb R}
\newcommand{\RealClosed}{\overline{\mathbb{R}}}
\newcommand{\Natural}{\mathbb N}

\newcommand{\eps}{\varepsilon}

\newcommand{\bigO}{\mathbf{O}}

\def\IMAGESPATH{.}
\def\BIBPATH{.}

\begin{document}

\title{The Complexity Of The NP-Class}
\author{Carlos Barr\'{o}n Romero \\
Universidad Aut\'onoma Metropolitana\\
Unidad Azcapotzalco, Unidad Cuajimalpa \\
Av. San Pablo No. 180, Col. Reynosa Tamaulipas, C.P. 02200, \\
Calle Artificios No. 40,  6$^\texttt{o}$ piso,  Col. Hidalgo, C.P. 01120,
 MEXICO
}


\date{2010}

\maketitle

\begin{abstract}
This paper presents a novel and straight formulation, and gives a
complete insight towards the understanding of the
complexity of the problems of the so called NP-Class. In
particular, this paper focuses in the Searching of the Optimal
Geometrical Structures and the Travelling Salesman Problems. The
main results are the polynomial reduction procedure and the
solution to the Noted Conjecture of the NP-Class.


Algorithms, NP, Numerical optimization.

\end{abstract}
\section{Introduction}

Over the previous century the problem of the complexity of the
NP-Class problems has been remain open. The broad types of
problems are quite important the science and the technology. Just to mention
nano-technologies, design of drugs, and process optimization have
driven many researchers to design algorithms, and to publish
thousands of papers devoted to improve and solve them. However
besides the equivalence of members of the NP-Class no unique
efficient algorithm has been designed and moreover the possible
determination of complexity has not
differentiate the P-Class from the NP-Class.

In order to study in details the complexity of the NP-Class,
the following points are presented in this paper.
\begin{enumerate}
    \item Research Space, Complexity, and Properties of two
    NP-Problems.
    \item Polynomial Reduction Procedure.
\end{enumerate}

The main result shows, that the complexity of the NP-Class is
exponential and lacks of a optimization property in their
structure for arbitrary and large instances of NP-Problems.
Also,
It is presented that the special case of the 2D Euclidian
Traveller Salesman Problem has polynomial complexity. This
modifies the frontier's spectrum of computational complexity~\cite{
siamCBMS-NSF:Tarjan1983}, in the sense that some problem
classified as intractable changes to tractable.

The next proposition is the key of this research.
\begin{prop}\nonumber
The problems of the NP-Class have not an polynomial algorithm for
checking their solution.
\end{prop}

To explain its importance, let us to assume the contrary, i.e., a
polynomial algorithm for checking the solution exists for any
arbitrary and large instances of a NP-Problem. Therefore, it is
possibly to use and joining it with a polynomial research
algorithm to come out with a joining polynomial algorithm to solve
any NP-Problems, furthermore, the properties used into the
checking algorithm could help to design and improve the joining
algorithm. On the other hand, if it exists the polynomial algorithm
able to solve any arbitrary and large NP-Problem it must has an
algorithm for verifying the solution, with complexity no more than
polynomial time, otherwise there is not support to accept its
solution. Therefore, it is a reasonable step towards the design of
a polynomial algorithm able to solve any NP-Problem to try to
design first the checking algorithm of the solution. Also, the
design of an algorithm is based in structures and properties from
the problem, and it is not always easy to relate the length's data
input with the execution time. It is necessary to explore problem,
data, and properties to understand and design an algorithm.

\begin{defn} {Notation}
\begin{enumerate}

\item The real number are denoted by $\Real$, and $\RealClosed = \Real \cup
\{-\inf, \inf\}$. Also, natural number are denoted by $\Natural$.

    \item  LJP$_n$ stands for the Problem of Searching Optimal
Geometrical Structures of $n$ particles under the Lennard-Jones
Potential.

    \item GAP$_n$ is the General Assign Problem of size $n$.
    \item TSP$_n$ is the traveller salesman problem of size $n$.

    \item Euclidian $m$D TSP$_n$ stands for the version of the TSP$_n$
where the vertices are like cities in a real metric space of
$m$-dimension, i.e., the vertices are points in $\Real^m$.

     \item The Research Space of a problem P$_n$ is all valid and
possible entries that comply with the properties of P$_n$.

    \item Integer formulation are denoted by IP, where P denotes
    GAP or TSP.

    \item The complexity is denoted by $\bigO(f(n))$.
Polynomial time means $n^p$, $p<n$ where $n$ is the length's data
input, and $p$ is a positive integer.
\end{enumerate}
\end{defn}

The NP-Problems are isomorphous and it is worth to study in deep
some of them for understanding and take advantage of their
properties, restrictions and structures. This is the reason to
analyze LJP$_n$ and GAP$_n$. The former is a particular case of
the Searching of the Optimal Geometrical Structures of clusters of
$n$ particles under the potential function of Lennard-Jones.
GAP$_n$ is the general formulation of the TSP$_n$. The traveller
salesman problem has two interesting cases Euclidian and
arbitrary. Here an arbitrary TSP$_n$ means that edges' cost
are not from a norm or distance of the vertices as a points
in $\Real^m$ but from arbitrary real values $\geq 0$. The two
next section cover an study of these NP-problems.
 Section~\ref{sc:Lennard-Jones} describes the state
of the art of LJP$_n$.  GAP$_n$ and its related problems are cover
in section~\ref{sc:GAP}.
 The section~\ref{sc:integer} has the integer formulation
of GAP$_n$. The section~\ref{sc:MTFiniteAutomataGAP} presents the
computational model of the Turing Machine and states a relation to
an algorithm for checking the solution of
GAP$_n$. The section~\ref{sc:Reducibility} describes the main
results of the complexity of GAP$_n$, and the last section contains
conclusions and future work.



\section{The Lennard-Jones Problem}~\label{sc:Lennard-Jones}

The chemistry, physics, and applied mathematics academic community
had dedicated thousand of articles because the importance of
understanding of how the nature builds
matter or beings from particles. The Searching of the Optimal Geometrical
Structures of $n$ particles consists to determine
under a pairwise potential the geometrical shape of this finite
group of $n$ particles, (also,
the result is called an optimal cluster of size $n$) that
correspond to the minimum potential. There are many pairwise
potential functions Buckingham (BU) Potential, Kihara Potential,
Lennard-Jones (LJ) potential, Morse Potential, and so on.
There are many methods to address this problem and some authors
had been found that for LJ clusters of $n$ ($ \leq $ 147) their
algorithms have polynomial complexity but no generalization have
claimed for large cluster's size.

The putative optimal LJ clusters (potential and particles' 3D
coordinates) had been shared by the Wales, Doye et al. in their WWW
site ``The Cambridge Cluster Database (CCD)''(see~\cite{http:CCD}),
Shao's articles(see~\cite{jcics:Cai2002, t:Cai2002, jpca:Jiang2003,
jcics:Shao2004, cp:Shao2004, pc:Shao2005, jpca:Xiang2004B}), and
Barr\'on et. al's articles (see ~\cite{aml:Barron1999,
cpc:Romero1999A}.

In my experience, the success of some methods are based in the use
of lattices to select an initial cluster of $n$ particles
and then, a minimization or relaxation procedure applied to it
to estimate its minimum potential.
 I share this idea with Shao in a
personal communication and he and his group achieve to repeat the
previous results for $n$ $\leq$ 309 and states the new putative optimal
clusters up to 1610 particles. Northby in his seminal article~\cite{
jcp:Northby1987} stated the conjecture of the existence of a growing
sequence in the lattice IC but the putative optimal clusters up to 1610
particles show that it is a local property without possibility to
generalize to other lattices. There is also possible that even clusters
with big magic number could be not longer the putative optimal cluster
because of the central vacancy. Finally, the experience given by the
putative optimal clusters up to 1610 do not provide general properties
to reduce the alternatives to find out optimal cluster in an efficient
way.

The Methods based in lattices drive my study towards this type of
research space, using a version of a genetic algorithm adapted to
the lattice IF  in the article ``Minimum search space and efficient
methods for structural cluster optimization''  was possible to
repeat Shao group's results and found out new clusters, and also
to state that the lack of central particle starts in the clusters
542,543,546,547,548 of the shell 309-561 and not as Shao states in
the shell 561-923.  The following proposition
from~\cite{arXiv:Barron2005}

\begin{prop}~\label{prop:DiscreteLattice_p1}
Exist a discrete set, $\Omega $, where $\forall j\in N$, $j\geq
2$, where the potential of the Searching Optimal Clusters of size
$j$ in a Discrete Space for an XX potential has the same optimal
value of Searching Optimal Clusters of size $j$ in a Continues
Space for the same potential. The XX potential function complies
with
\begin{enumerate}
    \addtolength{\topsep}{-\baselineskip}
    \addtolength{\itemsep}{-\baselineskip}
    \setlength{\parskip}{\baselineskip}
\item  $\lim_{r_{i,j}\rightarrow 0}\hbox{VXX}\left( r_{i,j}\right)
=\infty$. \item  $\nabla ^{2}\hbox{VXX}\left( x^{\ast }\right) $
semi-positive$,\left\| \nabla \hbox{VXX}\left( x^{\ast }\right)
\right\| \ll 1$ and $\frac{\left\| \nabla \hbox{VXX}\left( x^{\ast
}\right) \right\| }{\left| \hbox{VXX}\left( x^{\ast }\right)
\right| }<\delta _{0}$, where $0<\delta _{0}\ll 1$
\end{enumerate}
XX could be BU or LJ potential.
\end{prop}

The previous proposition depicts a very populated research space
where the process of the selection of  particles from $\Omega$ in
order to match of the results of a continuous minimization from an
initial cluster from a lattice does not seem to be a very
efficient option for designing an algorithm to solve the LJP$_n$.
Therefore
the lattice IF (which is combination of the lattice IC and the lattice FC)
was
proposed. Besides, that lattice IF unifies and avoid to use other
lattices, it is not a simple research space. Some putative
optimal clusters are not centered in the origin of the lattice IF.
 This cause that the selection of an initial cluster of size
$n$ from the lattice IF  is not a simple process and depends
where to put the center of the selection. In fact,  the conjecture
if all optimal clusters for any size
exists inside of the lattice IF is still opens.

On the other hand, the existence of $\Omega$ point out to other
possible lattices. One option is, the lattice CB  which is a cubic
lattice with edges of size $\frac{1}{2}d^\ast$ and it has a
centered particle in (0,0,0), where $d^\ast=\sqrt[6]{2}$.

\begin{prop}
Any shape of $n$ particles with edges $\approx d^\ast$ can be
approximated from the lattice CB.
\begin{proof}
Any shape is bounded by the well known Kissing Number in 3D. The unit
centered in the origin cube of edge $=d^\ast$ of the lattice CB has 27
particles, the Kissing Number in 3D states that 12 spheres can be
allocated around a centered particle.  By example, the icosahedron's
edges $=d^\ast$ has a centered particle with 12 neighbors and the other
external twelve particles have 6 particles (five externals and the
centered) and these particles can be selected from the 27 particles of
the unit CB cube. Therefore there are sufficient neighbors to
approximate any shape with edges $\approx d^\ast$ with $n$ particles
from the cubes centered in the origin of the lattice CB.
\end{proof}
\end{prop}

\begin{figure}
\centerline{ \psfig{figure=\IMAGESPATH/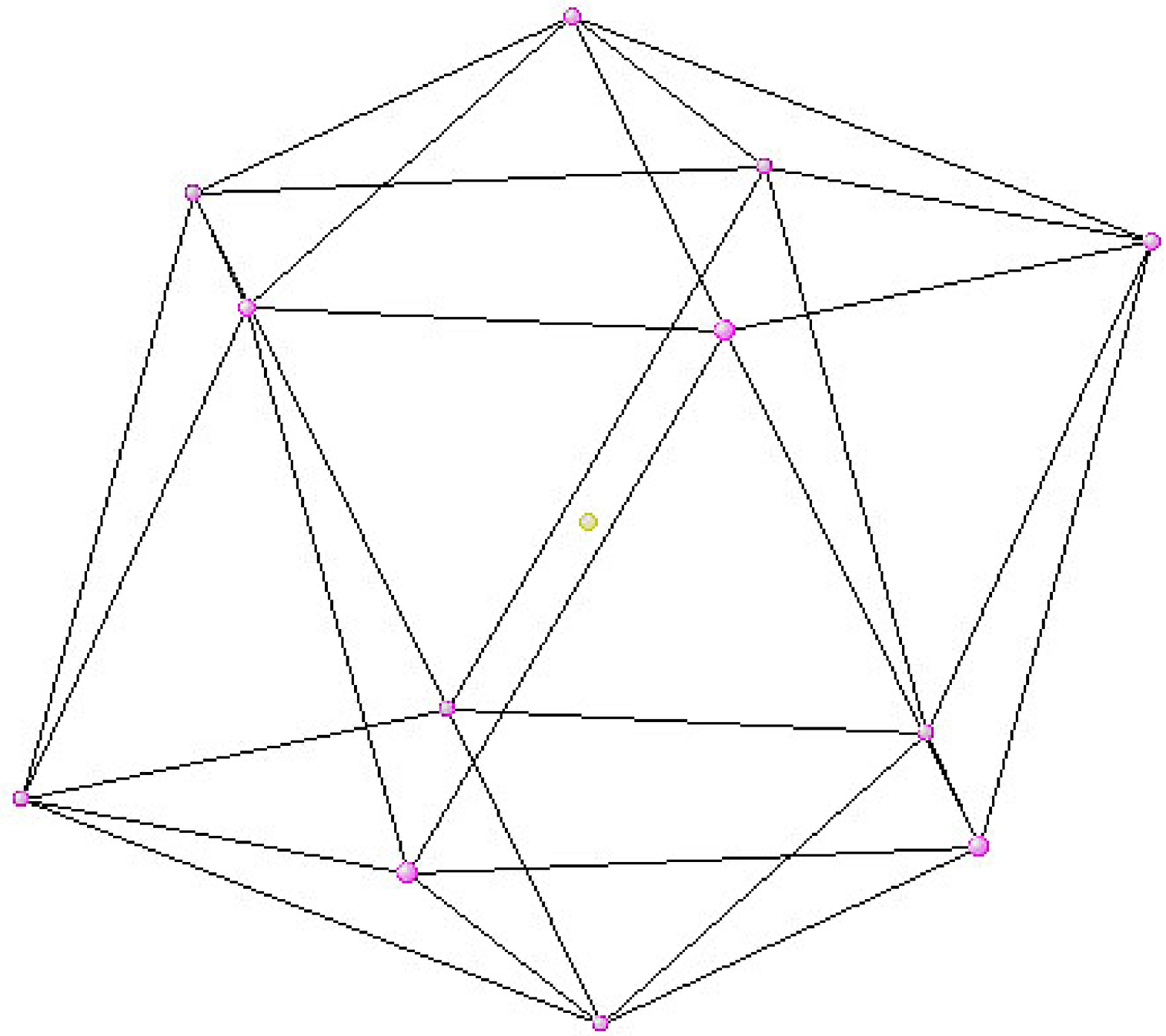, height=50mm}
\psfig{figure=\IMAGESPATH/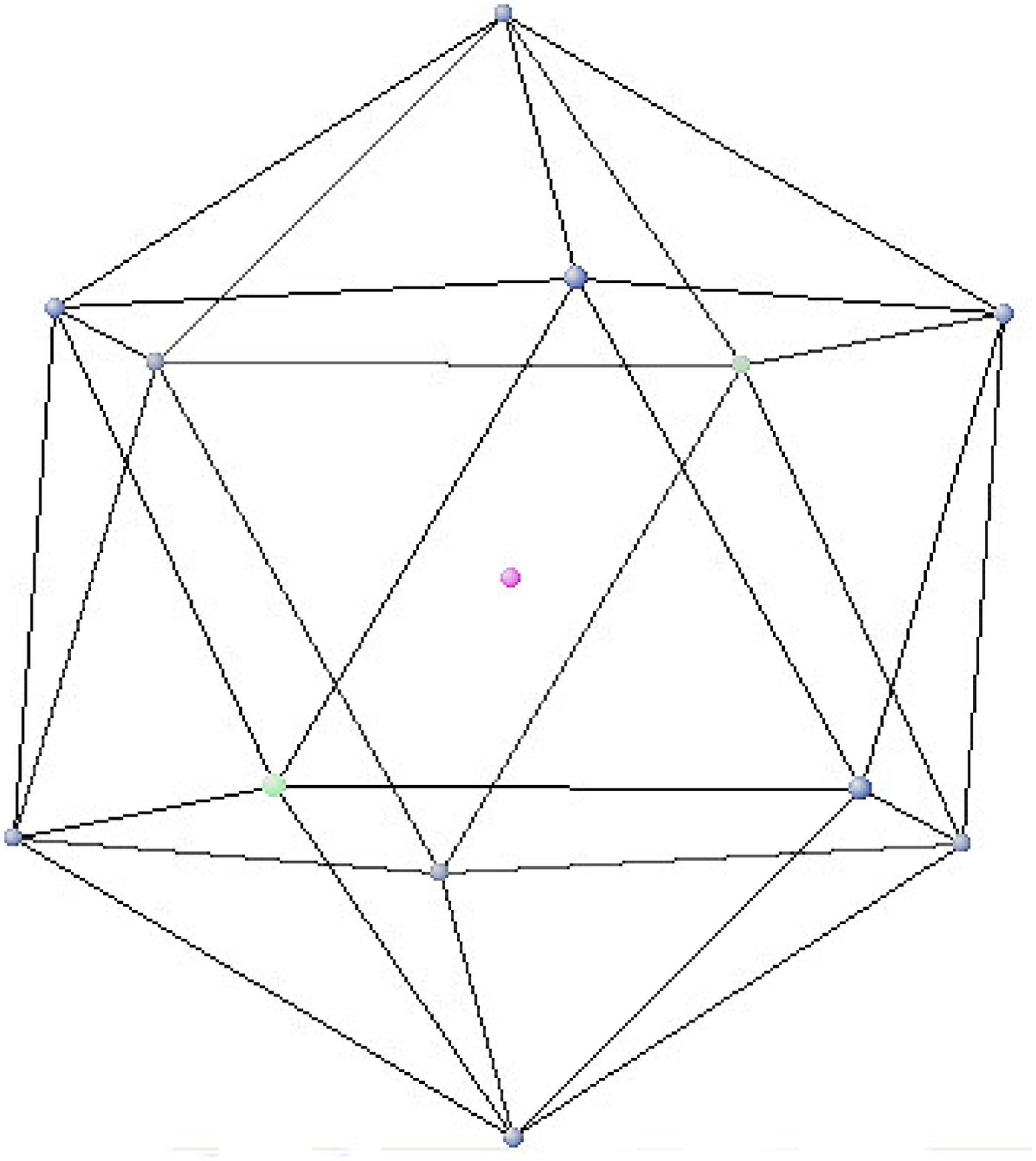,height=50mm} } \centerline{
\makebox[2.2in][c]{ a) }\makebox[2.2in][c]{ b) } } \caption{a) Initial
CB cluster, and b) Optimal cluster of 13 particles.}~\label{fig:cb013}
\end{figure}

The figure~\ref{fig:cb013} depicts the initial CB cluster, and
optimal cluster of 13 particles.

With the lattice CB  instead of the lattice IF a genetic algorithm
was able to repeat the known results. The complexity of such
algorithm is due to the selection of clusters from the lattice CB
procedure and the genetic minimization procedure. The later
procedure can guarantee by elitism that the cluster selected as
optimal is the best cluster for the set of initial clusters given
by the former procedure. However if this is the case there is not
longer needed to use, other mechanism than the minimization
procedure applied to all clusters given for the selection
procedure. Therefore the cost of minimization over a subset of
clusters selected from CB is proportional to the number of
clusters by the computational cost of the minimization procedure.
The minimization procedure complexity depends of selected
optimization algorithm, in our genetic algorithm the Conjugated
Gradient Method (see~\cite{ springerCP:Glowinski1984}) is used. It
can approximate the minimum in at most 3n iterations (the
complexity of minimization is due to the size of the cluster by 3,
the dimension of the 3D particles' coordinates).

\begin{figure}
\centerline{
\psfig{figure=\IMAGESPATH/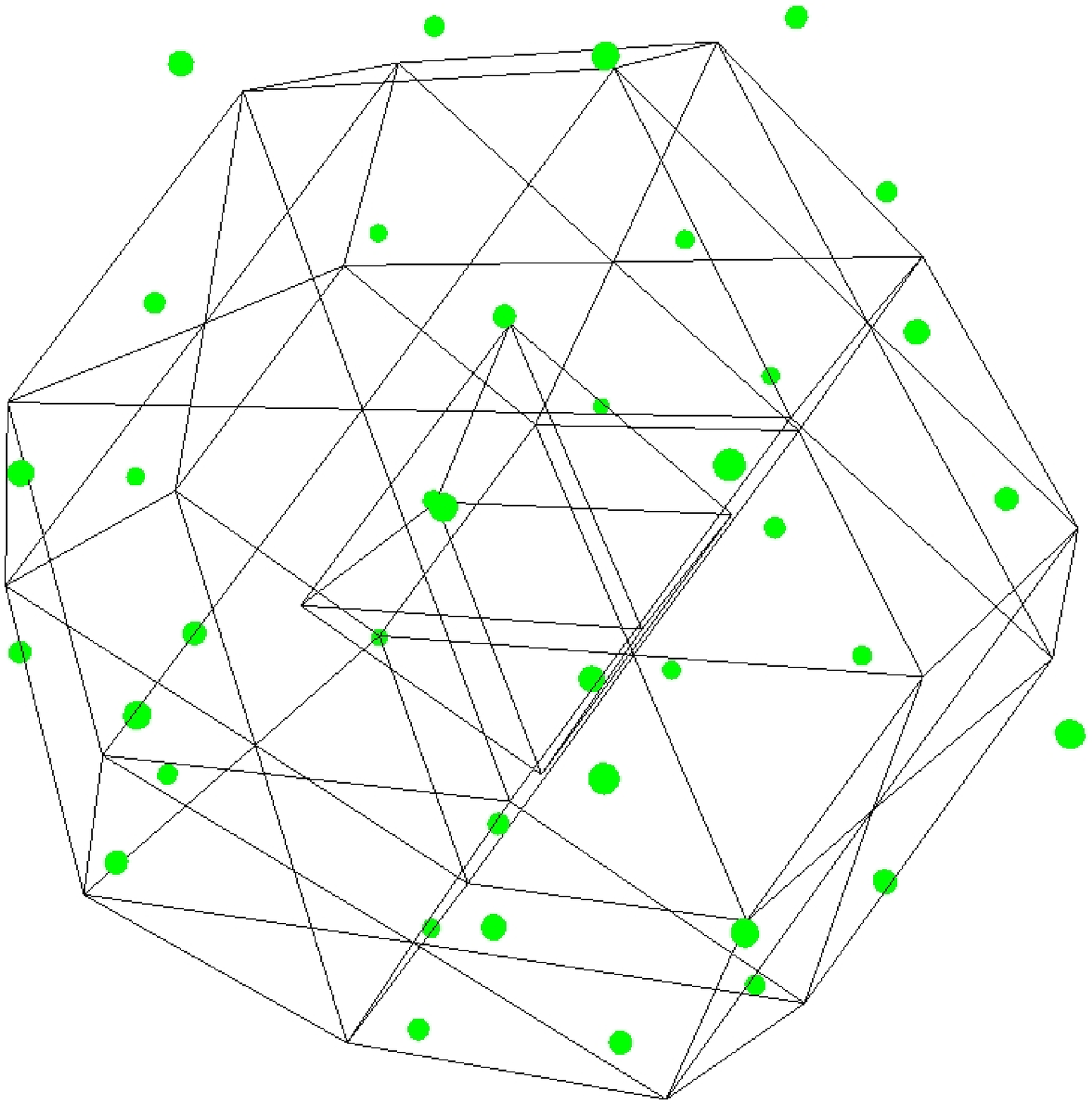,height=50mm}
\psfig{figure=\IMAGESPATH/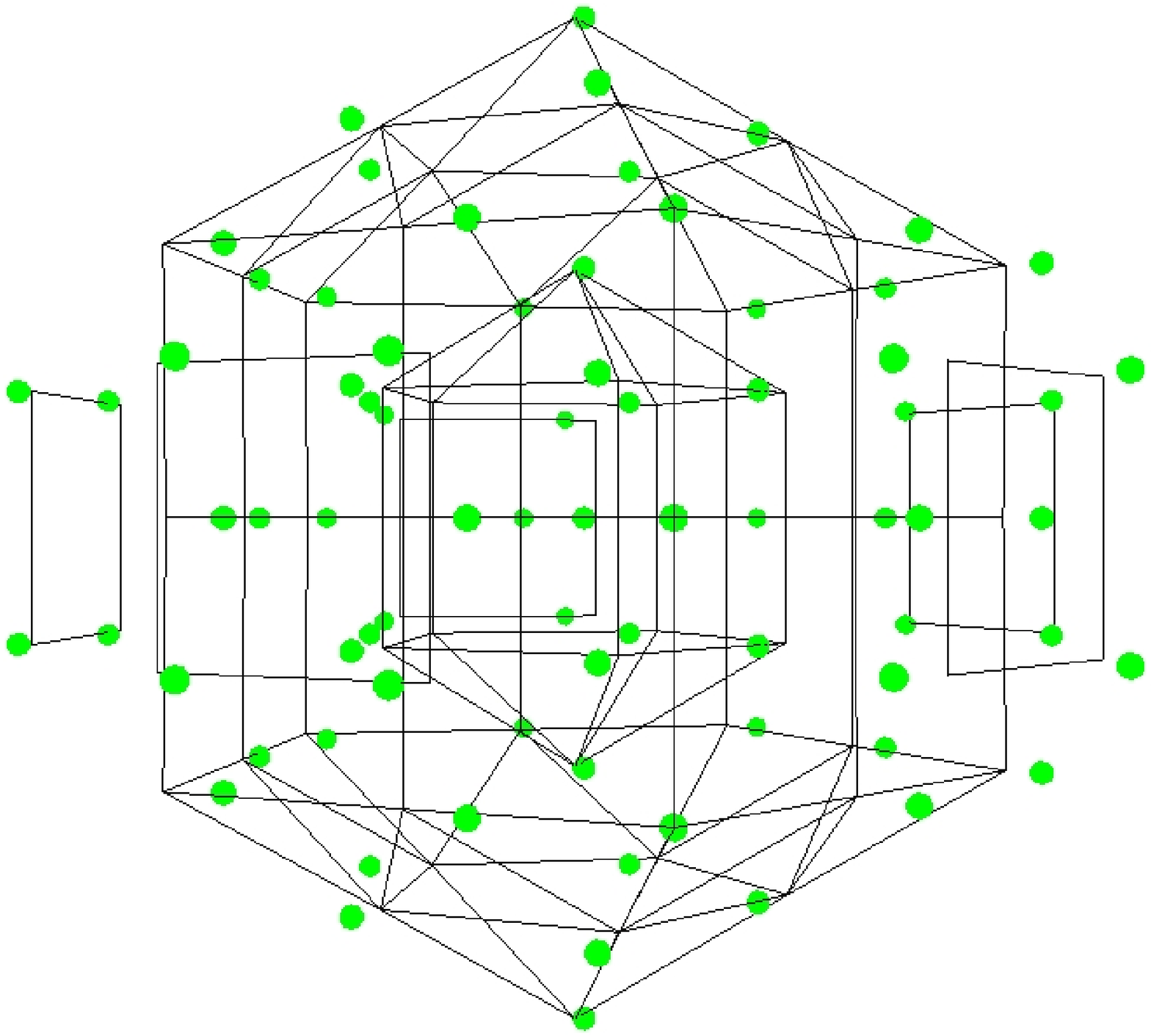, height=50mm}
\psfig{figure=\IMAGESPATH/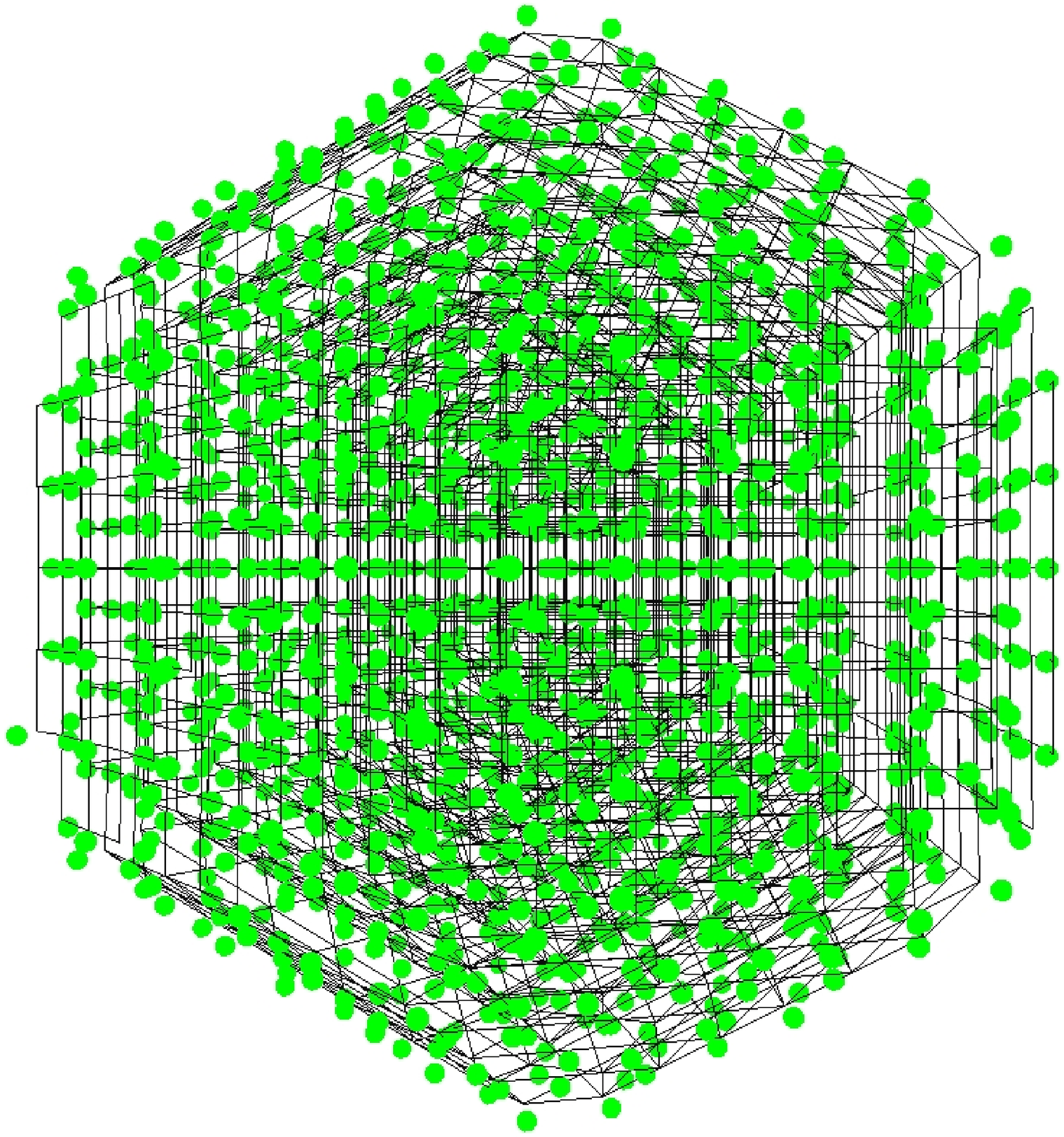, height=50mm} }
\caption{Optimal and initial clusters for 38, 75, and 1,600
particles.}~\label{fig:examplesCB_optimal38_75_1600}
\end{figure}

The lattice CB provides a search space where the selected clusters
are centered in the origin, this means that a centered sphere with
an appropriate ratio can give a sufficient research space to
determine inside the optimal clusters until some size.
Examples of optimal clusters for $38$, $75$, and $1,600$ as wired black figures
with the initial particles in green from
lattice CB are depicted in
Fig.\ref{fig:examplesCB_optimal38_75_1600}. The
convexity and local continuity of the LJ potential
guaranty that for the cluster with
$n=2,3,4$ with edges $=d^\ast$ are global minima cluster with
first and second optimal condition satisfies.
The gradient of the LJ potential
is equal $0$ and the Hessian of the LJ potential is definite positive
but for $n>4$ the gradient of the LJ potential is not null.

On the other hand, if the exploration is full in the appropriate
centered sphere of the lattice CB, the optimal clusters are not more
putative, they are global optimal clusters. The next
proposition define a research subspace of the lattice CB  for the
determination of the global optimal clusters inside of it.

\begin{prop}
A cluster of $n$ $(>4)$ particles with edges $\geq d^\ast$ has a
descend property and a minimum over the centered cluster's
diameter for the LJ potential in an centered in the origin CB sphere with double ratio
determined by the
 ratio of the IC cluster over the so called magic number.
\begin{proof}
Given $n$, the ratio of the cluster of the magic number can be
determined by the formula $n_r = \frac{10}{6}r(r+1)(2r+1)+2r+1$
for the first $r \in \mathbb{N}$ such that $n_r> n$. Then the
ratio of the sphere is $2rd^\ast$. Now, the diameter of a cluster
inside of the sphere could be as big as $4d^\ast$. A cluster with
large diameter has less edges of size $d^죞ast$ between particles
than a cluster with smaller diameter, therefore $LJ(C_n(d_1))>
LJ(C_n(d_2))$ if $d_1 \geq d_2$. Finally, the minimum LJ potential
of a cluster of $n$ particles is reached over the finite possible
selection of $n$ particles in the CB sphere of ratio $2rd^\ast$.
\end{proof}
\end{prop}

\begin{figure}
\centerline{ \psfig{figure=\IMAGESPATH/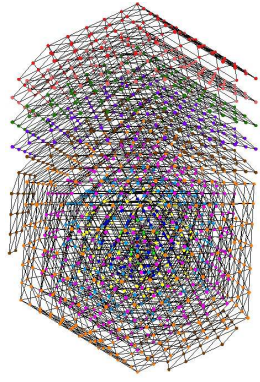,
height=50mm}}
 \caption{MIF1739 contain $C^*_n$, clusters with minimum LJ
 potential for $n=2,\ldots,1000$.}~\label{fig:min_lattice1739}
\end{figure}

This research subspace from the lattice CB  are quite different from
the minimum research space of the lattice IF, MIF1737 is depicted in
figure~\ref{fig:min_lattice1739}. Also, the other difference is that
all clusters can be selected with  one parameter, the ratio of the
centered CB sphere instead of two parameters, the ratio and the origin
of an sphere in the IF lattice. So far, only the centered CB spheres
with ratio $>0$ are needed for searching optimal clusters of size $n$
contained in a ratio of $2rd^\ast$, where $r$ is such that $n\geq n_r$.
This avoid the cone ice cream shape of the search space based on the
lattice IF. With CB spheres to build an initial cluster is only
necessary to pick up particles but even with trying to take advance of
the symmetry and the previous and posterior optimal clusters of a given
one, there is not possible to reduce the number of combinations to
explore clusters of size $n$ inside of a sphere with a guaranty that
the exploration is complete. Different shapes of clusters with and
without protuberances, the elasticity of the particles' position, and
the existence of holes makes impossible to reduce combinations by
symmetry.  The shapes of putative optimal clusters in the range of the
known putative optimal clusters from 2 to 1612, shows that they do not
share general geometrical characteristics. The special extreme cases of
selecting particles occur when the size of a cluster is around of the
value of a magic number. The following proposition shows that for these
cases, the complexity of generating initial cluster of $n$ particles is
exponential.

\begin{prop}
Given a set of M particles the selection of $n$ particles with $M
\gg n$ has exponential complexity.
\begin{proof}
$ 
 \left( \, ^M _n \right) = \frac{M!}{\left( m-n\right)
!n!}=\frac{M\left( M-1\right) \cdots 2\cdot 1}{n\left( n-1\right)
\cdots 2\cdot 1}\geq \left( \frac{M}{n}\right) ^{n}$,  $M \gg n$.
Taking $k=$fix$\left( \frac{M}{n}\right)$, $k$ is an integer and the
complexity is greater than $k^n$.
\end{proof}
\end{prop}

This proposition point out that the complexity grows exponentially
because there are many alternatives for picking initial
particles even if this selection is performed into an appropriate
reduced research space. This reduced space is a centered sphere of
the lattice CB.

In
the next section a similar analyzes to reduce the research space
is applied to NP problems with data structure given by a complete graph.

\section{The General Assign Problem}~\label{sc:GAP}

The well known Travel Salesman Problem (TSP)  is a case of an
 Assign General Problem(GAP). In this section the analysis of the GAP
 is presented.  First, the results and properties are
 presented.

The GAP$_n$ consists a complete graph, a
function which assigns a value for each
 edge, and objective function,
$\text{GAP} \_n$ $=$
 $(G_n,c,f)$, where $G_n =\left( V_n, A\right) $,  $V_n$  $\subset$ $\mathbb{N}$
 $A$ $=$
 $\left\{ \left( i,j\right) \, | \, i,j \in V\right\}$,
$c : V \times V \rightarrow  \RealClosed$, and $f$ is a real function
for the evaluation of any path of vertices. Solving a GAP means to look
for a minimum or a maximum value of $f$
    over all cycles of $G_n$.
Note
that $c(i,i)=\inf$, and $c(\cdot,\cdot)$ can be see as a matrix of  $\RealClosed^{\,n \times n}.$

\begin{rem} Some properties and characteristics are:
\begin{enumerate}
    \item The complete property of $G_n$ means that $\forall i, j \in V_n, \exists e=(i,j)\in A$.
    \item $G_n$ is a directed graph, which means that edges are a ordered pair.
    \item A path of vertices is a sequence of vertices of $V_n$.
    \item A cycle is a path of different vertices but the first and the last
    vertex are the same. Hereafter, a complete cycle is a cycle
    containing the $n$ vertices of $V_n$. Also, a complete cycle has $n$ edges.
    \item A TSP$_n$ is a case of a GAP$_n$ where
    $c(i,j)=c(j,i)$ (the cost matrix is symmetric) and $c(i,j) \geq 0, \forall i,j$, and finally
    $f$ is the sum of the edges' cost of a cycle.
    \item A $m$D Euclidian TSP$_n$ is a special case of a TSP$_n$ where
    $c(i,j) = \hbox{distance}(i,j)$ $\forall i,j \in V_n$, and the vertex $i \in
    \Real^m, m  \geq 2, \forall i=1,\ldots,n$.

\end{enumerate}
\end{rem}

\begin{prop}
Any GAP$_n$ has a solution.
\begin{proof}
Without lost of generality, it is assumed to look for a minimum.
$f$ is evaluated on all the cycles of $G_n$. The cycles corresponds to
the finite permutations of the $n$ vertices of $V_n$.
Then, the image of $f$ is
discrete, therefore $f$ reach its minimum in a cycle.
\end{proof}
\end{prop}

\begin{prop}
For $f$ as minimum or maximum edge압 cost, then
the cycle of the GAP$_n$ which is the solution can be found in polynomial time.
\begin{proof}
These are trivial cases where the complexity is at most $n(n-1)$
which corresponds to search into the entries of the cost matrix $c(i,j)$.
\end{proof}
\end{prop}

Notes. The last proposition defines in clear way what an
algorithm must look for, but how to calculate and prove the
optimality is quite different. Nevertheless, for any objective function $f$,
if we have the  discrete image of
$f$, GAP$_n$ is a computable problem and the minimum selection procedure gives the
solution. The minimum selection procedure in mathematical notation
is:
$$
y^{ \ast } = \arg \min_{y=(v_1,v_2,\ldots,v_n,v_1),
v_1 \in V_n, v_i \in V_n, v_i \neq v_j, 1\leq i < j \leq n} \left\{ f(y) \,| \,
\hbox{GAP}_n=(G_n,c,f), G_{n}=(V_n,A) \right\}
$$

Properties of GAP$_n$ derived from its graph's
properties are presented in the following propositions.

\begin{prop}~\label{Prop:GAP_basic_properties}
Any GAP$_n$ has
\begin{enumerate}
    \item $n(n-1)$ edges.
    \item $(n-1)!$ complete cycles. The complete cycles can be
    enumerated in $n$ different ways. Hereafter, a cycle means a
    complete cycle.
    \item For $n\geq 4$, the maximum number of coincident edges of
    two different complete cycles is (n-3).
\end{enumerate}

\begin{proof}
\begin{enumerate}

    \item By induction, it is immediately. But the next data structure helps to prove. For
    computational purposes a function to identify the edges with integer values, $(1,2) \leftrightarrow  1$,
        $(2,1) \leftrightarrow -1$,      $(1,3) \leftrightarrow 2$,
         $(3,1) \leftrightarrow -2,$ $\ldots$,
    is also depicted in the following matrix:
    \begin{eqnarray}
\begin{array}{c|ccccc}
  Vertex   & 1               & 2                 & 3      & \hdots & n
  \\ \hline
  1        & X               & 1                 & 2      & \hdots & (n-2)(n-1)/2+1 \\
  2        & -1              & X                 & 3      & \hdots & (n-2)(n-1)/2+2 \\
  3        & -2              & -3                & X      & \hdots & \vdots         \\
  \vdots   & \vdots          & \vdots            & \vdots & \ddots & \vdots         \\
  n        & -(n-2)(n-1)/2+1 & -(n-2)(n-1)/2+2   & \hdots & \hdots &  X             \\
\end{array}~\label{eq:mat_identify_edgesbyInteger}
    \end{eqnarray}

     The matrix's elements correspond to the edges,
      but the diagonal's elements, which are marked by $X$,
     therefore the number of edges are $n^2-n=n(n-1)$.

    \item The first vertex and the last vertex are the same, therefore there are $(n-1)!$ different sequences of vertices (paths)
    of the remained (n-1) vertices. On the other hand, the first vertex can be
    selected in $n$ different ways, therefore the complete cycles
    can be enumerated in $n$ different ways.
    \item Without lost of generality the complete cycles can be depicted in descent order.
    Then two consecutive paths are $(n,n-1,\ldots,4,\textbf{3,2,1,n})$, and
    $(n,n-1,\ldots ,4,\textbf{3,1,2,n})$ which match on $(n-3)$ edges.
\end{enumerate}
\end{proof}
\end{prop}

Some important differences between GAP$_n$ and TSP$_n$ are in the
following proposition.

\begin{prop}~\label{Prop:GAP_vs TSP}
The objective function $f$ is $c$, and let $c(p)$ be the cost function of $p$, i.e.,
it is  given by the summation of edge압
cost of the consecutive pairs of vertices of $p$.
Let $l$ be the length path function, i.e., $l(p)=k$ where $p$ is a path,
 and k is the number of vertices of $p$.

\begin{enumerate}
\item For TSP$_n$, $c$ is monotonically increasing in the following way, $c(p_1)
\leq c(p_2)$ where the sequence of vertices of $p_1$ is a subsequence of
$p_2$ (hereafter, $p_1$ is a sub-path of $p_2$), and $l(p_1) < l(p_2)$.
\item For GAP$_n$, $c$ is not monotonically increasing as in TSP$_n$.

\item Let $c' \in \Real $, then for TSP$_n$, $P(c(p_2) \geq c'|
c(p_1) = c')=1$ where $P$ is the probability function, $p_1$ is a sub-path of $p_2$.

\item Let $c' \in \Real$, then for GAP$_n$. $P(c(p_2) \leq c' |
c(p_1) = c')>0$ where $P$ is the probability function, $p_1$ is a sub-path of $p_2$.
\end{enumerate}
\begin{proof}
\begin{enumerate}
\item It is monotonically increasing because $c(i,j) \geq 0$, and
$c(p_2)$ has some edges` cost more than $c(p_1)$. \item For an
arbitrary GAP$_n$, if $c(\cdot)$ is monotonically increasing then it is
a TSP$_n$. \item For TSP$_n$, the monotonically increasing of
$c(\cdot)$ implies that if a sub-path reach a value $c'$ for sure any
path containing it has a greater or equal value than $c'$. \item
Because the previous case, $c(\cdot)$ for GAP$_n$ is not  monotonically
increasing. Then the probability of a descend from a reference value
$c(p_1) = c'$ on a path $p_2$, where $p_1$ ia sub-path of $p_2$ is not
zero.
\end{enumerate}
\end{proof}
\end{prop}

\begin{definition}~\label{eq:funtionEdgesToInteger}
The function to identify the edges is $e:\mathbb{N} \times
\mathbb{N} \to \mathbb{Z}$  (it was depicted in
equation.~\ref{eq:mat_identify_edgesbyInteger})

$$e\left( i,j\right) =
\left\{
\begin{array}{cc}
  i<j & (j-2)(j-1)/2+i \\
  \text{otherwise} & -(i-2)(i-1)/2+j \\
\end{array}%
\right. $$
\end{definition}

The Research Space of GAP$_n$ is finite and numerable, and it has
$(n-1)!$ elements and can be enumerated in $n$ different ways (see
Prop.~\ref{Prop:GAP_basic_properties}). An example of an enumeration in
descendent order of the vertices is the following.
$$\begin{array}{ccc}
  \mathbb{N} &                 & \text{Cycle} \\
  1          & \leftrightarrow & (n,n-1,\ldots, 2, 1, n) \\
  2          & \leftrightarrow & (n,n-1,\ldots, 1, 2, n) \\
  \vdots     & \vdots          & \vdots \\
  (n-1)!-1  & \leftrightarrow  & (n,1, \ldots,  n-1, n-2, n) \\
  (n-1)!    & \leftrightarrow  & (n,1, \ldots,  n-2, n-1, n) \\
\end{array}
$$

Hereafter, this numeration is assumed.
For any GAP$_n$, given $n$ and $j>0$, a natural number
between 1 and $(n-1)!$, the next algorithm computes the
corresponding cycles assuming descending order.

\begin{algorithm}~\label{alg:cylclesNumerationDescendig}
\textbf{Input:} j, n. \textbf{Output:} $y$ cycle of GAP$_n$ (array of
$\Natural^{n+1}$)

\begin{enumerate}
    \item ~ V$_a$:=$(n-1, n-2,\ldots,2,1);$ (array of
    $\Natural^n$)
    \item ~ $y(1)=n;$
    \item ~ $y(n+1):=n;$
    \item ~ $j_a := j-1;$
    \item ~ $n_r$ := $n-1;$

    \item ~ \textbf{for} $k:= 1$ \textbf{to} $n-3$ \{

    \item~\hspace{0.5cm}     $d := (n-k-1)!;$

    \item~\hspace{0.5cm}    $r := \hbox{fix} (j_a / d) + 1;$

    \item~\hspace{0.5cm}     $j_a := \hbox{mod}(j_a, d);$

    \item ~\hspace{0.5cm} $y(k+1) := V_a(r);$
    \item ~\hspace{0.5cm} $V_2 := (1,\ldots,n_r-1);$ (array of
    $\Natural^n_{r}-1)$)
    \item ~\hspace{0.5cm} $k_2 := 1;$

    \item~\hspace{0.5cm} \textbf{for} $k_3$ := 1 \textbf{to} $n_r$ \{
            \item~\hspace{1.0cm} \textbf{if} $r$ $\neq$ $k_3$
            \textbf{then}
            \item~\hspace{1.5cm} $V_2(k_2) := V_a(k_3);$
            \item~\hspace{1.5cm} $k_2 := k_2 + 1;$
    \item~\hspace{0.5cm} \} (end for $k_3$)
    \item~\hspace{0.5cm} $n_r := n_r - 1;$
    \item~\hspace{0.5cm} $V_a := V_2;$

\item ~ \} (end for $k$)

\item ~ \textbf{if} $j_a$ == 0 \textbf{then}

\item ~\hspace{0.5cm} $y(n-1) := V_a(1);$

\item ~\hspace{0.5cm} $y(n)  := V_a[2];$

\item ~ \textbf{else}

    \item~\hspace{0.5cm} $y(n-1) := V_a(2);$
    \item~\hspace{0.5cm} $y(n)   := V_a[1];$
\end{enumerate}
\end{algorithm}

The next propositions shows that the solution of any GAP$_n$ can
be mapped to first position of a given enumeration of the research
space of an equivalent GAP$_n$.

\begin{prop}~\label{prop:Cycle_for_numering}

Let be $[i_k]_{k=1}^{n+1}$ a given cycle  of a GAP$_n$. Then
$\exists$! $m:[1,\ldots,n]\rightarrow [1,\ldots,n]$ such that the
cost of the cycles do not change for an equivalent GAP'$_n$ with
cost function given by $ c \circ  m^{-1} = c(m^{-1}(\cdot))$ and
$[m(i_k)]_{k=1}^{n+1}=[n,n-1,\ldots,1,n]$, which corresponds to
the first cycle of the descent enumeration of the vertices
beginning with $n$.

\begin{proof}
Note that the first $n$ vertices of the given cycle
$[i_k]_{k=1}^{n+1}$ correspond to a permutation of
$[1,2,3,\ldots,n]$. Therefore the unique $m$ is the mapping $i_1
\leftrightarrow n$, $i_2 \leftrightarrow n-1$, $\ldots$, $i_n
\leftrightarrow 1$. By construction the cost in the equivalent
GAP'$_n$ is the original cost of the GAP$_n$. Let be
$[l_k]_{k=1}^{n+1}$ an arbitrary cycle of the equivalent GAP'$_n$,
its cost is given by
$f(c(m^{-1}([l_k]_{k=1}^{n+1})))=f(c([j_k]_{k=1}^{n+1}))$, where
$m^{-1}([l_k]_{k=1}^{n+1})$ = $[j_k]_{k=1}^{n+1}$.
\end{proof}
\end{prop}

\begin{prop}~\label{prop:enumeraCycles_descending_n_first}

Let be $[i_k]_{k=1}^{n+1}$ the solution cycle  of a GAP$_n$. Then
$\exists$! $m:[1,\ldots,n] \rightarrow [1,\ldots,n]$ such that
$[n,n-1,\ldots,1,n]$ is the solution of the equivalent GAP$_n$,
which corresponds to the first cycle of the descent enumeration of
the vertices beginning with $n$.

\begin{proof}
Let be $m(\cdot)$ constructed as in the previous proposition. The cost
function of the equivalent GAP$_n$ is $ c(m^{-1}(\cdot))$. Let be
$[l_k]_{k=1}^{n+1}$ an arbitrary cycle in the equivalent GAP$_n$,
with $m^{-1}([l_k]_{k=1}^{n+1})$ = $[j_k]_{k=1}^{n+1}$. Therefore
the optimality follows from
$f(c(m^{-1}([n,n-1,\ldots,1,n])))=f(c([i_k]_{k=1}^{n+1}))$,
 and $f(c([i_k]_{k=1}^{n+1}))$ $\leq$
$f(c([j_k]_{k=1}^{n+1}))$ $\forall [j_k]_{k=1}^{n+1}$ cycle of
GAP$_n$.
\end{proof}
\end{prop}

\begin{prop}
Let be GAP$_n$ such that $c(i,j)$ is given by the following matrix
 $$%
\left\{ \begin{array}{ c c c c c}
          & 1                  &     2          & \cdots   & n-1          \\
   n      &                    &  n \cdot 2     & \cdots   & n \cdot n-1  \\
  \vdots  & \vdots             &  \vdots        & \vdots   & \vdots       \\
  n^{n-2} & n^{n-2}\cdot 2     &  \cdots         &          & n^{n-2}\cdot (n-1)  \\
  n^{n-1} &  n^{n-1}\cdot 2    &  \cdots       & n^{n-1}\cdot (n-1) &   \\
\end{array}%
\right\}
$$
then it has a unique solution and all cycles have different cost.

\begin{proof}
The cycles of GAP$_n$ correspond to a unique combinations of its
edges. The elements of the matrix $c(i,j)$ are all different
monomials on the numerical base $n$. Therefore, any cycle has a
unique cost given by its unique polynomial combination of $n$
terms $c(i,j)$ on the numerical base $n$.
\end{proof}
\end{prop}

\section{Integer Problem Formulation of GAP}~\label{sc:integer}

A given a GAP$_n$ can be formulated as an integer problem
(IPGAP$_{n}$) as follow:
\begin{eqnarray*}
& \min \sum_{i=1}^n \sum_{j=1}^n c_{ij}x_{ij} \\
\hbox{s.t.} & \\
& \sum_{j=1}^n c_{ij}x_{ij}=1, \forall i=1,\ldots,n, \\
& \sum_{i=1}^n c_{ij}x_{ij}=1, \forall j=1,\ldots,n, \\
& \sum_{i=1}^n c_{ii}x_{ii}=0,\\
& \{0, 1\} \in x_{ij}, \forall i,j,
\end{eqnarray*}
where $c_{ii}$ are the entries of matrix of cost, and $x_{ij}$ are
the variables associated to an edge with vertices $i$, $j$.

\begin{prop} A cycle of GAP$_n$ $\Leftrightarrow$ a feasible point of IPGAP$_{n}$.
\begin{proof}
$\Rightarrow$. Let $v_{1},v_{2},\ldots,v_{n},v_{1}$ the vertices a
cycle of  GAP$_n$. The corresponding point IPGAP$_{n}$ is given by
$$x_{ij}=\left\{%
\begin{array}{cl}
  1 & \hbox{if\ }  i=v_{k} \hbox{\ and \ } j=v_{k+1}, k=1,\ldots,n \\
  0 & \hbox{otherwise.} \\
\end{array}%
\right. .$$

For any $i=1,\ldots,n$, it exists a unique $v_{k}=i$, then
$\sum_{j=1}^n  x_{ij} = x_{v_{k}v_{k+1}}=1$. For any
$j=1,\ldots,n$, it exists a unique $v_{k+1} =j$, then
$\sum_{i=1}^n  x_{ij} = x_{v_{k} v_{k+1}}=1$. Finally, a cycle has
different consecutive vertices, $\sum_{i=1}^n c_{ii}x_{ii}=0$
Therefore $\{ x_{ij}\}$ is feasible.

$\Leftarrow$. Let be $\{ x_{ij}\}$ a feasible point of
IPGAP$_{n}$. This means that there are zeros in the diagonal and
there is a unique $1$ by column and by row in $\{ x_{ij}\}$. Then
the corresponding cycle is constructed by taking
$$v_{i_k}=\left\{%
\begin{array}{cl}
  v_1=1 &   \\
  v_{k+1}=j & \hbox{where\ } x_{v_{k}j}=1, k=1,\ldots,n  \\
\end{array}
\right. $$ By induction over $n$. For $n=2$ the unique cycle of
GAP$_2$ corresponds $1,2,1$ given by the previous formula on the
matrix of the feasible point of IPGAP$_2$
$\left(%
\begin{array}{cc}
  0 & 1 \\
  1 & 0 \\
\end{array}%
\right) $.

For $n=3$ the two cycles of GAP$_3$ corresponds $1,2,3,1$, and
$1,3,2,1$ are given by the previous formula on the matrices of the
feasible points of IPGAP$_3$
$\left(%
\begin{array}{ccc}
  0 & 1 & 0\\
  0 & 0 & 1\\
  1 & 0 & 0
\end{array}%
\right) $, and $\left(%
\begin{array}{ccc}
  0 & 0 & 1\\
  1 & 0 & 0\\
  0 & 1 & 0
\end{array}%
\right)$. Also the number of feasible points for IPGAP$_3$ is
$2*1=2!$. The hypothesis  for $n$ is that all $(n-1)!$ feasible
points give all the cycles. For $n+1$, first a feasible point of
IPGAP$_n$ is taken, then a cycle is constructed
$v_{1},v_{2},\ldots,v_{n},v_{1}$, also a column, and a row are
added. On the resulting matrix $\{ x_{ij}
\}_{i,j=1,\ldots,n,(n+1)}$, the unique entry for which $x_{in}=1$
is set to zero, and for that $i$, the entry $x_{i(n+1)}$ is set to
$1$, and finally the entry $x_{(n+1)n}$ is set to 1. With these
changes the point  $\{ x_{ij} \}_{i,j=1,\ldots,n,(n+1)}$ is a
feasible point of IPGAP$_{(n+1)}$, and it has a  corresponding
cycle of GAP$_{(n+1)}$
$v_{1},v_{2},\ldots,i,n+1,n,\ldots,v_{n},v_{1}$. The last step can
be done $n$ times because $x_{(n+1)(n+1)}=0$. Therefore $n!$ is
the total number of feasible points of IPGAP$_{(n+1)}$.
\end{proof}
\end{prop}

\begin{rem}
By the previous proposition, the minimum number of vertices in order to
change a cycle are two, this means 3 edges, or in term of IPGAP$_n$, 3
variables of $x_{ij}$ need to be replaced to pass from one feasible
point to another. This means that a pivot method for solving IPGAP$_n$
needs to change at the same time at least three active variables. For
TSP$_n$ the cost function is monotone, this can help to solve the
problem with pivot's strategies with limited variables (see
proposition~\ref{Prop:GAP_vs TSP}). For GAP$_n$ the possibility of
missing the solution using pivot's strategies limited by the number of
variables makes hard to solve GAP$_n$ by integer programming methods.

\end{rem}

\begin{prop} The GAP$_n$ and IPGAP$_{n}$ are equivalent.
\begin{proof}
It is immediately from the previous proposition.
\end{proof}
\end{prop}

To visualize the cost matrix, two linear transformation are used to map any cost matrix of
GAP$_n$ into $[-1,1]^n$, or $[0,1]^n$. Scale transformation:
$$ c'(i,j)=\left\{%
\begin{array}{cl}
  1 &  \hbox{if\ } c(i,j) = \inf \\
  c(i,j)/s^+ & \hbox{if\ } c(i,j) \geq 0\\
   c(i,j)/s^- & \hbox{otherwise} \\
\end{array}%
\right.
$$
where $s^+ = \max_{i,j=1,\ldots,n, c(i,j) \neq \inf} c(i,j)$, and
$s^- = \min_{i,j=1,\ldots,n} c(i,j).$

Scale and translation transformation:
$$c'(i,j) = \left\{%
\begin{array}{cl}
  1 &  \hbox{if\ } c(i,j) = \inf \\
  (c(i,j)-m)/s &\hbox{otherwise} \\
\end{array}%
\right.
$$
where $m = \min_{i,j=1,\ldots,n} c(i,j)$, $M =
\max_{i,j=1,\ldots,n, c(i,j) \neq \inf} c(i,j)$, and $s=M-m$.

Section~\ref{sc:Reducibility} contains images of matrices using the last
transformation. The idea is to study the cost matrix and an associate vertex's index matrix
to explore pivot's strategies. Because, the diagonal of the cost matrix is $\inf$
 the diagonal was omitted for an easy interpretation on a gray scale of color
$[0,1]$, where minimum values correspond to black (0), and maximum
values correspond to white(1). A sorted matrix $\mathcal{M}$ depicts a
relation between edge's cost and vertex's number. By the moment,
pivot's strategies are not proposed but
algorithm~\ref{alg:VerificationSol} can be modified for integer
programming approach. The next proposition presents a version of a
necessary optimality condition for IGAP$_n$.
\begin{prop}
Given IGAP$_n$. $x_{ij}^*$ is its solution, if for any pivot's strategy
that suggest any set of $x_{ij}$ feasible (i.e., a sub-path or a cycle of the equivalent GAP$_n$)
with cost less than $t^*=\sum_{i=1}^n \sum_{j=1}^n c_{ij}x_{ij}^*$, they can
not be inserted to replace appropriately the variables  of $x_{ij}^*$ with diminish $t^*$.
\begin{proof}
It is immediately by the global optimality of $y^*$.
\end{proof}
\end{prop}

The next section presents a well known computational model to
apply for defining an algorithm to verify the solution of GAP$_n$.

\section{Turing Machine and Finite Automata for GAP$_n$}~\label{sc:MTFiniteAutomataGAP}
This section relates two simple computational models to resolution
of GAP$_n$. The section~\ref{sc:Reducibility} will present the
algorithm ``The Cycle generator'' that is a  Turing Machine (TM).
To justify and analyze the complexity of
an algorithm for checking the solution of  GAP$_n$ a simple variations of the problem
to locate a mark on an infinity or finite tape are presented. This
is done in detailed way to distinguish the importance of the
existence of properties to analyze  problems, constrains, data, and
structures.

P1. It consists to locate a mark in an infinity tape (or memory).

The importance of properties and well defined hypothesis is quite
important. What if for P1 there is no mark! No matters Finite Automata (FA) or TM the
problem is not computable. Hereafter, the existence of a mark is
assumed. Other practical matter is how the mark is putted, i.e., how
the problem can be constructed, what process or person can manages an
infinity tape. But this is done by a assuming and accepting,
 that this problem is a mental experiment.

P1 is perhaps the ideal, easy and simple problem that a TM
can compute but a FA can not. There are two reason for the success of the TM
 over the FA model for this problem, the first
one is the ability to move forward and backward, the second is the
interaction with the tape or memory that allow it to remember what memory has been explored.
It can be argued that a FA can not be used for this problem but for the problem that the mark is on the
direction that the FA can analyze. If FA depends from an special configuration that need to
know in what direction is the mark is not the same problem, knowledge has been added in order that the FA can solve a similar
but it is a different problem to locate a mark in a given direction.

P2. It consists to locate a mark from the origin of an infinity tape.

Now, for solving P2, a FA is sufficient. A TM can solve the problem but
no memory interaction and backward are needed.

P3. It consists to locate a mark from the origin of an finite tape.
Again, for solving P3, a FA is sufficient. A TM can solve the problem
but no memory interaction and backward are needed.

A TM has a halting property no mater where the mark is located for the
three problems. However, there is a crucial difference, because
computability, i.e., that it is for sure that a TM can always give a
solution for these problems. A FA can only solve P2 and P3 with
assuming that there is a mark. It can be also showed that the TM has
not advantages using its ability to interact with the tape than a FA.
The simple algorithm to locate a mark is:

\begin{enumerate}
    \item if the mark is found then ``here is the mark'' stop,
    \item otherwise if there is an available memory's space then skip,
    \item otherwise stop ``there is not mark''.
\end{enumerate}

P4. This problem consist in determine the minimum of a finite set
of marks (each one has a associate an integer value) on a finite
memory.

For P4, there is a  TM, that looks the memory, then if there is a mark
then stores and keep this mark when it corresponds to the minima, until
all memory is reviewed. The main hypothesis are a finite number (but
unknown number) of marks, and each mark correspond to an integer value.
The existence of the minima is guaranty by elemental Mathematical
Analysis, therefore P4 is computable. Without more assumptions this
last problem can not finish before to look for all the memory. Its
complexity depend of the length of the tape. Moreover, the previous TM
that can compute P4, and there is no properties to gain efficiency. If
an  TM claims to solve P4 in efficient time  for any instance of P4. A
practical numerical test can be formulated  for verification where the
claimed TM must has a probability of finding the solution equal one. If
this is the case, any arbitrary set of arbitrary instances of P4
selected for this test are related but arbitrary! What if with
appropriate assumptions or properties it is possible to formulate
algorithms with high probability to determine a putative solution
founded in efficient time. This opens an interesting line of research
for probabilistic algorithms, they trade off between putative or
approximate the solution and early stop determined by efficiency (no
more than polynomial iterations), this type of algorithm must be
addressed and studied but this is out of the scopes of this paper.

\begin{figure}
\centerline{
\psfig{figure=\IMAGESPATH/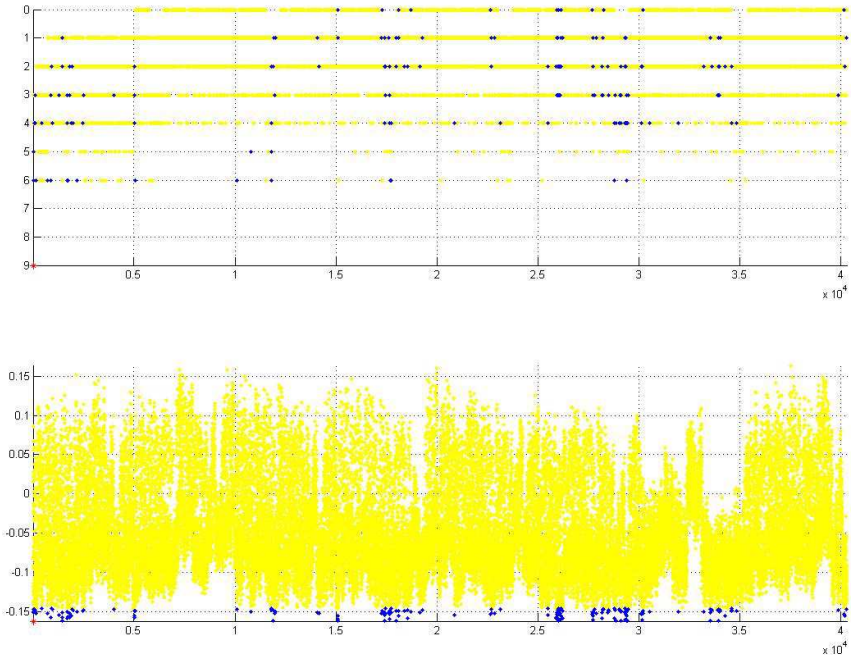,
height=100mm}}
 \caption{Edges' cost that coincide with the edges' cost of the optimal cycle,
 and cost each cycle numerated from 1 to 8! for an example of TSP$_9$.
 The first cycle is the solution, and it is depicted in red.
 Blue dots depicts cycles with cost closed to the optimal cost.
 }~\label{fig:coin_var_cities_09_land_cost_vs_Cycles}
\end{figure}

\begin{figure}
\centerline{ \vbox{
\hbox{\psfig{figure=\IMAGESPATH/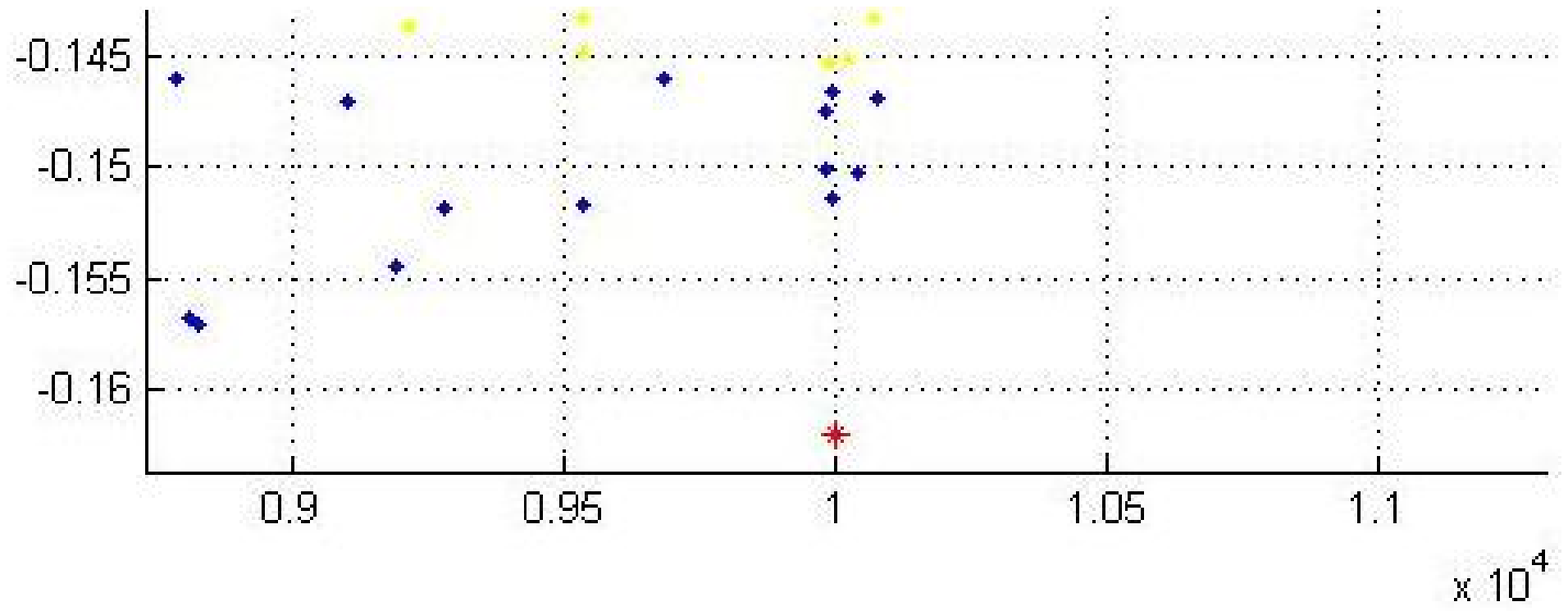, height=35mm}}
\hbox{\psfig{figure=\IMAGESPATH/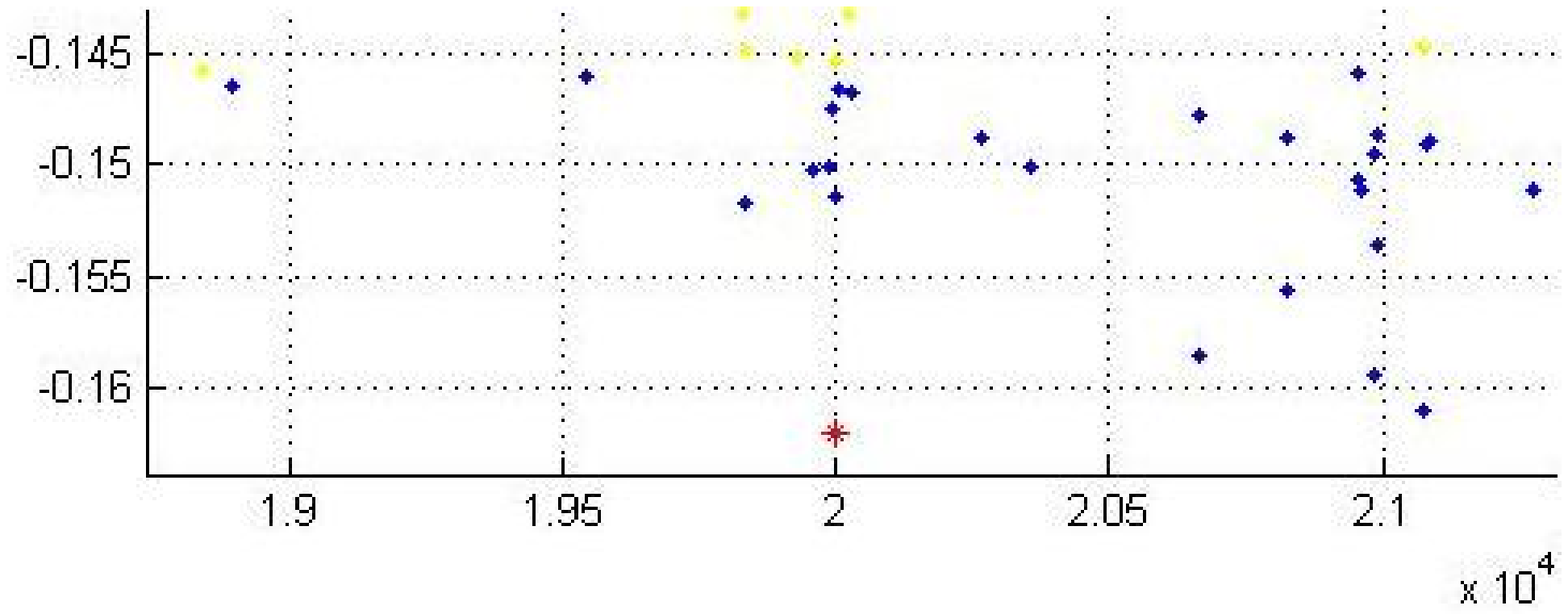, height=35mm}}
\hbox{\psfig{figure=\IMAGESPATH/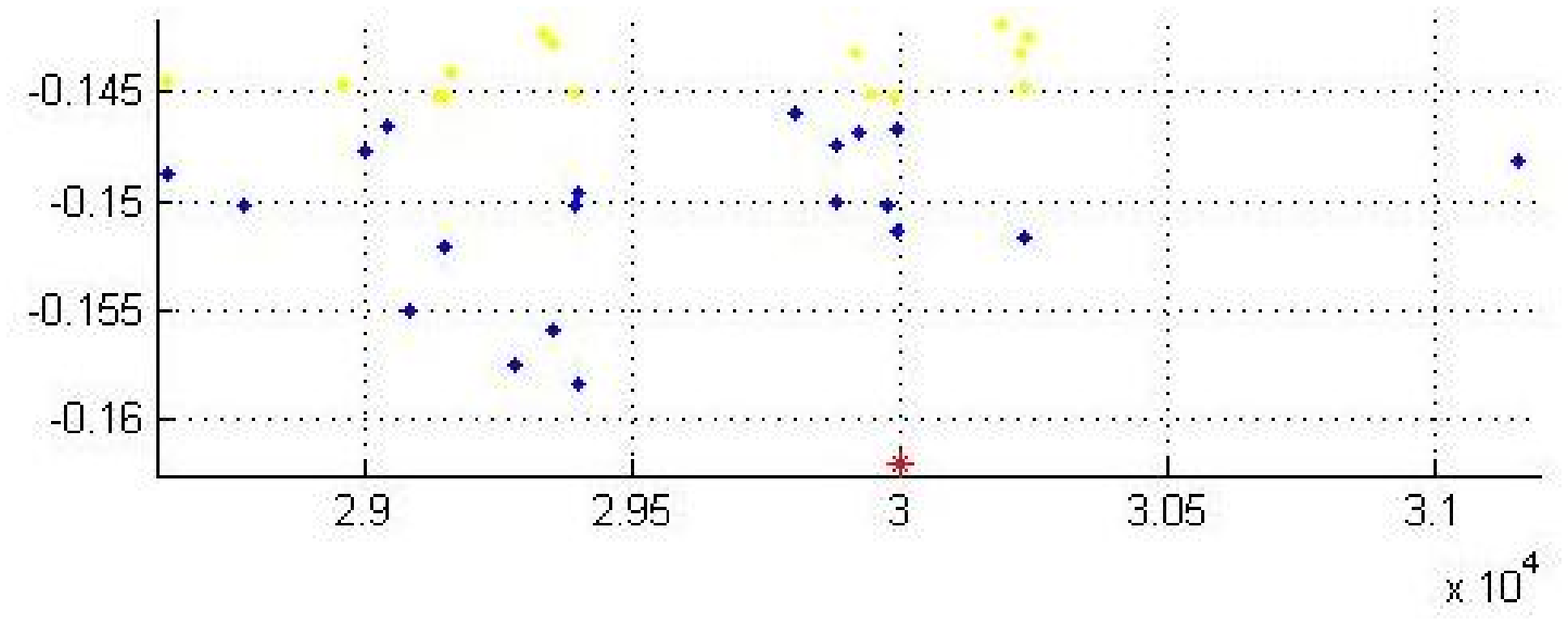, height=35mm}}}
 }
 \caption{Optimum cycle in positions = 10,000, 20,000, 30,000 of the
  numeration from 1 to 8! corresponding to cycles on descending order
  of the algorithm~\ref{alg:cylclesNumerationDescendig}
  for an example of TSP$_9$.
 The solution is depicted in red.
 Blue dots depicts cycles with cost closed to the optimal cost.
 }~\label{fig:sol_1000_200_300_cities_09_land_cost_vs_Cycles}
\end{figure}

\begin{enumerate}
    \item if the mark is found then ``here is the mark'' stop,
    \item otherwise if there is an available memory's space then skip,
    \item otherwise stop ``there is not mark''.
\end{enumerate}

\begin{figure}
\centerline{ \psfig{figure=\IMAGESPATH/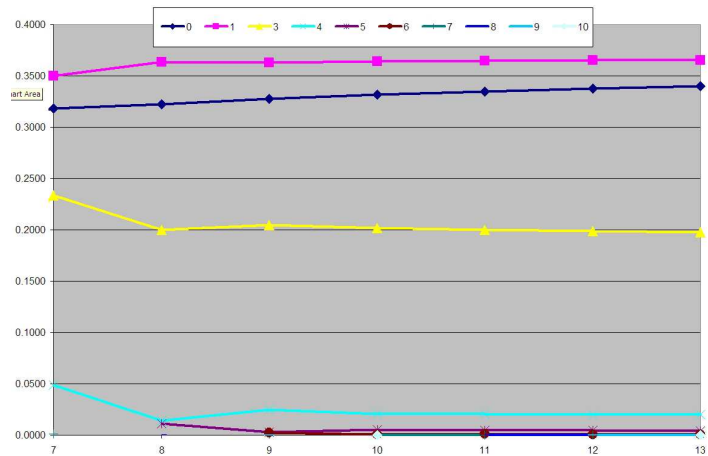,
height=60mm}}
 \caption{Edges' cost that coincide with the
 edges' cost of the optimal cycle for GAP$_n$, $n=7,8,9,10,11,12,13$.
 }~\label{fig:edge cost_vs_n 7 8 9 10 11 12 13}
\end{figure}

These computational models are without complicated instructions
but simple transitions driven by the input data. Its complexity is
not related to the number of transitions but by the length of
the tape or memory until finish to reviewing it.
Without properties or conditions, they can
be used instead of a more elaborated  algorithm.

Here, in order to solve GAP$_n$ an algorithm could keeps a record
of the generated cycles  but the number of marks requires exponential memory's size.
This is elaborated in following paragraphs.
First,  the lack of a local vicinity
optimal property on the cycles' cost makes
not necessary to keep a
record of the putative optimal cycles but to keep the best one.
The information of a local optimal cycle could not be useful to
predict possible paths to pursue the global optimal. Therefore a
TM capable to generate cycles and keep the putative optimal cycle
is sufficient.

On the other hand, for an special
cost matrix, i.e., matrix with properties, it is highly possible
to build an ad-hoc algorithm, that could solve and demonstrate the
solution, and if matrix's properties are related to the putative
solution then a record of the algorithm steps could improve the
research, and provides the justification of the solution.
For GAP$_n$ any
property or relation in the cost matrix can be nullified by an
arbitrary combination of its coefficients for large and arbitrary problems of it.

Figure~\ref{fig:coin_var_cities_09_land_cost_vs_Cycles} depicts
the landscape of TSP$_n$, i.e, the cycle's cost versus cycle's number
in descending order (see proposition~\ref{prop:enumeraCycles_descending_n_first}).
 Blue dots depict cycles
with closed values to the optimal. Notes that the blue dots are
spread, and there are some blue dots with zero coincidences with
the edges' cost of the optimal solution (examples of this are
depicted in the upper figure over the same vertical position of the cycles
(marked on blue) with cost closed
to the solution).
The optimal cycle was
mapped to first position (see proposition~\ref{prop:enumeraCycles_descending_n_first})
and its is depicted by a red dot in the
origin. It is noted that the colored dots (including the optimal)
are not a local minima and they could be closed or far away
between them. Also,
fig.~\ref{fig:sol_1000_200_300_cities_09_land_cost_vs_Cycles}
depicts how the optimum압 vicinity change when the same optimum
cycle is mapped on different positions of the descending order of
the algorithm~\ref{alg:cylclesNumerationDescendig},
the cycles' cost are the same but in different positions. The optimum압
vicinity is not the same besides  the cost matrix $C(i,j)$ is
permuted according to these cycles' numeration.
The change affects the
neighbors' cost of the optimal cycle because,
they share sub-paths but in different orders of the vertices
by the descending numeration.

The number of cycles grows exponentially and each cycle inherent
by the enumeration of the branches edges' cost but not necessarily
to form a global minima as it is depicted in the landscape of the
fig.~\ref{fig:coin_var_cities_09_land_cost_vs_Cycles}, and
fig.~\ref{fig:sol_1000_200_300_cities_09_land_cost_vs_Cycles}.

It is possible to solve $GAP_n$ until $n=14$ in a home made
computer (These numerical experiments run in a AMD Athlon 64X2
Dual core processor 5200+, 2.6GHz, 3.25 GB of RAM) by a direct
evaluation of the $13!$ different cycles computed by
algorithm~\ref{alg:cylclesNumerationDescendig} (some adjust are
need to use large integer numbers) but the execution's time
grows as it expects exponentially from few seconds (
$n=4,5,6,7,8,9$), 1.1 minutes  ($n=10$), 13.7 minutes ($n=11$),
2.70 hours (n=$12$), 1.5 days ($n=13$), 0.66 months ($n=14$), and
0.82 years ($n=15$). The solutions are verifying at the same time
by this exhaustive approach but it is completely impractical.

However, studying these small cases fig.~\ref{fig:edge cost_vs_n 7
8 9 10 11 12 13} is constructed. It depicts an remarkable relation
between the edge's cost of the different cycles respect to the first cycle.
It is interesting
to note that the number of edge's cost that no coincide at all or
coincide with only one of the edges cost of the first cycle has a
tendency approximately of 33\% and 36\% respectively with a
slightly growing(zero coincidence is the dark blue line, one
coincidence is the fuchsia line in the fig.~\ref{fig:edge
cost_vs_n 7 8 9 10 11 12 13}). This means 69\% of the cycles of
GAP$_n$ could not share or they are not related to the current
putative optimal cycle. To keep track in the memory of the visited cycles
requires to have a memory's size of 69\%$(n-1)!$, which is a huge
requirement. This
requirement makes not possible to keep track of the visited
cycles.

If it is possible to cut branches of the complete graph to reduce
the research space of a GAP$_n$, then it is not necessary to use
``if'' to cut branches that can not provide the solution. Even
more, there is not property or justification to define an early stop
after reach a putative solution for an arbitrary and large GAP$_n$ without
special properties.
In fact, because GAP$_n$'s objective cost
function is not monotonically increasing but oscillating, skip
cycles could drive to miss the solution, it is probable that a reference cost
diminish on a large path of
a given sub-path (see proposition~\ref{Prop:GAP_vs TSP}). For
TSP$_n$ an ''if얎 can be useful to cut cycles with sub-path's cost
above the current putative optimal cycle's cost. Also, this
descending property is the track to prove and justify the optimality of
the last stored optimal cycle.

It is a TM the appropriate computational model for a simple algorithm
to explore at full the GAP$_n$'s research space or a reduced research
space of it. In the next section the generator of putative cycles for
GAP$_n$ is a TM. It generates by enumeration of the vertices all cycles
remaining in the reduced research space, and the complexity only
relates to the number of them.

\section{Reducibility}~\label{sc:Reducibility}
\begin{figure}
\centerline{ \psfig{figure=\IMAGESPATH/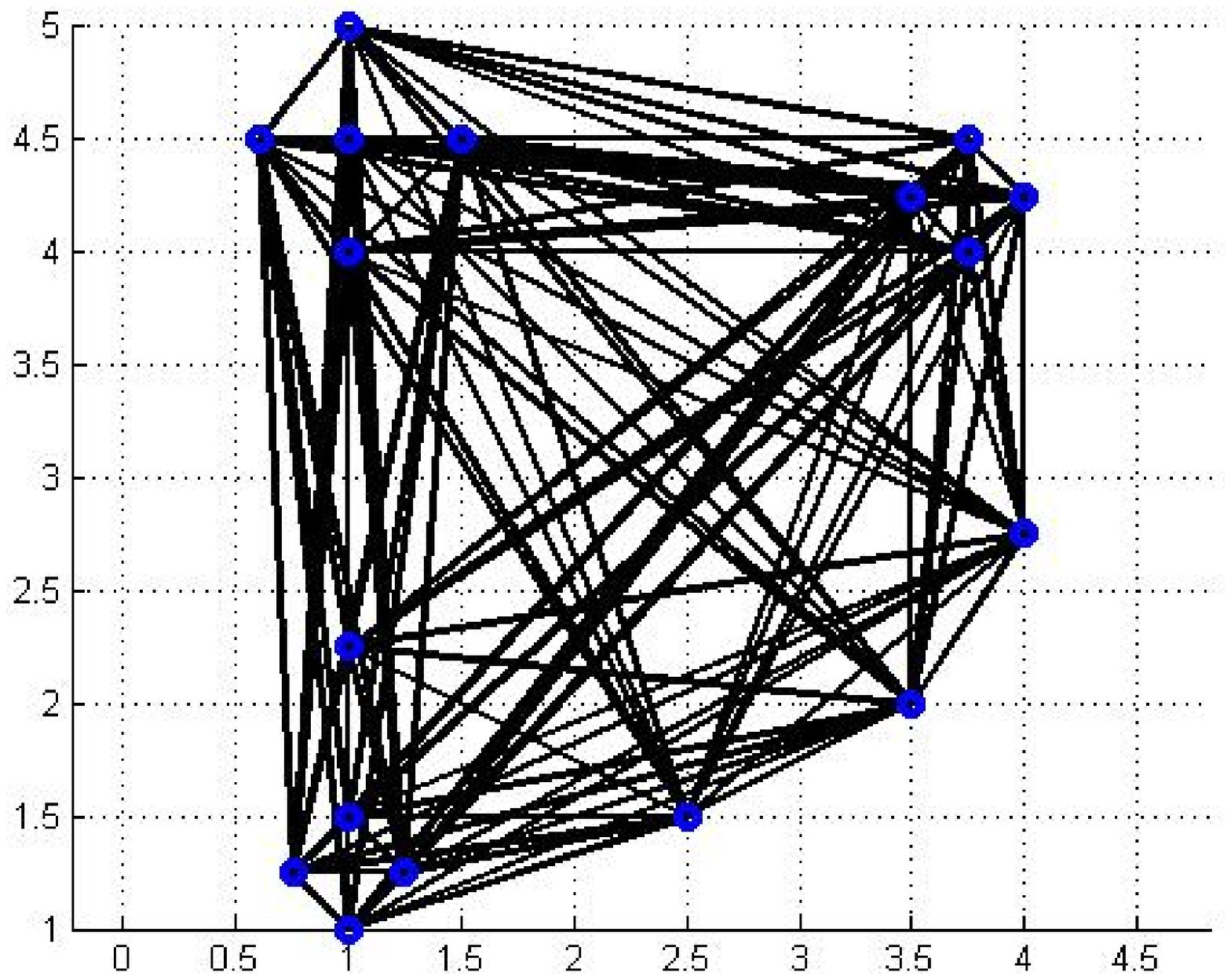,
height=50mm}
\psfig{figure=\IMAGESPATH/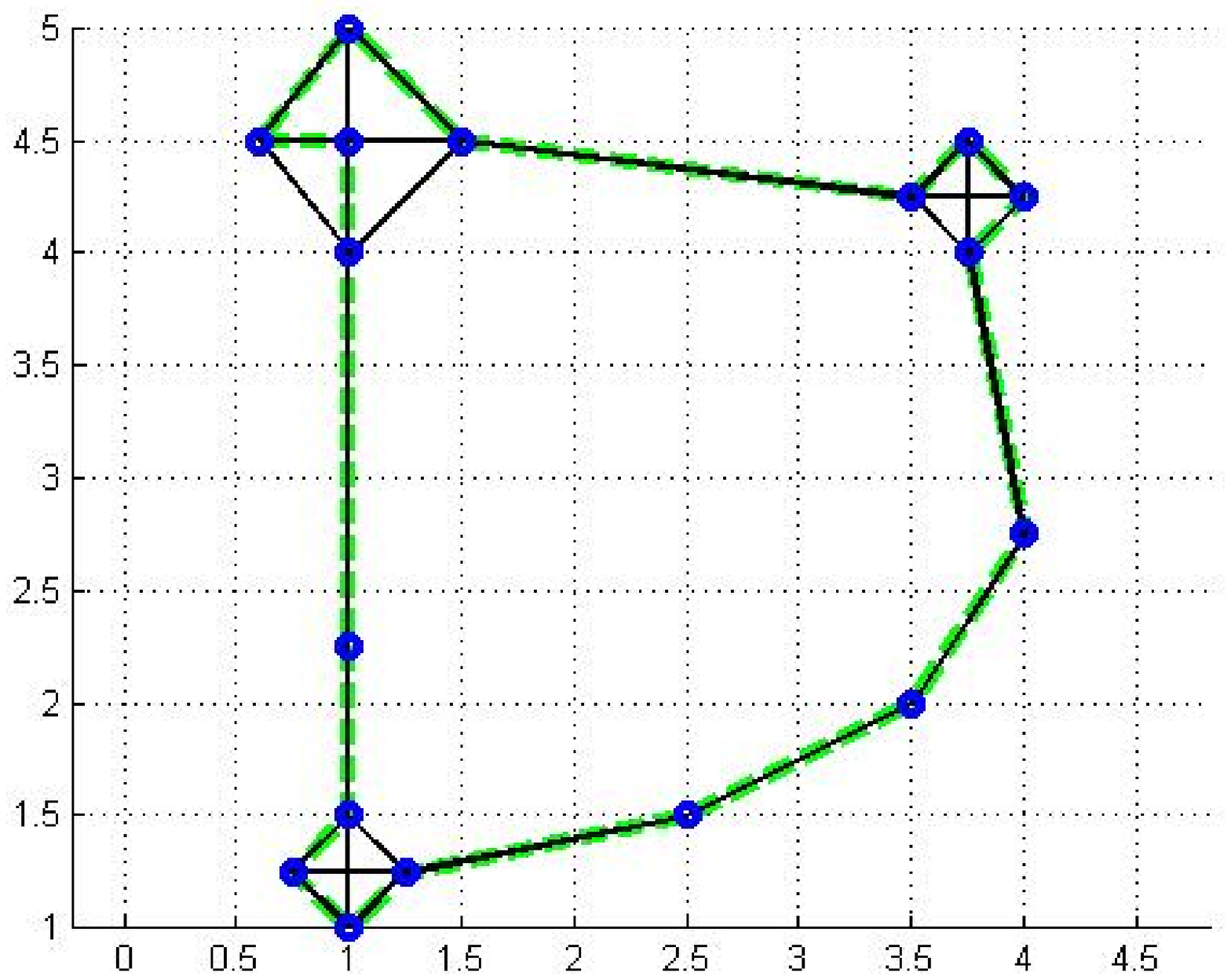,
height=50mm}}
 \caption{Complete and reduced equivalent graph of an 2D Euclidian
TSP$_{17}$.}~\label{fig:cmp_red_equ_graph}
\end{figure}

\begin{figure}
\centerline{
\psfig{figure=\IMAGESPATH/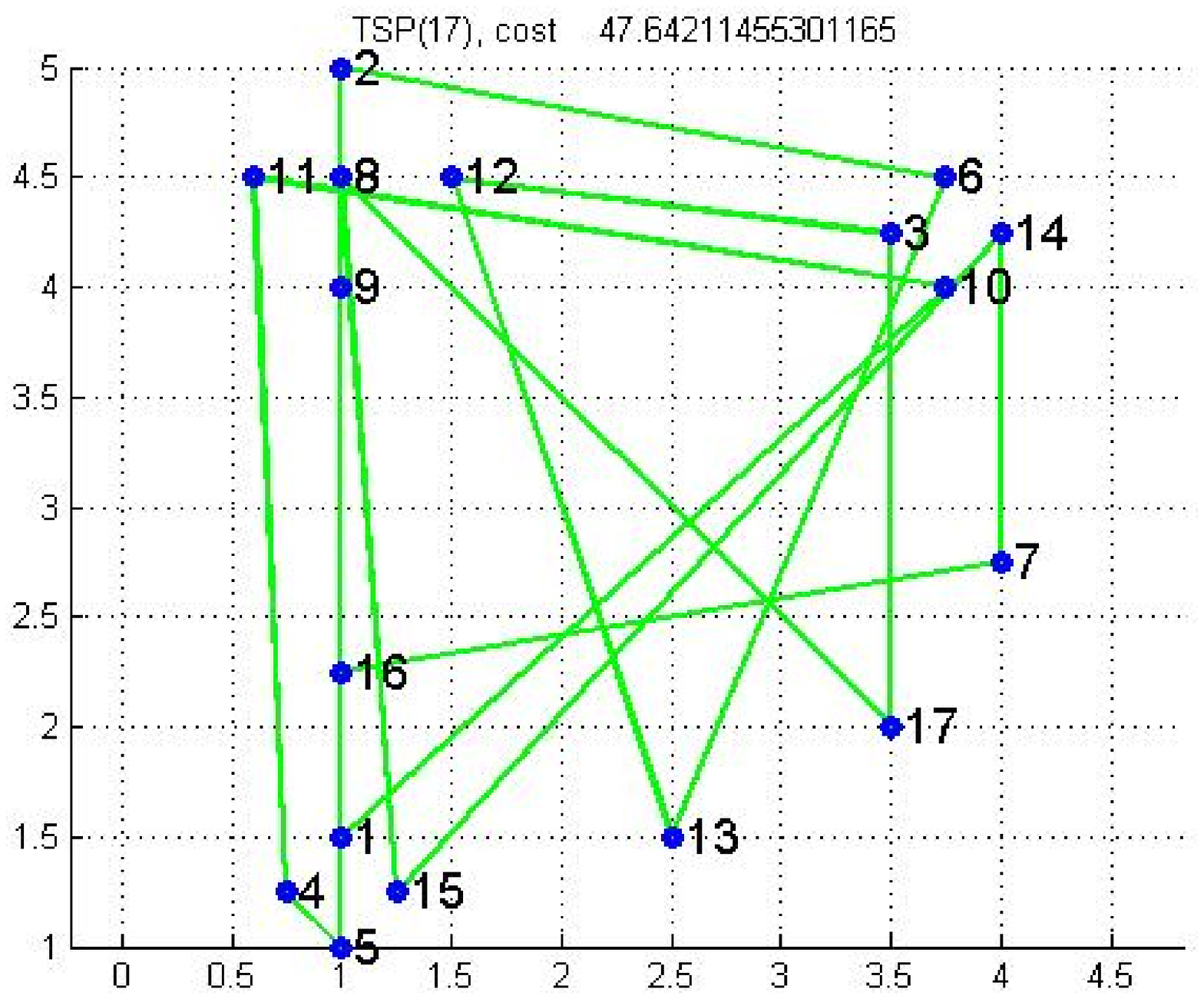,
height=50mm}
\psfig{figure=\IMAGESPATH/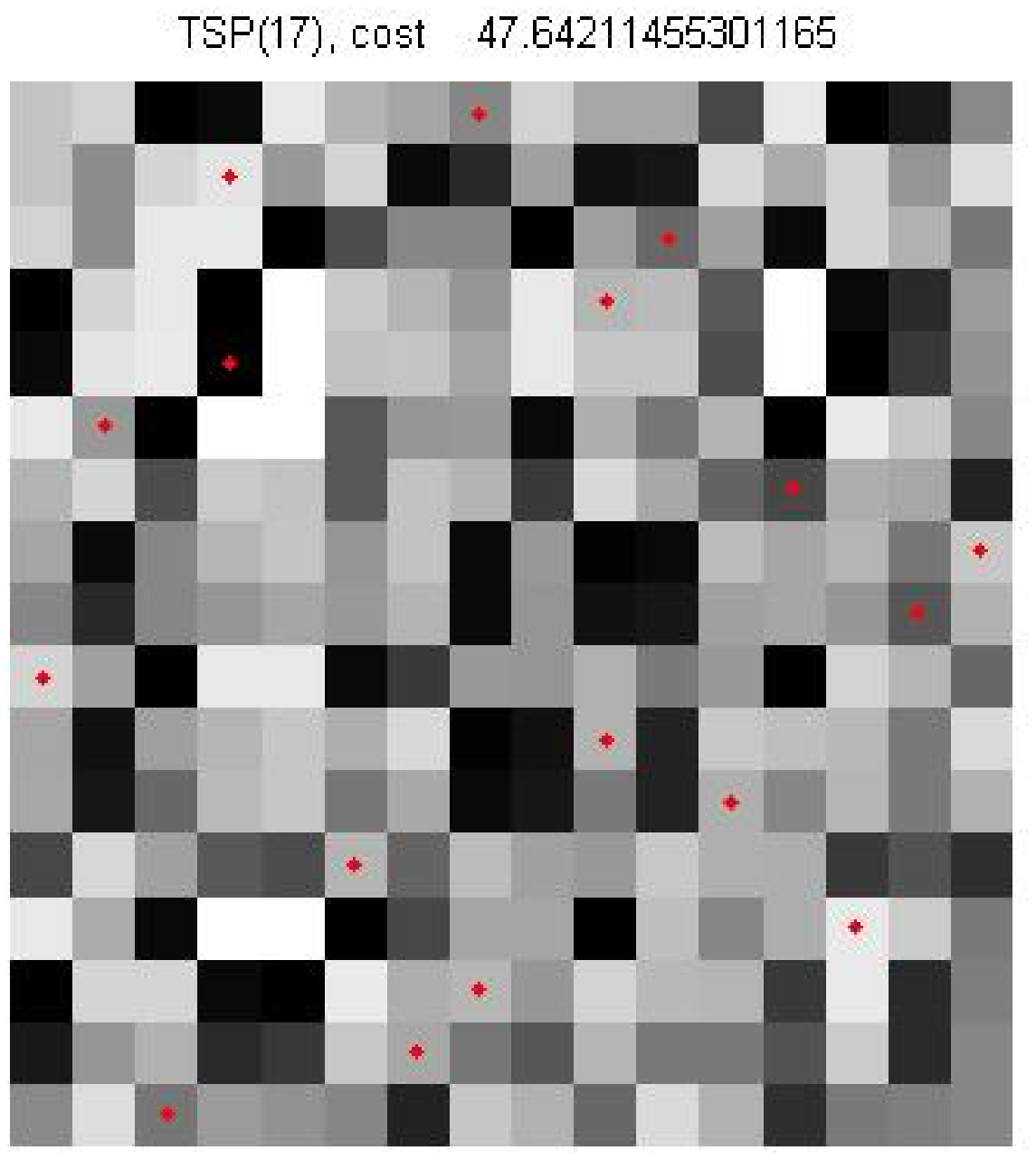,
height=50mm}}
 \caption{Initial numeration, an initial cycle (green lines),
 and cost matrix with initial cycle (red dots) for the example 2D Euclidian
 TSP$_{17}$.}~\label{fig:mat_red_equ_graph}
\end{figure}

\begin{figure}
\centerline{
\psfig{figure=\IMAGESPATH/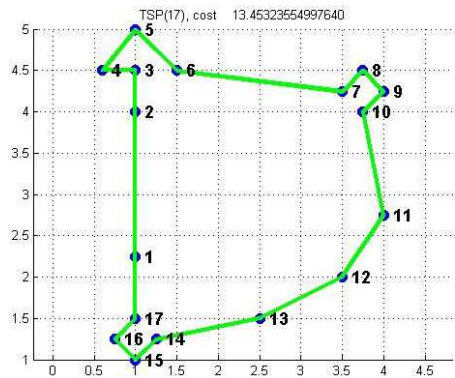,
height=50mm} }
 \caption{Solution (green lines) of the example 2D Euclidian
 TSP$_{17}$.}~\label{fig:sol_TSP17 num_mat_red_equ_graph}
\end{figure}

\begin{figure}
\centerline{
\psfig{figure=\IMAGESPATH/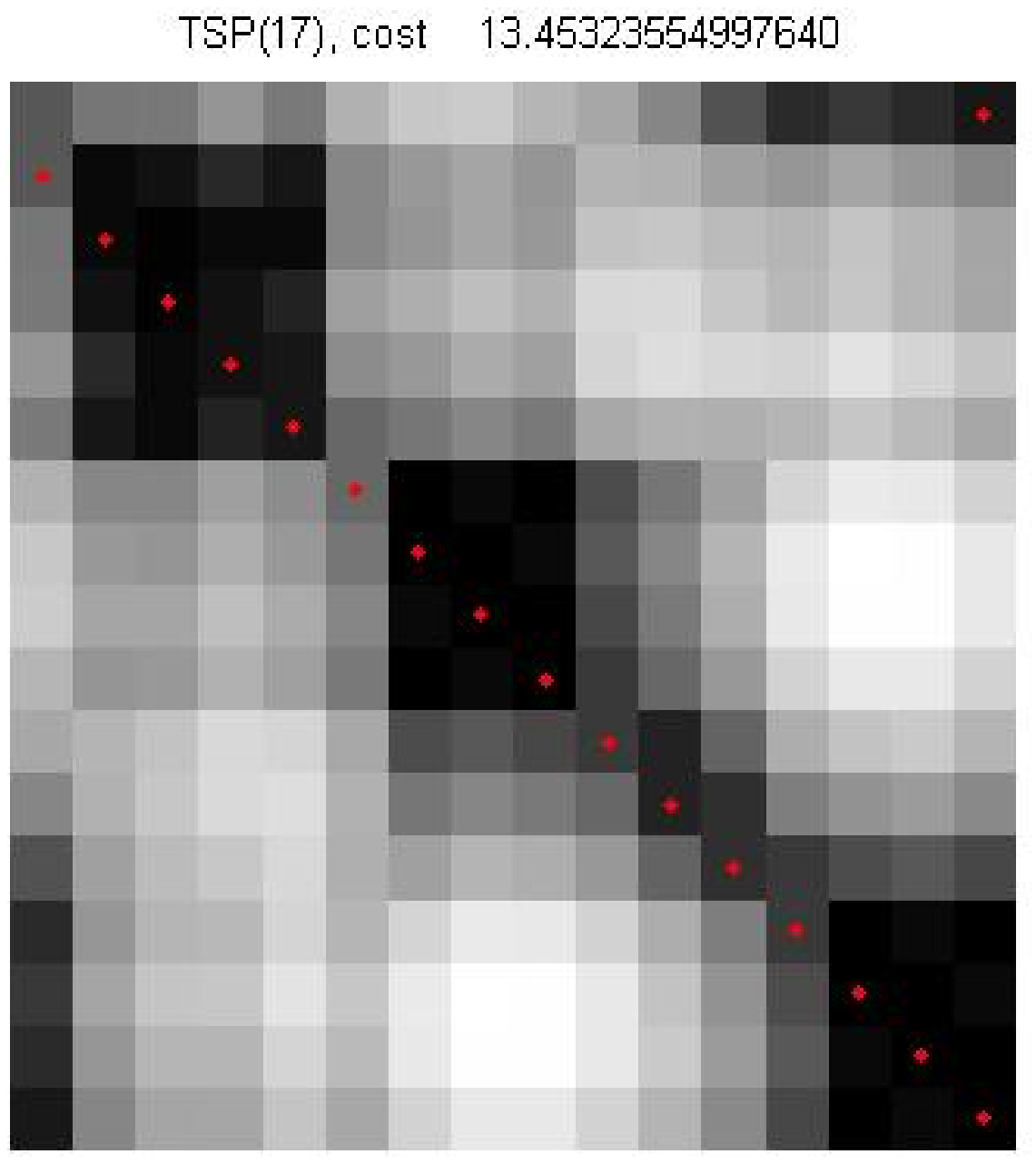,
height=50mm}
\psfig{figure=\IMAGESPATH/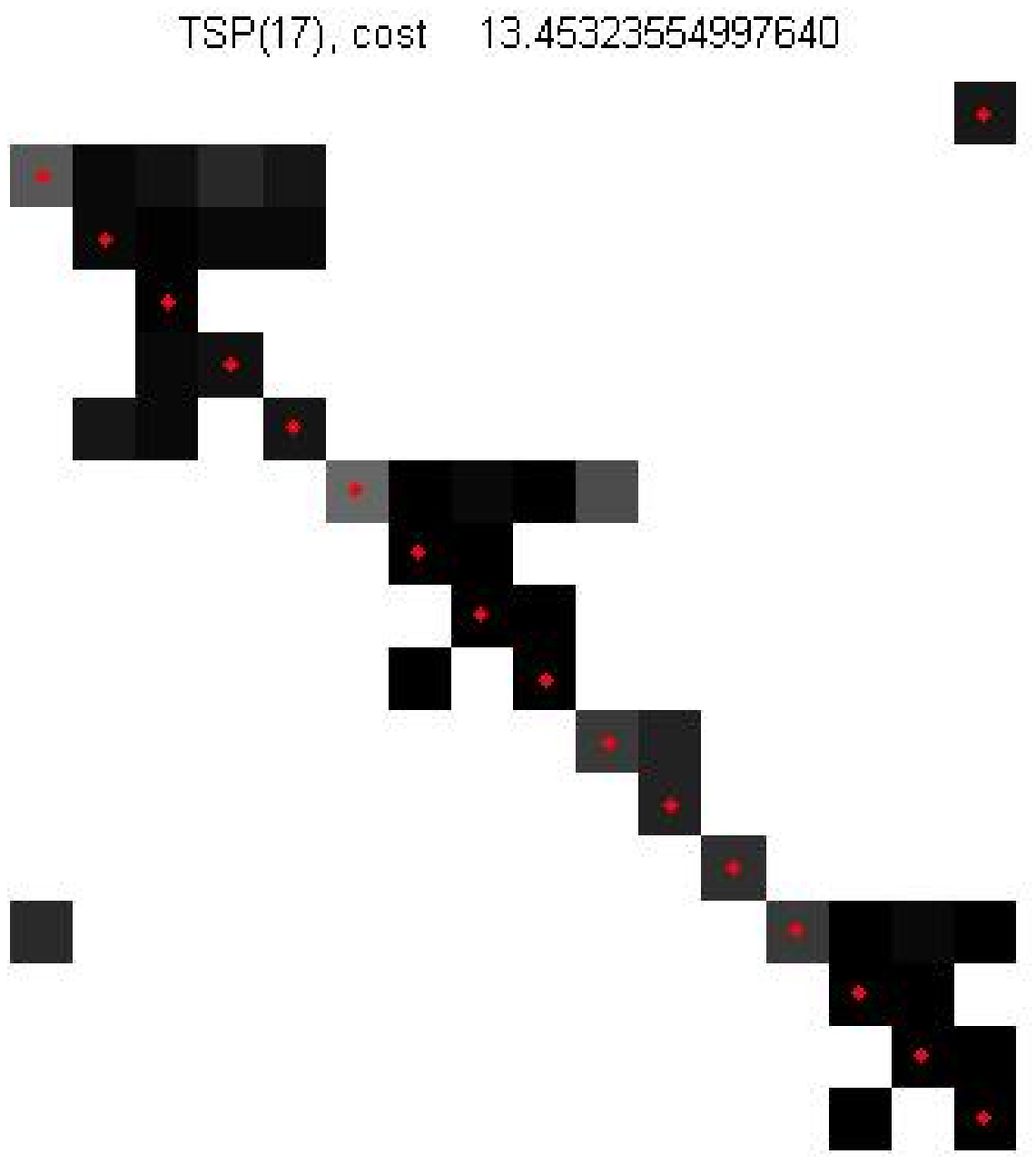,
height=50mm}}
 \caption{Cost matrix enumerated by the solution (red dots) and
 reduced equivalent cost matrix for the example 2D Euclidian
 TSP$_{17}$.}~\label{fig:cost_matrix_reduced cost matrix TSP 17}
\end{figure}

The main concerns in designed an efficient algorithm is to take
advantage of the properties of the problem with an appropriate
data structures.
The proposition~\ref{prop:enumeraCycles_descending_n_first} shows that a GAP'$_n$ can
be isomorphically transformed in an equivalent GAP$_n$ with its solution
on the first position under a given enumeration
of the cycles, i.e., the research space of GAP$n$ is finite,
numerable, and the optimal cycle could be on any position
of the natural interval $[1, (n-1)!]$,
by example, the first cycle
(see section~\ref{Prop:GAP_basic_properties}).

For the following it is assumed that a computer machine has
sufficient memory and processors. Also, it is able to perform
sum of arbitrary size as one operation, and it has simple flow
control structures such as ``while'', and  ``if''.

\begin{definition}
Given a GAP$_n$ and $\{ a_{i} \in \Natural \}_{i=1}^n$ where $a_i$
is the number of edges consider relevant that connect the vertex
$i$ with the other vertices. This collection is called
alternatives.

A tube is a set of vertices of a given a GAP$_n$ where the number
of edges consider relevant that connected them and the rest of the
vertices are less than 3. $T$ denotes the number of vertices of
the tubes of a given GAP$_n$.

The next definition comes from the Multiplication Rule.

If exists $p \ll n$ such that
\begin{eqnarray}
    \frac{\sum_{i=1}^{n} \log_2(a_{i})}{ \log_2(n-T)} & \leq &
    p~\label{eq:gradoP_alternatives}
\end{eqnarray}
then GAP$_n$ is reducible and its research space has polynomial
size.

\end{definition}

The figure~\ref{fig:cmp_red_equ_graph} corresponds to an instance of
the 2D Euclidian TSP$_{17}$, where a vertex represents a city in a
given 2D position and the edges' cost are the Euclidian distance
between them. The left figure depicts the complete graph of the 2D
Euclidian TSP$_{17}$. The right figure depicts the reduced graph where
the alternatives (lines in black) for verifying the solution (lines in
green)  are defining
 by appropriately using  the triangle inequality to avoid
considering large edges. Therefore, this picture has three tubes that
connect three clouds of vertices. The introduction of $T$ is necessary
to support in clear way that the complexity depend strongly of the
alternatives, the tubes do not increase the complexity for
verifying the solution. They are vertices connected by the best possible edge's cost,
 i.e., these sets of vertices can be considered connected without an
exploration of alternatives because they  do not have other choice.
In this example using the numeration of the fig.~\ref{fig:sol_TSP17
num_mat_red_equ_graph},
the edges fixed, i.e., the tubes are $(17, 1, 2, 3)$, $(6, 7)$, and
$(11, 12, 13, 14)$. In this case for verifying the solutions the parts to be
considered are the three clouds $\{3, 4, 5, 6\}$, $\{7, 8, 9, 10\}$, and
$\{14, 15, 16, 17\}$. This motivate the next definition for
parallel computing.

\begin{definition}[The parallel complexity version of GAP$_n$]
GAP$_n$ is
reducible if exists $p \ll n$ such that
    $$
    \frac{\max_{k=1,\ldots,K} \sum_{i=i_k}^{j_k} \log_2(a_{i})}{ \log_2(n-T)} \leq p.
    $$
    where $K$ is the number of clouds, $\{i_k,\ldots,j_k\}$ are the
    vertices of cloud $k=1,\ldots,K$.

\end{definition}

For the 2D Euclidian TSP$_{17}$ the grade for verifying the solution in
sequential mode is $\approx 2.934$ and in parallel is $\approx 1.386$,
where $n=17$, $T=11$, and the clouds' total alternatives are $12, 4, 4$
respectively.

The exploration of the irrelevance of the alternatives by properties
helps to define an algorithm for solving NP problems. An example of
this is the triangle inequality, and the monotony of the cost function
(as the distance between cities) for the $m$D Euclidian TSP$_n$. This
is not aim in this article because we are exploring to build an
algorithm for verifying the solution of NP problems.

Let be $\mathcal{M}$ an array structure as follows
    $$%
\begin{array}{|c|c|c|}
  \hline
 c_{i_1, j^1_1}, v^1_1  & \cdots  & c_{i_1, j^1_{n-1}}, v^1_{n-1} \\ \hline
 \vdots            & \vdots  & \vdots \\ \hline
  c_{i_n,j^n_1}, v^n_1  & \cdots  & c_{i_n, j_{n-1}}, v^n_{n-1} \\ \hline
\end{array}%
$$
where the columns have the edge's cost and edge's vertex, they are in ascending order by the values of
$c(i_k,j)$ (edge's cost) for each row of the vertex $i_k$.
Hereafter, $\mathcal{M}(i,j).c$ denotes the corresponding edge's cost, and
$\mathcal{M}(i,j).v$ denotes the corresponding vertex.

An algorithm to build $\mathcal{M}$ is

\begin{algorithm}~\label{alg:SortM}
\textbf{Input:} Unsorted $\mathcal{M} = \left[ c(i, j), j  |
c(i, j)\neq \inf \right], i=1,\ldots, n; j=1,\ldots, n.$

\textbf{Output:} Sorted $\mathcal{M}$, $y$ the optimal cycle when
the first column's indexes of the sorted $\mathcal{M}$ conform a
cycle.

\begin{enumerate}
\item ~ \textbf{for} i:= 1 \textbf{to} n
    \item~\hspace{0.5cm}  sort in ascending order the row $i$ of $n-1$ elements of
    $\mathcal{M}$ by $\mathcal{M}(i,\cdot).c$.

\item ~ $y :=0;$ (array of $\Real^{n+1}$)
\item ~ $v :=0;$ (array of $\Natural^{n}$)
\item ~ $i_d : =0;$ (cycle's next vertex)
\item ~ \textbf{for} $j:= 1$ \textbf{to} $n$ \{
    \item~\hspace{0.5cm} $i_d$ $:=$ $\mathcal{M}(1,j).v;$
    \item~\hspace{0.5cm} \textbf{if} $v(i_d) == 0$ \textbf{then}
        \item~\hspace{1.0cm} $v(i_d) := 1;$
        \item~\hspace{1.0cm} $y(j) := i_d;$
    \item~\hspace{0.5cm} \textbf{else}
        \item~\hspace{1.0cm} ``No optimal cycle'';
        \item~\hspace{1.0cm} stop;
\item ~ \} (end for $j$)
\item ~ $y(n+1)  := y(1);$
 \item~ ``Optimal cycle is'', $y$;
\end{enumerate}
\end{algorithm}

Using the Quick Sort the complexity of the previous algorithm is
$\bigO(n^2\log_2(n))$.

The next algorithm is a greedy algorithm to compute an initial cycle using sorted
$\mathcal{M}$.

\begin{algorithm}~\label{alg:GreedyAlgorithm_initialCycle}
\textbf{Input:} Sorted $\mathcal{M} = \left[ c(i, j), j \right], i=1,\ldots, n; j=1,\ldots, n.$
\textbf{Output:} $y$ a cycle.
\begin{enumerate}
\item ~ $y :=0;$ (array of $\Real^{n+1}$)
\item ~ $v :=0;$ (array of $\Natural^{n}$)
\item ~ $i_d : =0;$ (cycle's next vertex)
\item ~ \textbf{for} $j$:= 1 \textbf{to} $n$ \{
    \item~\hspace{0.5cm} $i_d$ $:=$ $\mathcal{M}(1,j).v;$
    \item~\hspace{0.5cm} \textbf{if} $v(i_d) == 0$ \textbf{then}
        \item~\hspace{1.0cm} $v(i_d) := 1;$
        \item~\hspace{1.0cm} $y(j) := i_d;$
    \item~\hspace{0.5cm} \textbf{else}
        \item~\hspace{1.0cm} break;
\item ~ \} (end for $j$)
\item ~ $r$ := $n$ - sum($v$);
\item ~ \textbf{if} $r$ == $0$ \textbf{then}
\item  ~\hspace{0.5cm} $y(n+1)  := y(1);$
 \item ~\hspace{0.5cm}``The optimal cycle is'', $y$;
\item ~\hspace{0.5cm} stop;

\item~ \textbf{for} $k$:= $1$ \textbf{to} $r$ \{

    \item~\hspace{0.5cm} \textbf{for} $l$:= $1$ \textbf{to} $n-1$ \{
        \item~\hspace{1.0cm} $i_t$ $:=$ $\mathcal{M}(i_d,l).v;$
        \item~\hspace{1.0cm} \textbf{if} $v(i_t) == 0$ \textbf{then}
            \item~\hspace{1.5cm} $i_d := i_t;$
            \item~\hspace{1.5cm} $v(i_d) := 1;$
            \item~\hspace{1.5cm} $y(j) := i_d;$
            \item~\hspace{1.5cm} $j := j+1$;
    \item~\hspace{0.5cm} \} (end for $l$)
\item ~ \} (end for $k$)
\item  ~ $y(n+1)  := y(1);$
 \item ~``An initial cycle is'', $y$;
\end{enumerate}
\end{algorithm}

The complexity of the previous algorithm is $\bigO(n^2)$. It takes
advantage of the sorted $\mathcal{M}$ picking first the vertices with
small edge's cost than the vertices with large edge's cost. The
algorithm's behavior depends of the numeration of the vertices that
affect the cost order of the rows. It is no invariant, different
equivalent GAP$_n$ (see proposition~\ref{prop:Cycle_for_numering})
could give different initial cycles.

\begin{prop}
Given GAP$_n$, if the sorted $\mathcal{M}$ has in its first column
all vertices then these vertices conform a cycle, which is the
solution. Furthermore, this is done in polynomial time using
the previous algorithm or the algorithm~\ref{alg:SortM}.

\begin{proof}
First, the optimality of the  cycle comes from the vertices  of first
column of $\mathcal{M}$.
 Given any cycle of vertices $(v_1,v_2,\ldots,v_n,v_1),$
 $v_i \neq v_j$, $1 \leq i < j \leq n$, its cost is given by
$$
\sum_{k=1}^{n} c(v_k, v_{k+1}).
$$

But each term $c(v_k, v_{k+1}) \geq c(v_k,j^*_k)$, where $j^*_k$
is a vertex on the first column of $\mathcal{M}$. Then
$$
\sum_{k=1}^{n} c(v_k, v_{k+1}) \geq \sum_{k=1}^{n} c(v_k,j^*_k).
$$

The right side is the cost of the cycle that corresponds to the
vertices  of the first column of $\mathcal{M}$.

The complexity of the algorithm~\ref{alg:SortM}, or the algorithm~\ref{alg:GreedyAlgorithm_initialCycle}
is bounded by $\bigO(n^3)$.
\end{proof}
\end{prop}

\begin{prop}
Given GAP$_n$, a sufficient condition for computing in polynomial
complexity its solution is that the matrix cost $c(i,j)$ has a
minimum by row and the corresponding column압 index of each row
cover all vertices.
\begin{proof}
This follows from the previous proposition.
\end{proof}
\end{prop}

\begin{rem}
Because the first column of the vertices' number of sorted
$\mathcal{M}$ determines that there is only one alternative by row to
define the optimal cycle, the right side of the inequality~\ref{eq:gradoP_alternatives}
(tubes are not considered, i.e., $T=0$) imply $p \geq  0$ $=$
$\frac{\sum_{i=1}^{n} \log_2(1}{ \log_2(n)}$=$\frac{\sum_{i=1}^{n} \log_2(a_{i})}{ \log_2(n)}.$
Therefore, the verification of the optimal cycle is not necessary,
it is given by the dominance property of the first column edge`s cost
and vertex's number of the matrix $\mathcal{M}$. Zones of this type
in a GAP$_n$ are considered tubes.
\end{rem}

\begin{figure}
\centerline{
\psfig{figure=\IMAGESPATH/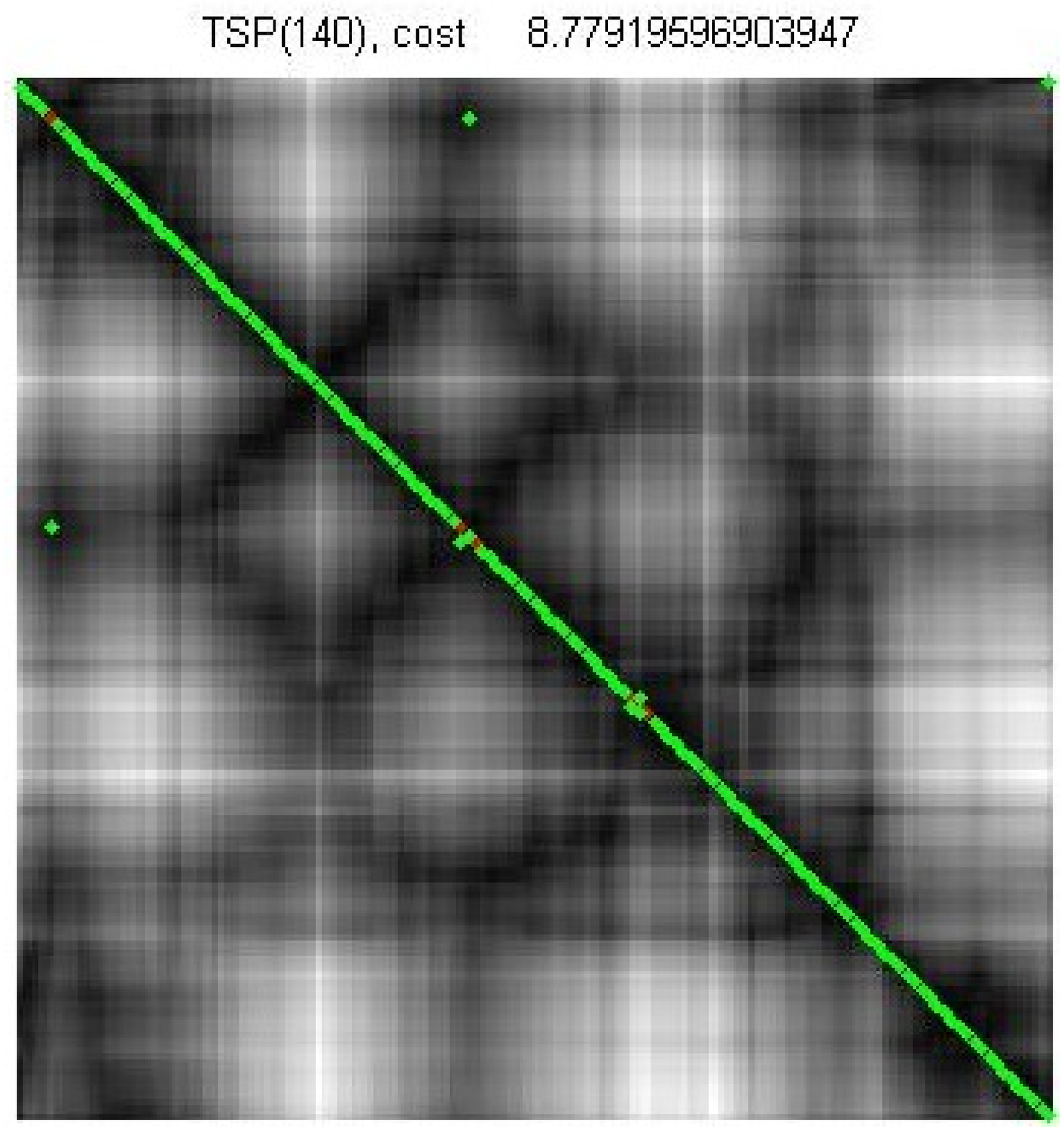,
height=50mm}
\psfig{figure=\IMAGESPATH/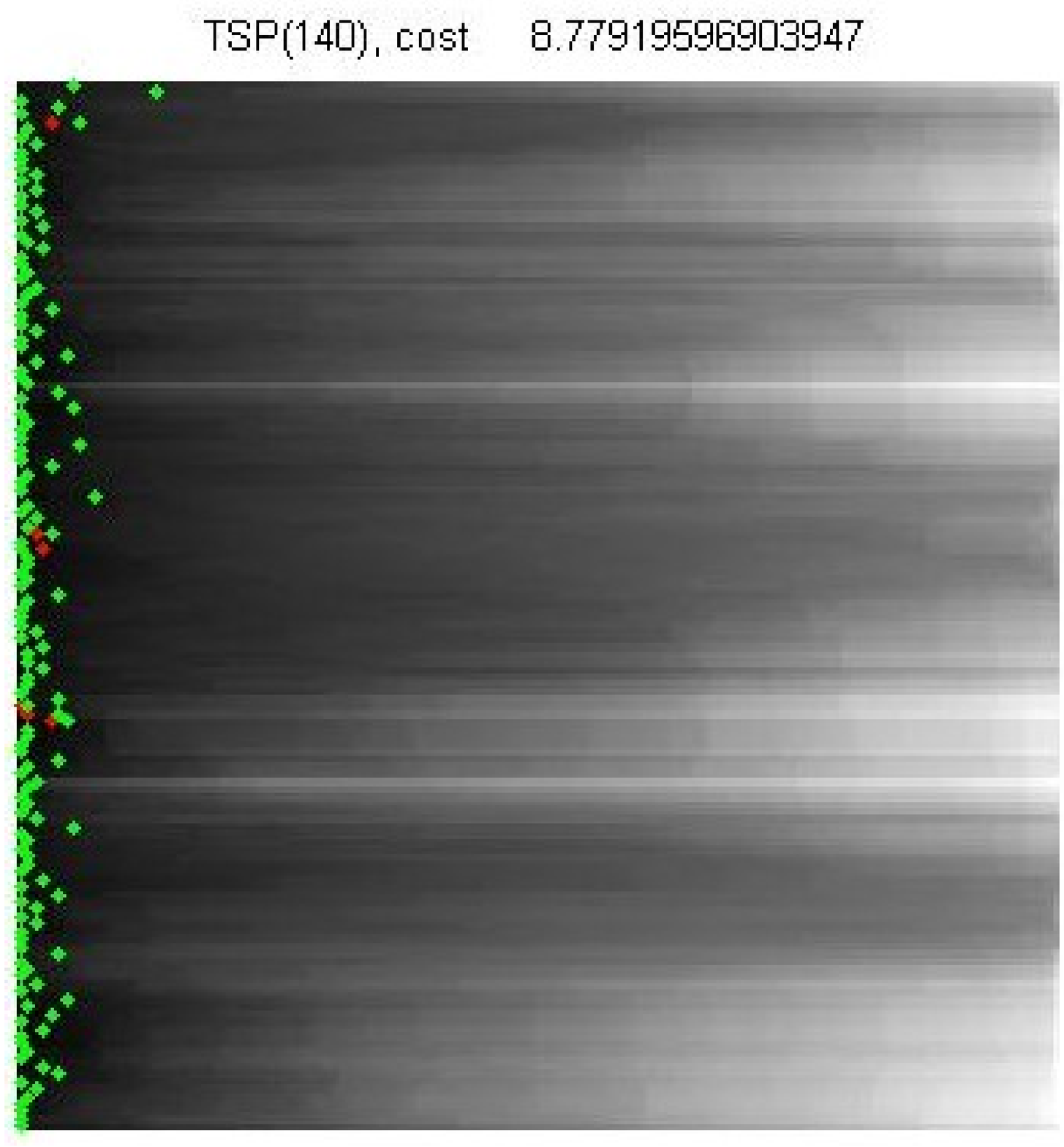,
height=50mm}}
 \caption{On the left cost matrix,
 and on the right sorted cost of $\mathcal{M}$
 for the example of the 2D Euclidian TSP$_{140}$.
 The red dots are the cycle's frontier (with cost 8.7791),
 and the green dots are an above cycle with cost
 8.8880.}~\label{fig:TSP140_abovecycle}
\end{figure}

\begin{definition}
Given a GAP$_n$, and after computing the sorted $\mathcal{M}$, the
vertices of any cycle define a frontier on $\mathcal{M}$.
\begin{enumerate}
    \item  A cycle is below if all its costs are equal or less than their corresponding cost of the frontier,
    i.e., they are on the left side of the frontier.
    \item A cycle is above if all its costs its costs are equal or greater than their corresponding cost of the frontier,
    i.e., they are on the right  side of the frontier.
    \item A cycle is oscillating if it has costs on both sides of
    the frontier.
\end{enumerate}
\end{definition}

Figure~\ref{fig:TSP140_abovecycle} depicts an example of the 2D
Euclidian TSP$_{140}$. Taking in consideration, that when the vertices
coincide only the green dots are visible. The left picture only shows
that a cycle (green dots) does not coincide with the diagonal (the
optimal cycle) but the right picture implies that the green dots are an
above cycle, therefore its cost is greater than the cycle's cost of red
dots. The right picture has the sorted cost of $\mathcal{M}$ where the
frontier (red dots) is below of the cycle with green dots. Then the
green dots are an above cycle because there are not any green dots
below of the red dots. Finally, the red dots are the optimal cycle of
the 2D Euclidian TSP$_{140}$ because there is not below cycles of it as
frontier.

\begin{prop}
Given GAP$_n$, and after computing the sorted $\mathcal{M}$. A
necessary condition for a cycle to be a possible solution of
GAP$_n$ is that there is not cycles strictly below of it as a
frontier.
\begin{proof}
By the definition, if there is a cycle strictly below, then the
frontier cycle has a cost greater than it. Therefore the frontier
cycle is not a candidate to be a minimum, i.e., it is not longer a
solution.
\end{proof}
\end{prop}

\begin{figure}
\centerline{
\psfig{figure=\IMAGESPATH/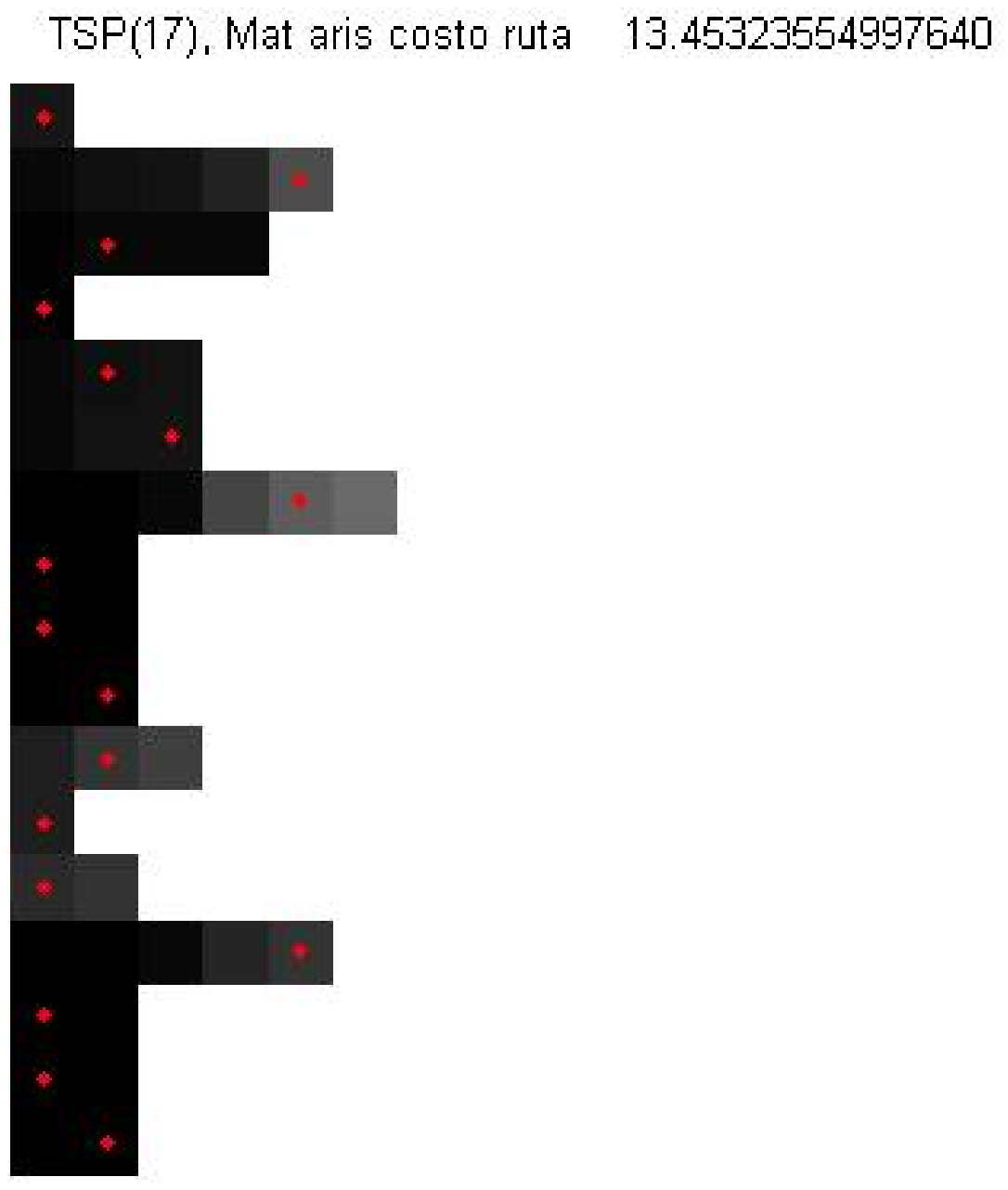,
height=50mm}
\psfig{figure=\IMAGESPATH/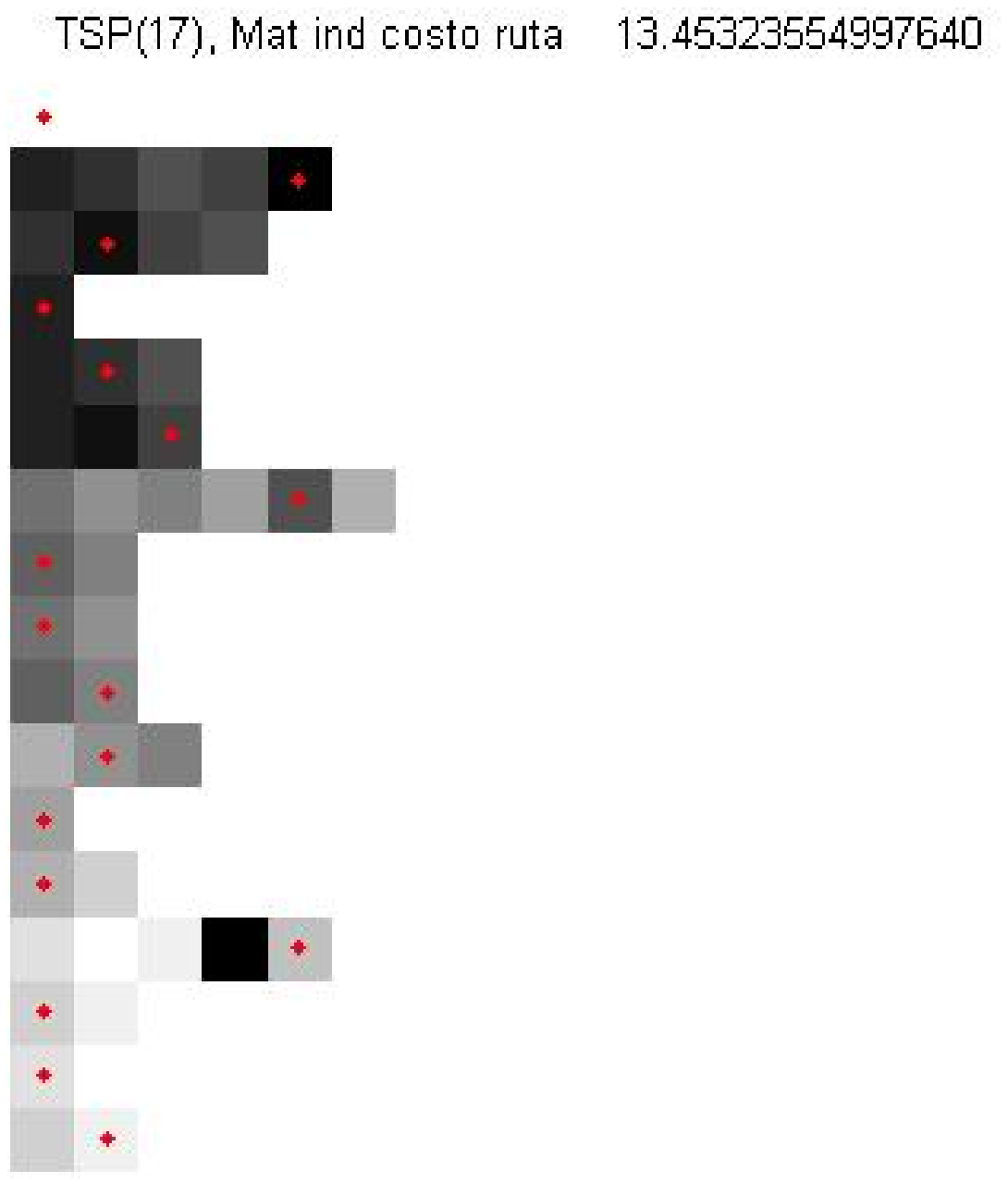,
height=50mm}}
 \caption{On the left the sorted cost matrix,
 and on the right their vertices of the sorted $\mathcal{M}$
 for the example of the 2D Euclidian TSP$_{17}$.}~\label{fig:M_cost_indFor17_cities_graph}
\end{figure}

\begin{figure}
\centerline{
\psfig{figure=\IMAGESPATH/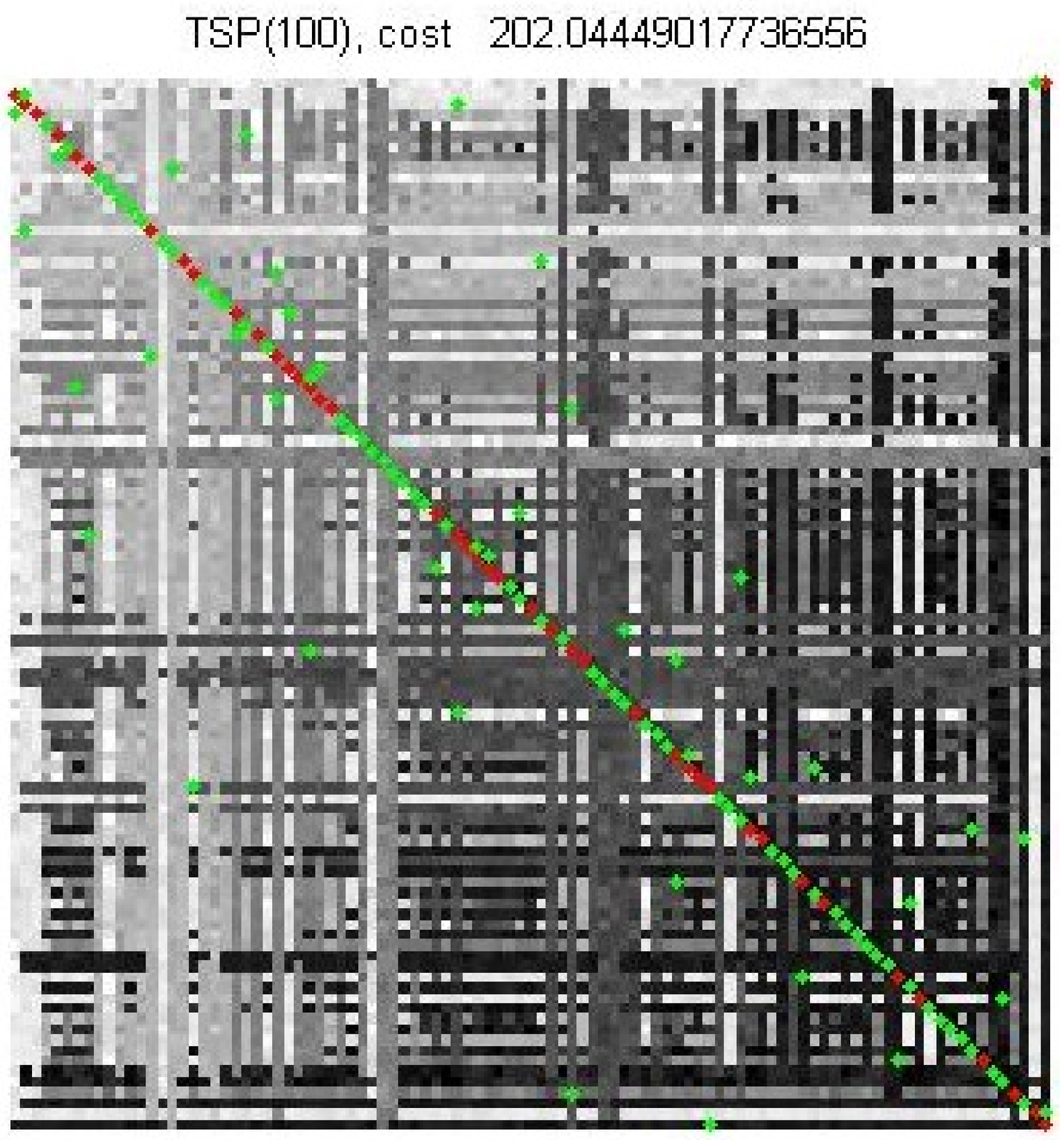,
height=50mm}
\psfig{figure=\IMAGESPATH/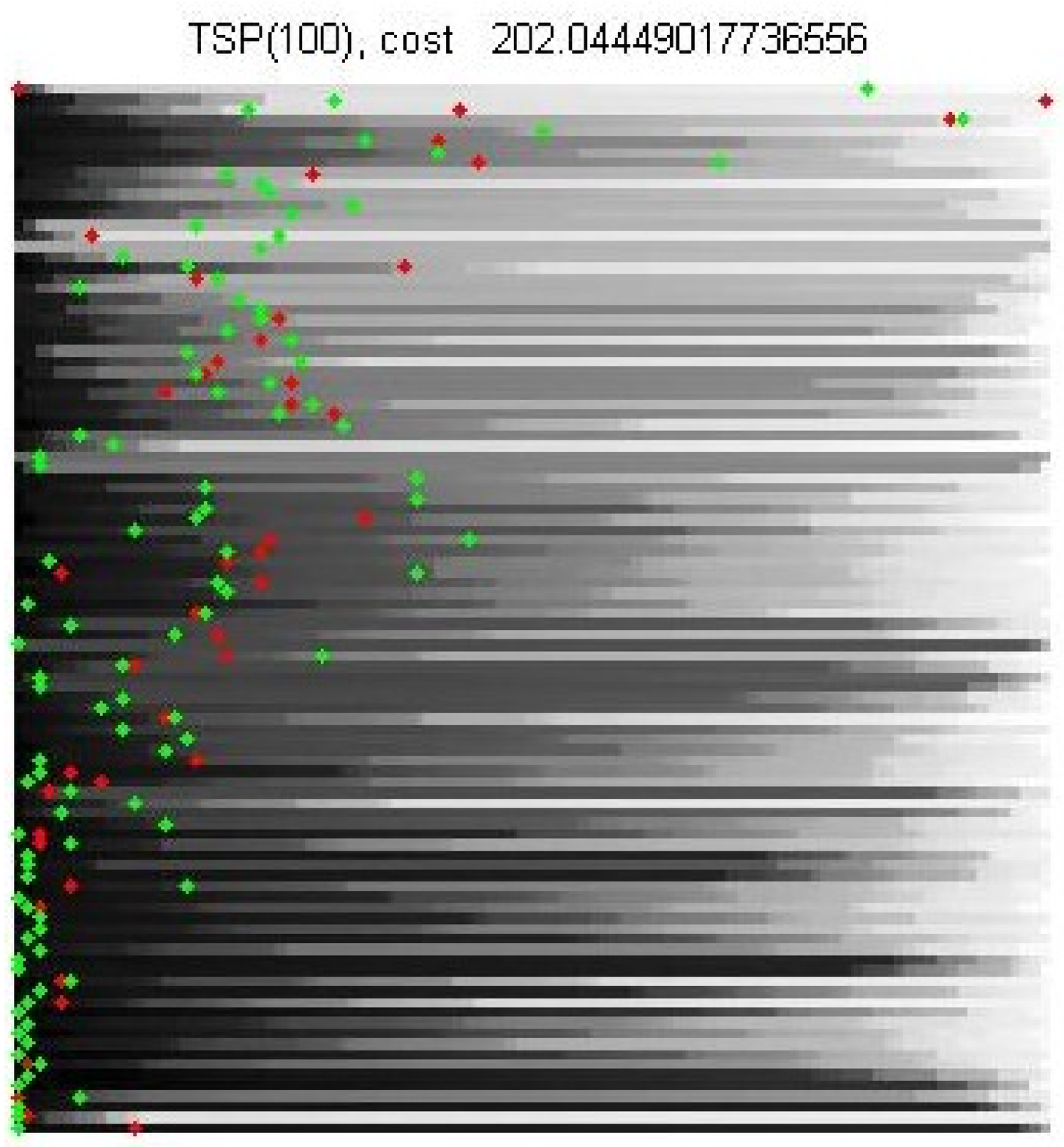,
height=50mm}}
 \caption{On the left cost matrix,
 and on the right sorted cost of $\mathcal{M}$
 for the example GAP$_{100}$. The red dots are the cycle's frontier (with cost 202.0444),
 and the green dots are a oscillating cycle with cost
 204.2770.}~\label{fig:GAP100_oscillanting}
\end{figure}

\begin{rem}
The importance of the previous proposition is that it is possible
to use this necessary condition on the oscillating cycles of a
given frontier. Figure~\ref{fig:GAP100_oscillanting} depicts the
frontier (red dots), and oscillating cycle (green dots) with cost
204.2770. Where the cost coincide, only the green dots are visible.
This example shows how sensitive is GAP$_n$, many green dots are
above, on, and  below of the red dots, and cost of these cycles is
similar.The main cause of this is that the frontier is shifted to the right position
in the image of the sorted $\mathcal{M}$,
and the frontier cycle contains edges with cost negative and positive.

In short, above cycles must be discarded, below cycles must not
exist, therefore, an algorithm needs only to focuss in  the oscillating
cycles
of the sorted $\mathcal{M}$, and if the putative optimal cycle is
not, then it discards the previous and keeps the best cycle.
This means that the complexity of solving or checking the solution
of
a GAP$_n$ is related to the maximum between the number of
oscillating cycles (as the reduced research space) and
$\bigO(n^3)$ (the cost of computing sorted $\mathcal{M}$).

It is also possible to combine the necessary condition with
properties of special matrix $c(i,j)$ but this is not the general
case. Hereafter, the matrix $c(i,j)$ is arbitrary for any large
and arbitrary GAP$_n$.
\end{rem}

For the matrix $c(i,j)$ we can compute $\mathcal{M}$ and in order
to estimate a reduce research space for the cycles around the
candidate a solution that define the current frontier this can be
done by focus to the cost around and below of the cost of frontier
cost in $\mathcal{M}$. By taking the vertices in $\mathcal{M}$
around the frontier we get a reduced research space that it is
possible to contain the solution of the complete problem GAP$_n$.
For special matrices $c(i,j)$ could be possible to determine the
reduced research space with the guaranty that the solution is in
it. This means only to take in consideration the edges (and their
vertices) such that

\begin{eqnarray}
\nonumber  c(v,j) & \leq &  c_{vk} ~\label{eq:inequality_eps_r}
\end{eqnarray}
where $c_{vk} = \max \left\{ c(v,k)(1-\eps_{r_v}), c(v,k)(1+\eps_{r_v}) \right\}$,
 $v, k$ are consecutive vertices in the optimal cycle defining
 the frontier, and $-1 \leq \eps_{r_v}$, it depends of the vertex's vicinity.
 In the examples, $-0.9 \leq \eps_{r_v} \leq 0.6$.
The  algorithm~\ref{alg:Estimate_alternatives eps_r} estimates the
vertices' alternatives and $\eps_{r_v}$.
Figure~\ref{fig:TSP_37_triangulation} depicts alternatives of each
vertex as magenta dotted lines.

\begin{rem}
The previous inequality keeps vertices that could belongs to
oscillating cycles. $v_{vk}$ is shifted towards the positive direction
to accept all vertices with cost below of the corresponding frontier압
vertex that could form a cycle with cost less or equal than frontier's
cost. An example of an special matrix is a cost matrix of an $m$D
Euclidian TSP$_n$ where the cost correspond to a distance or a norm.
The discrete distance gives trivial TSP$_n$ problems with matrix's cost
full of zeros and ones. They are trivial because the solution is
rapidly and easily founded by the
algorithm~\ref{alg:GreedyAlgorithm_initialCycle}, only two columns
needed to consider to found a the optimal cycle, i.e, the cycle with
the maximum number of edges' cost zero. The verification of the
solution by the algorithm~\ref{alg:VerificationSol} is on linear time
to prove in this case that there are not cycles below of the frontier.
The dimension of the space is significative for the number of
alternatives to consider for updating a given path. Even, that there
are a good algorithms for building a triangulation for sets of points
in $\Real^m$, $m>2$, It becomes more difficult to build a good initial
vertices' numeration as the $m$ grows, and it is not easy to see what
vertex is the next one as in the plane.
\end{rem}

The next algorithm estimates a lower bound of the size of the reduced
research space for a given frontier.

\begin{algorithm}~\label{alg:sizeofResearchEps_r}
\textbf{Input:} $y^*$ the putative optimal cycle that defines the
current frontier. $c(i,j)$ the edges' cost matrix. \textbf{Output:}
minimum $p$, a possible polynomial grade (see
inequality~\ref{eq:gradoP_alternatives}), and $A$ lower bound of  the
number of alternatives, i.e., lower bound of the reduced research space
for a given frontier and a given set of $\{ \eps_{r_v}, v=1,\ldots, n
\}$.

\begin{enumerate}
\item ~ $t :=0;$ (array of $\Real^{n \times n})$
\item ~$t(1,n) := 1$;
\item ~ \textbf{for} $i$:= 2 \textbf{to} $n$ \{
\item ~\hspace{0.5cm} \textbf{for} $j$:= i \textbf{to} n-1 \{
    \item~\hspace{1.0cm} $k$ := $y^*(i);$
    \item~\hspace{1.0cm} $v_{ik}$  $:=$ $\max \left\{ c(i,k)(1-\eps_{r_i}), c(i,k)(1+\eps_{r_i})
    \right\};$
    \item~\hspace{1.0cm}  \textbf{if} $c(i,j) \leq v_{ik}$ \textbf{then} (see inequality~\ref{eq:inequality_eps_r})
    \item~\hspace{1.5cm}  $t(i,j) := 1$
\item  ~\hspace{0.5cm} \} (end for $j$)
\item \} (end for $i$)

\item ~ $a$ $:=$ $0.0;$ (array of $\Real^n$)

\item ~ \textbf{For} i:= 1 \textbf{to} n
    \item~\hspace{0.5cm}  $a(i) = \max \{ 1.0, \sum_{j=1}^{n}
    t(i,j)\};$
\item ~ $A$ $:=$ $1.0;$
\item ~ $p$ $:=$ $0.0;$
\item ~ \textbf{For} $i$:= 1 \textbf{to} $n$ \{
\item~\hspace{0.5cm}  $A$ $:=$ $A*a(i);$
\item~\hspace{0.5cm}  $p$ $:=$ $p + \log_2(a(i));$
\item ~ \} (end for $i$)
\item ~ $p:= \frac{p}{\log_2(n)}.$
\end{enumerate}
\end{algorithm}

Note that $y^*=[n,n-1,\ldots,1,n]$ by the numeration of
the cycles on descending order algorithm~\ref{alg:cylclesNumerationDescendig},
$t(1,n) := 1$, i..e., the
cycles starts on $n$ and finish on $n$ for counting the cycles in
descending order. Because $a(i) \neq 0, \forall i=1,\ldots,n$ by construction,
and because the graph $G_n$ of GAP$_n$ is complete,
this algorithm gives always an estimation of $A \neq 0$ and $p \neq 0$.
Also, it is possible to consider the lower edges' cost matrix.
In the case where is no reduction of alternatives,
the previous algorithm gives $A=(n-1)!$ and $p=\frac{\log_2(A)}{\log_2(n)}$.
For future work, it is possible to create a more precise
algorithm. For example, taking $a(i)$ as the number of vertices that comply
inequality~\ref{eq:inequality_eps_r} without repeating them
provides other estimation of $A$ and $p$.

\begin{prop}
Given GAP$_n$, and after computing the sorted $\mathcal{M}$. A
necessary condition for a cycle to be a possible solution of
GAP$_n$ is that there is not a cycle below of it as a
frontier.
\begin{proof}
By the definition, if there is a cycle with some vertices below and other
on the frontier, then the
frontier cycle has a cost greater than it. Therefore the frontier
cycle is not a candidate to be a minimum, i.e., it is not longer a
solution.
\end{proof}
\end{prop}

\begin{rem}
For GAP$_n$ and the solution's frontier, strictly below and above
cycles could not exists, all cycles could be oscillating. For
TSP$_n$ is possible that exists oscillating cycles.
\end{rem}

Figure~\ref{fig:M_cost_indFor17_cities_graph} depicts the case of the
example of the 2D Euclidian TSP$_{17}$ (Fig.~\ref{fig:sol_TSP17
num_mat_red_equ_graph} depicts the solution with green lines) where the
frontier for the solution is marked by red dots. Because many dots of
the frontier are on the left border the alternatives of many vertices
are less than 2 implying that the solution is easy to verify and
estimate, i.e., this is a case of very few alternatives to explore.
Note that also the tube given by the vertices $3,2,1,17$ is easily to
identify on the picture. The cost of the edge $1,17$ is marked by the
red dot on upper left. It is on the left of the first row because this
cost is less than all the edges' cost of the vertex $1$. The cost of
the edge $2,1$ is on the second row, four column because it is greater
than the cost of the edges $3,2$, $4,2$, $6,2$, and $5,2$. The red dot
of the cost of the edge $3,2$ on the third row is on second column
because it is greater than edge's cost $4,3$. On the other hand, the
picture on the left denotes also the frontier of the solution on the
gray image of the vertices' number. The first row has the red dot of
edge's cost $1,17$ over a white square that correspond to the color of
the vertex $17$. On the second row there is four gray squares
corresponding to the color of the vertices $3,4,6,5$ and the red dot of
the edge cost is on a black square that corresponds to the vertex $1$.
Note that over the columns the red dots goes from white-black gray
fading to white because the color of the vertex of the solution goes
from $(17, 16, 15, \ldots, 2, 1, 17)$. These two pictures depicted the
relation between the edges' cost and the vertices' number of the
solution. There is a possible efficient verification of the solution
when the cost of the solution are on closed to the left column and the
vertices' number creates a pattern white-black gray fading to white,
this combination point out that the alternatives for most of the
vertices are very few. For the example of the 2D Euclidian TSP$_{17}$
the other solution  is the inverse cycle $(1,2, 3, \ldots, 16, 17, 1)$.
These solutions are equals because the cost matrix is symmetric.

\begin{figure}
\centerline{
\psfig{figure=\IMAGESPATH/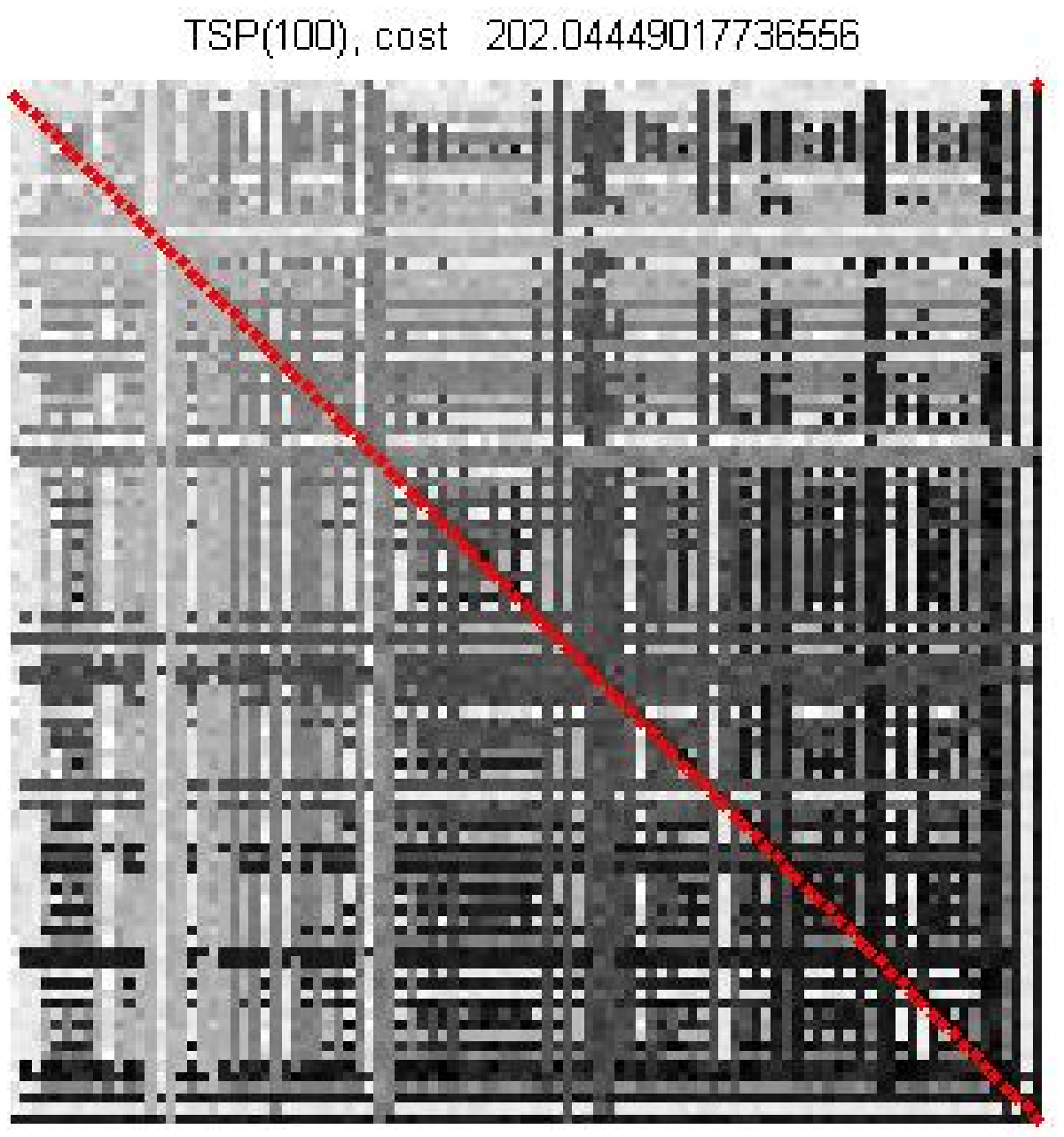,
height=50mm}
\psfig{figure=\IMAGESPATH/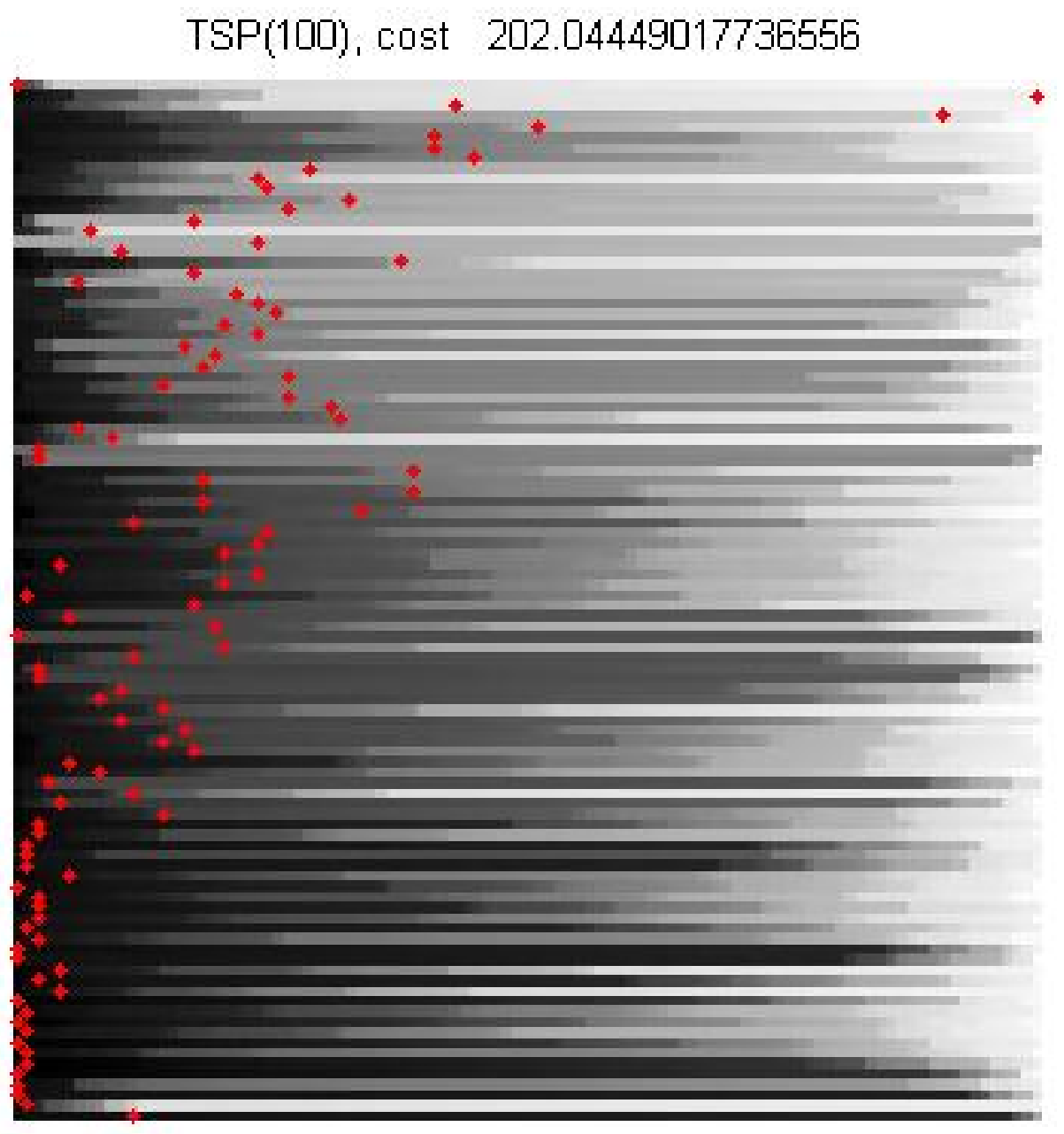,
height=50mm}
\psfig{figure=\IMAGESPATH/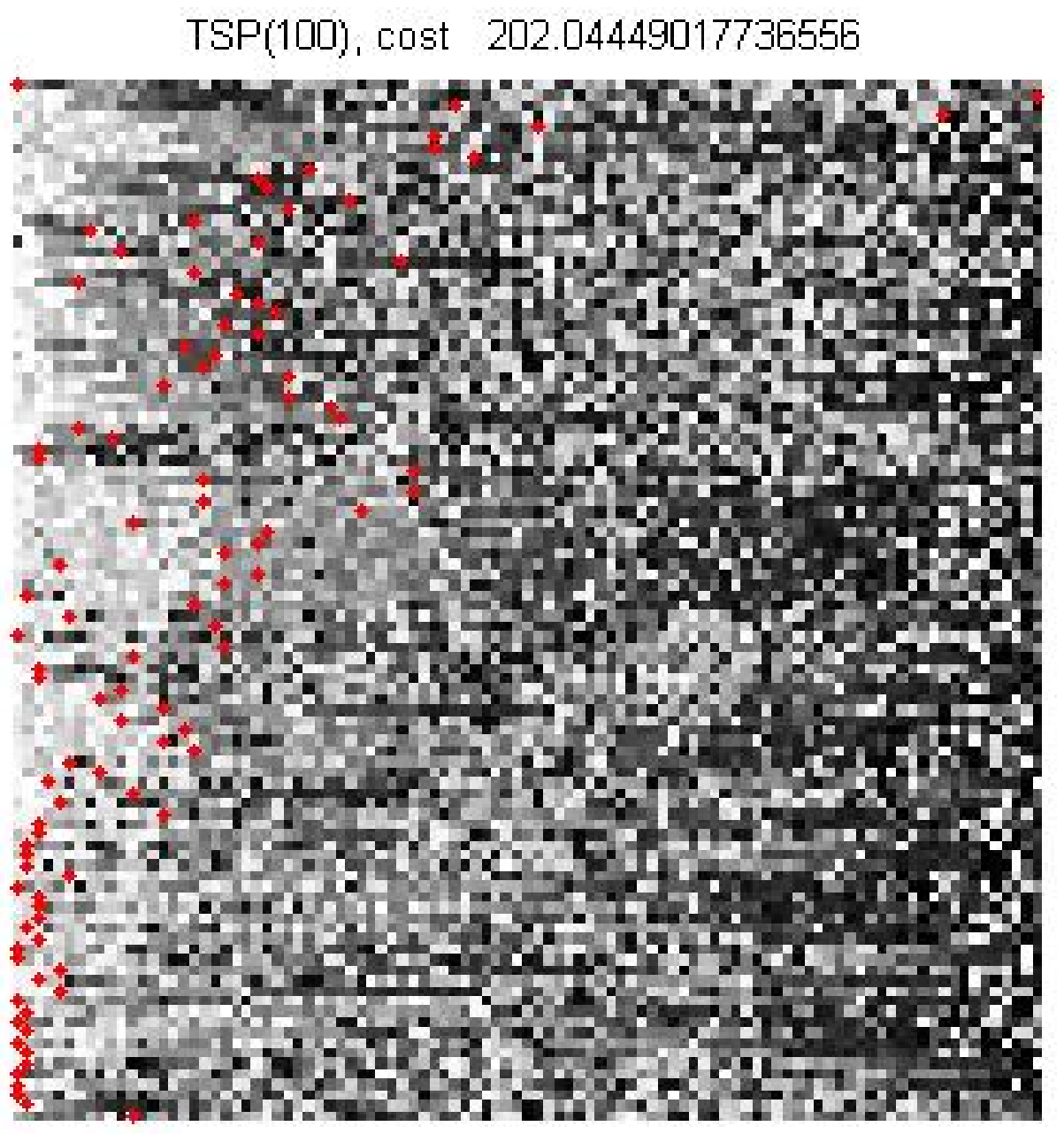,
height=50mm}}
 \caption{Example of an GAP$_{100}$ with arbitrary cost matrix,
  by construction there is not pattern to point out a reduction of vertices' alternatives
 From left to right, Cost matrix, sorted cost matrix,
 and their vertices of the Sorted $\mathcal{M}$.}~\label{fig:ex_100_vert_mat_order_mat}
\end{figure}

\begin{figure}
\centerline{
\psfig{figure=\IMAGESPATH/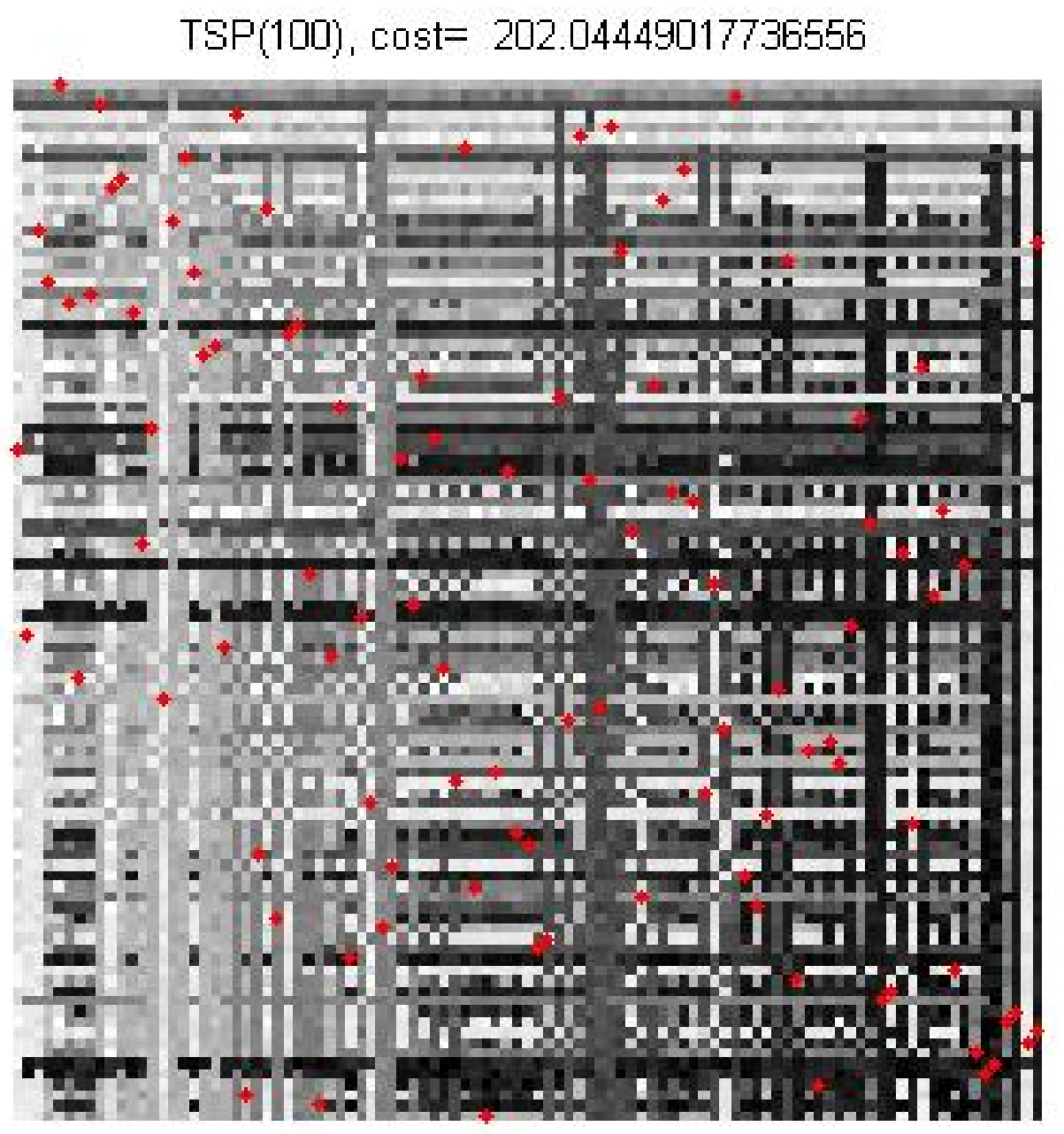,
height=50mm}
\psfig{figure=\IMAGESPATH/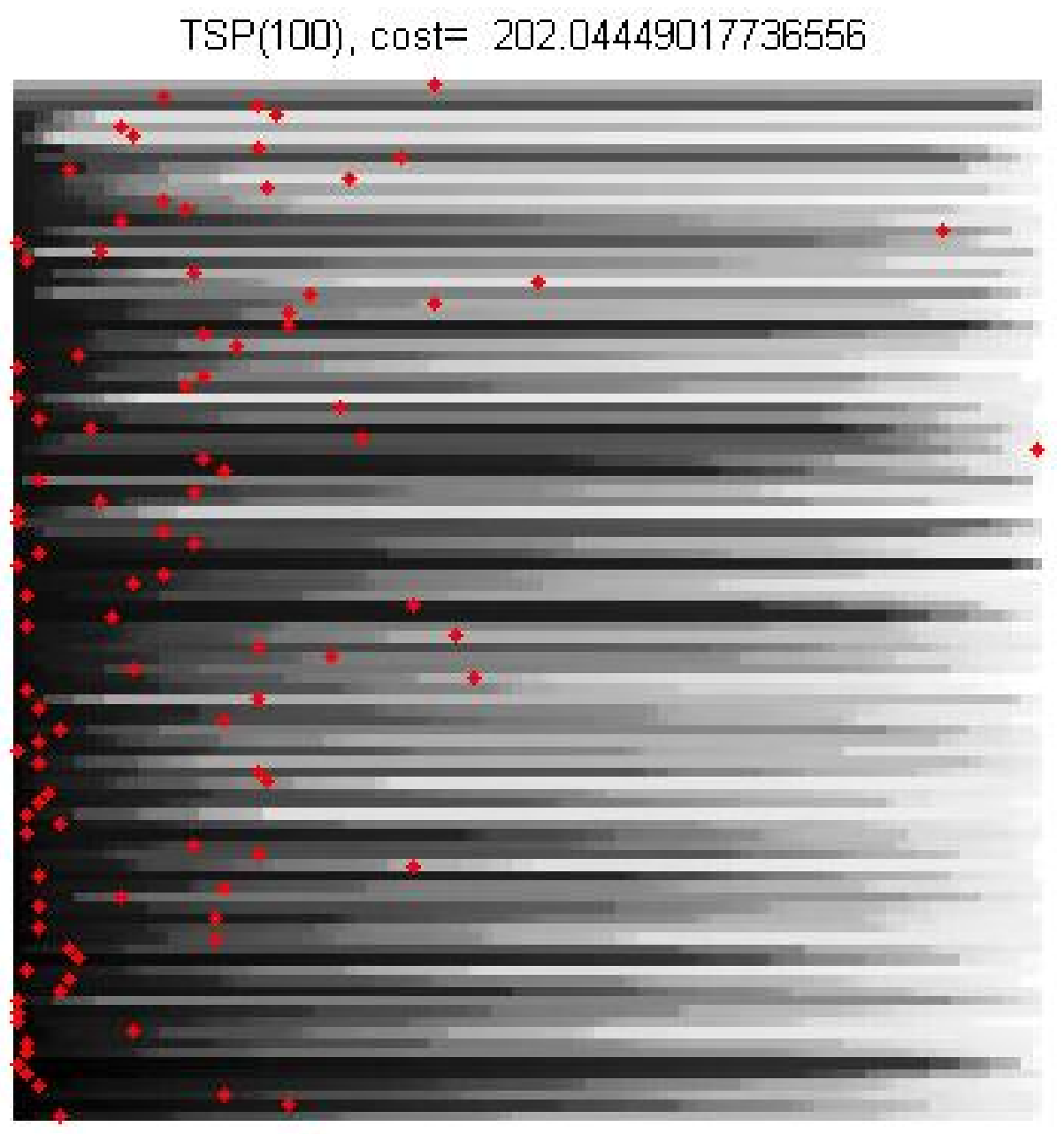,
height=50mm}
\psfig{figure=\IMAGESPATH/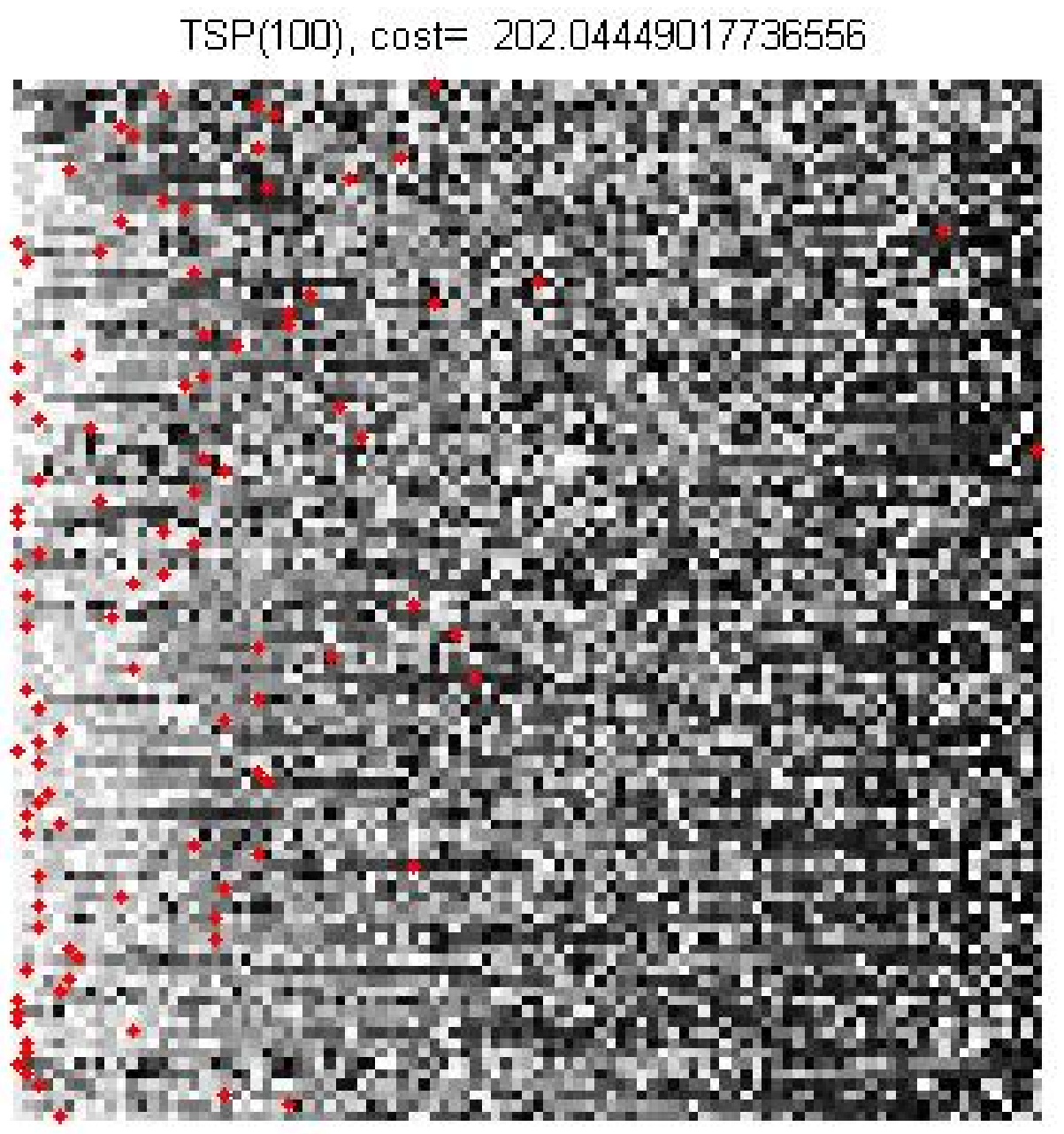,
height=50mm}}
 \caption{Example of an GAP$_{100}$ with arbitrary cost matrix.
 Here the matrices are sorted by columns and by row taking the first column's cost,
 from left to right, Cost matrix, sorted cost matrix, and their vertices of
 $\mathcal{M}$.}~\label{fig:ex_100_ren_vert_mat_order_mat}
\end{figure}

The Figure~\ref{fig:ex_100_vert_mat_order_mat} depicts an example of
GAP$_{100}$ where the frontier given by the solution is scattered. The
image of the cost matrix on the left depicts on many rows an
oscillation of gray squares from black to gray, this means that many
vertices have edge's cost worthwhile to explore. There are very few row
with uniform gray color. On the images of the sorted $\mathcal{M}$,
there are many red dots of the frontier to the right with a fading
sequence of gray squares, it means edges' cost less than the cost of
the edge's cost of the solution. Finally the vertices' number picture
depicted a random mosaic, it means that many vertices are related to
many vertices that could give a reduction of the cycle cost but the
research space is not reduced. The
Figure~\ref{fig:ex_100_ren_vert_mat_order_mat} depicts the previous
matrices sorted by the first column's cost. These figures makes more
clear the existence of oscillating cycles around the solution and also
depicts that the frontier in sorted $\mathcal{M}$ is not on the left
side but on the middle.

\begin{figure}
\centerline{ \psfig{figure=\IMAGESPATH/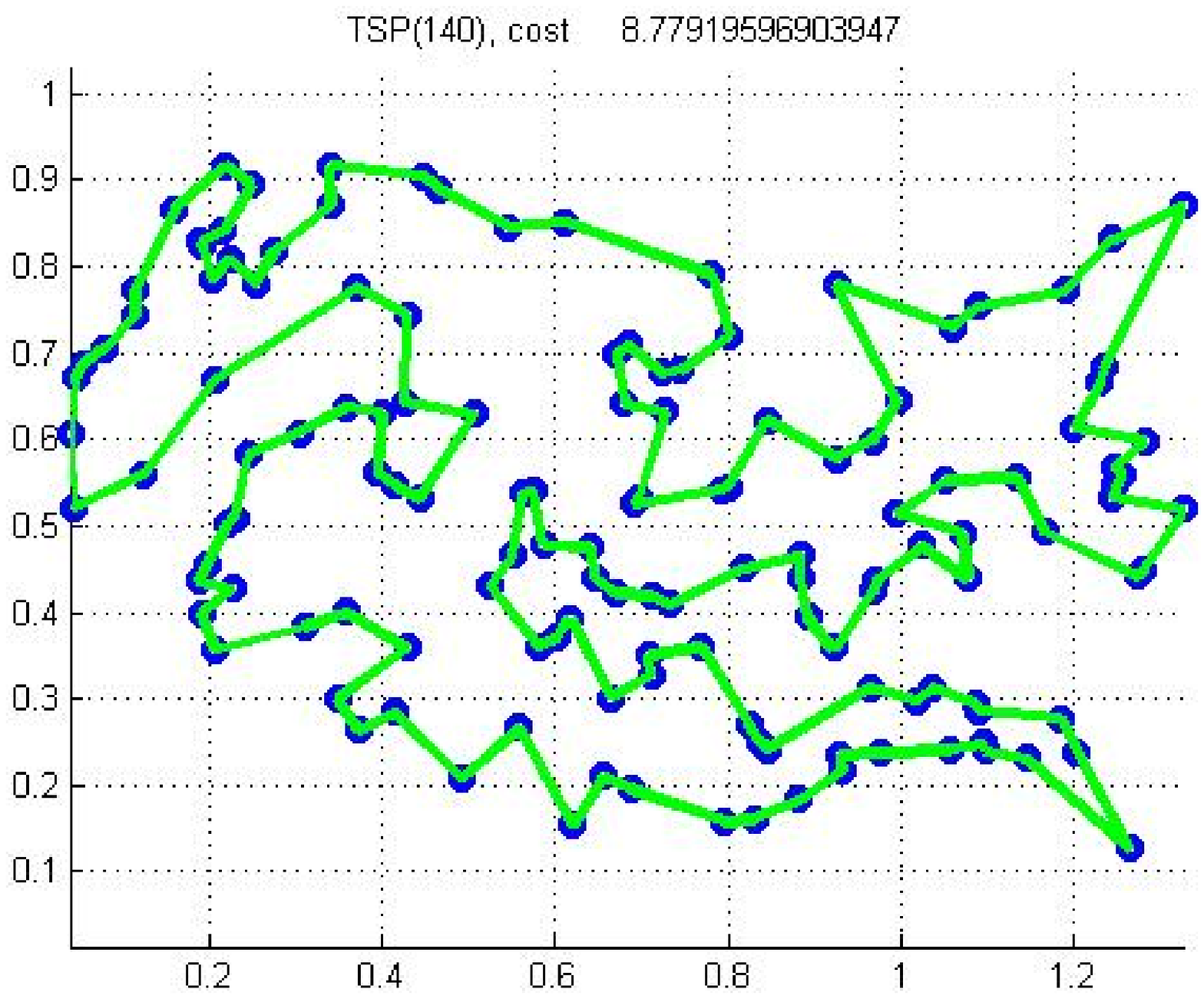,
height=50mm}
\psfig{figure=\IMAGESPATH/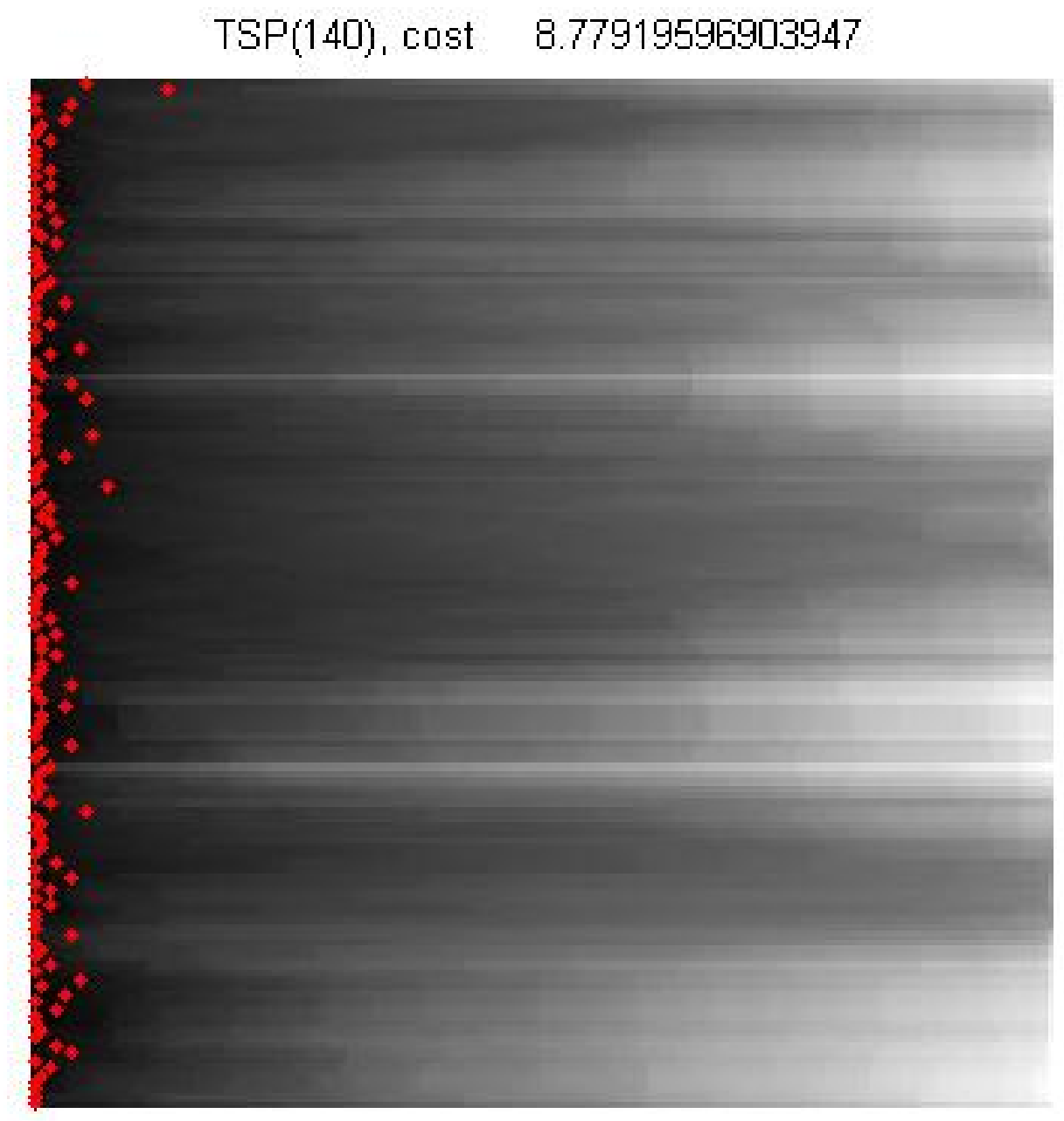,
height=50mm}
\psfig{figure=\IMAGESPATH/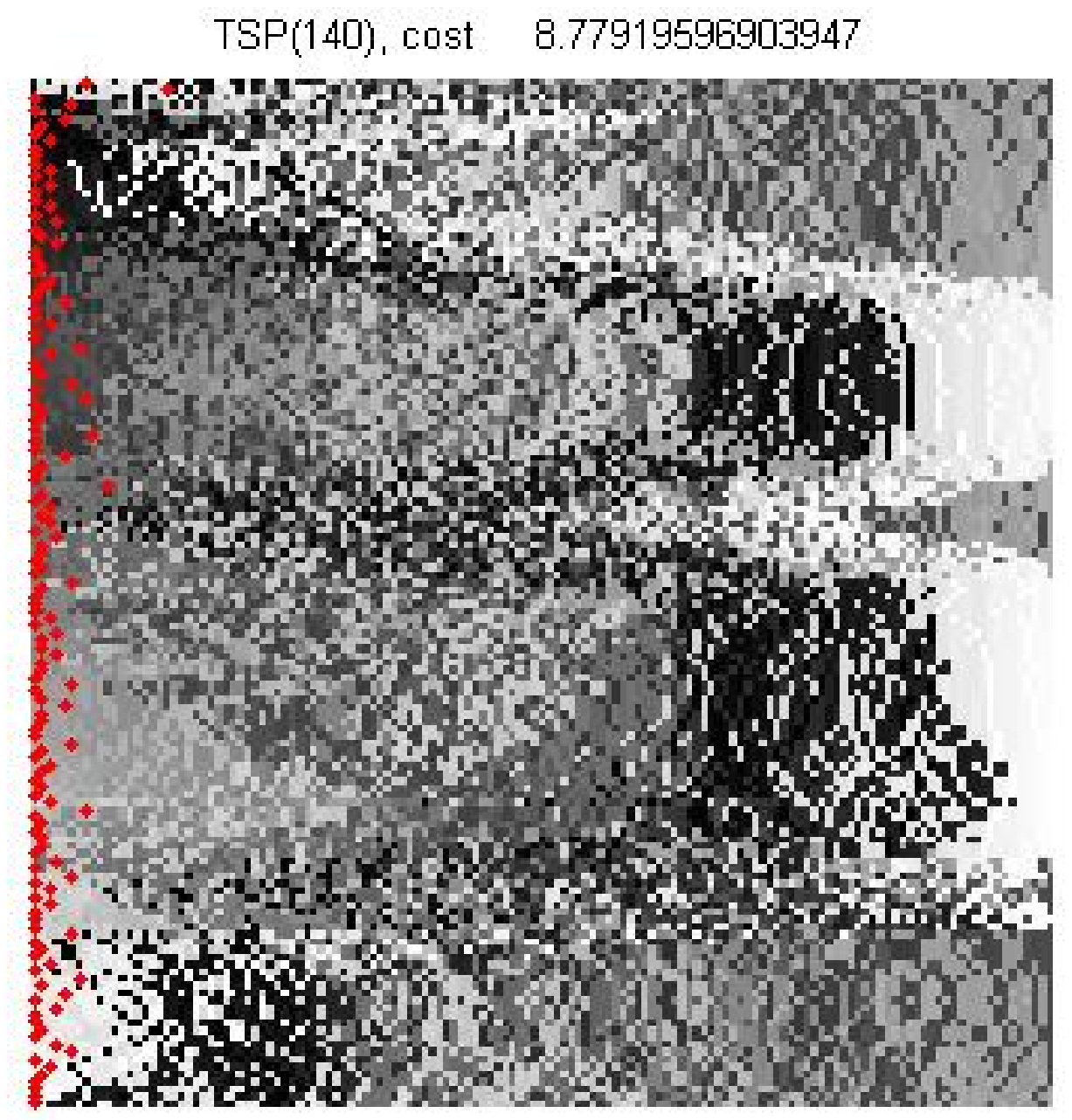,
height=50mm}}
 \caption{Example of an 2D Euclidian TSP$_{140}$.
 There is a pattern to point out a reduction of vertices' alternatives
 From left to right, cycle with the minimum cost,
 sorted cost matrix, and their vertices of the
 sorted $\mathcal{M}$.}~\label{fig:ex_140_cities}
\end{figure}

\begin{figure}
\centerline{\psfig{figure=\IMAGESPATH/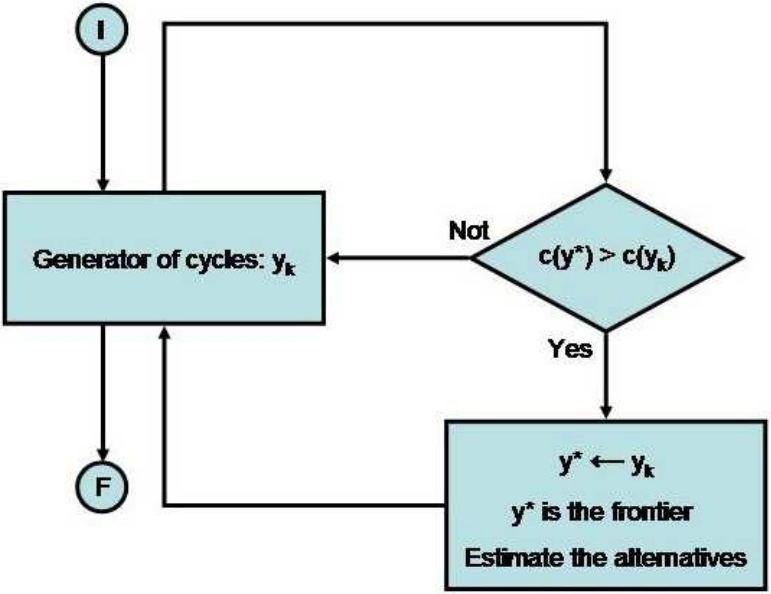, height=80mm} }
 \caption{Flow chart for the verification and estimation of the solution on a reduced research
space.}~\label{fig:Generator_test_cycles}
\end{figure}

An example not small is 2D Euclidian the TSP$_{140}$ depicted in the
Fig.~\ref{fig:ex_140_cities}. The pattern of the frontier압 solution
(red dots) is to left on both images of the sorted cost matrix, and
their vertices of the Sorted $\mathcal{M}$. This means a reduction of
the research space to verify (or even to possibly estimate the true
solution) the solution. To compute the solution of this example the
algorithm is the following. The main flow chart structure is depicted
in the Fig.\ref{fig:Generator_test_cycles}.

\begin{algorithm}~\label{alg:VerificationSol}
\textbf{Input:} GAP$_n$, $c(\cdot)$ cost function for cycles. \textbf{Output:} $y^*$ cycle solution,
and sorted matrix $\mathcal{M}$.
\begin{enumerate}
    \item ~Estimate for a greedy algorithm a possible solution.
    Let be $y^*$ a putative cycle solution.
    \item ~Reorder and renumbering the vertices and cost matrix for the solution. The first
    cycle $n,n-1,\ldots,2,1,n$.
    \item ~ Compute sorted $\mathcal{M}$.
    \item ~ Set the frontier, i.e., Compute the reduced research
    space by estimate the vertices' alternatives for the current minimum cycle
    by equation~\ref{eq:inequality_eps_r}.
    \item ~ \textbf{repeat} for the current frontier generate $y_k$ (a
    cycle)
    \item~\hspace{0.5cm} \textbf{if} $c(y^*)$ $>$ $c(y_k)$ \textbf{then}
    \item~\hspace{1.0cm} $y^*$ $:=$ $y_k$
    \item~\hspace{1.0cm} Update the frontier and estimate the vertices' alternatives.
    \item ~\hspace{0.5cm}\textbf{until} there is not changes and all possible cycles for the last reduced research space
    are generated. (end repeat)
\end{enumerate}
\end{algorithm}

An example of a greedy algorithm for step 1 is given in
algorithm~\ref{alg:GreedyAlgorithm_initialCycle}. The complexity of the
previous algorithm for verifying the solution (assuming that it is the
solution) is $\bigO$ $(\max$ greedy algorithm, sort of $\mathcal{M}$,
one estimation of the reduced research space, while for the generator
of cycles $)$. All but the while for the generator of cycles are
polynomial. The generator of cycles is a TM for keeping the minima
memory's interaction needed only to preserve the optimal solution,
therefore this loop corresponds to a TM when the candidate cycle does
not change, there is not necessary to repeat reorder and sort of
$\mathcal{M}$ for the new solution, and the estimation of the
alternatives of the new solution, ie.e, the estimation of the reduced
research space. Nevertheless its complexity depend of the number of
cycles generated. The computation of the cycle's cost is assumed in one
machine's operation. The only way to guaranty the algorithm압
complexity as polynomial is  if the reduced search space does not
change and it has no more $n^p$ possible cycles. It is easy to compute
cycles by arrays containing the number of the vertices of the reduced
research space, also it is possible to estimate cycles by a
backtracking algorithm. Nevertheless, the complexity of this loop of
the generator of cycles has a lower bound given by the product of the
alternatives that conform different cycles (i.e., paths with $n+1$
vertices without repeating vertices but the first and the last one). It
is possible to define a practical early stop over an accepted reduction
given by an empirical selection of $\eps_{r_v}$ of this algorithm,
i.e., for a given $\eps_{r_v}$ if there is not changes of the frontier
then, the frontier is the local solution accepted as the global
putative solution. It corresponds to the local minimum of the reduced
research space for all cycles  generated by the alternatives given by
such $\eps_{r_v}$ and such frontier. The next proposition exploits such
idea together with the 2D Euclidian properties, and it is
proven that the reduced research space keeps the global solution.

\begin{prop}
The 2D Euclidian TSP$_n$ of the NP-Class have an polynomial algorithm for
checking their solution.
\begin{proof}
For few vertices less than 10 is trivial to find the solution in
polynomial time, and no matter if the vertices are closed or disperse
the sorted $\mathcal{M}$ has the solution's frontier closed to the
left, and $\eps_{r_i}$ have appropriate values. Let assume an instance
of a problem 2D Euclidian TSP$_n$ and its solution. Using the
algorithm~\ref{alg:SortM} the sorted $\mathcal{M}_n$ is computed. Now,
a vertex is given and the cost matrix is updated for an 2D Euclidian
TSP$_{n+1}$. Three cases are need to be considering in 2D for any  flat
figure limited by vertices. The new vertex is in or out of the
vertices' hull. if it is out, by the minimum distance from this point
to vertices' hull in 2D this vertex can only add at most 3 cost
relevant at most 3 vertices which are closed to the minimum distance to
it. If the vertex is inside, then considerer a triangulation of the
vertices' hull. The new vertex must be in a triangle or on triangle's
edge. Any way,in the three cases, it can be added to keep a
triangulation between the vertices of the vertices' hull of the given
2D Euclidian TSP$_n$. Now the solution of the 2D Euclidian TSP$_n$ is marked
on the triangulation of the $n+1$ vertices and this solution is
altering to include the new vertex changing the closed pair of vertices
to include the new one without crossing edges and the numeration is
changed to include the new vertex to reflect its vicinity and also if
it is necessary adjust $\eps_{r_v}$ to keep this vicinity. Now the
algorithm~\ref{alg:VerificationSol} can be used to estimate the
solution for this 2D Euclidian TSP$_{n+1}$ with the updated
$\mathcal{M}_{n+1}$ and the suitable initial solution is the vertices'
numeration including the new vertex. The alternatives to verify the 2D
Euclidian TSP$_{n+1}$ are limited to the new vertex's alternatives of
its vicinity vertices of the previous solution of the 2D Euclidian TSP$_n$.
\end{proof}
\end{prop}

\begin{figure}
\centerline{\psfig{figure=\IMAGESPATH/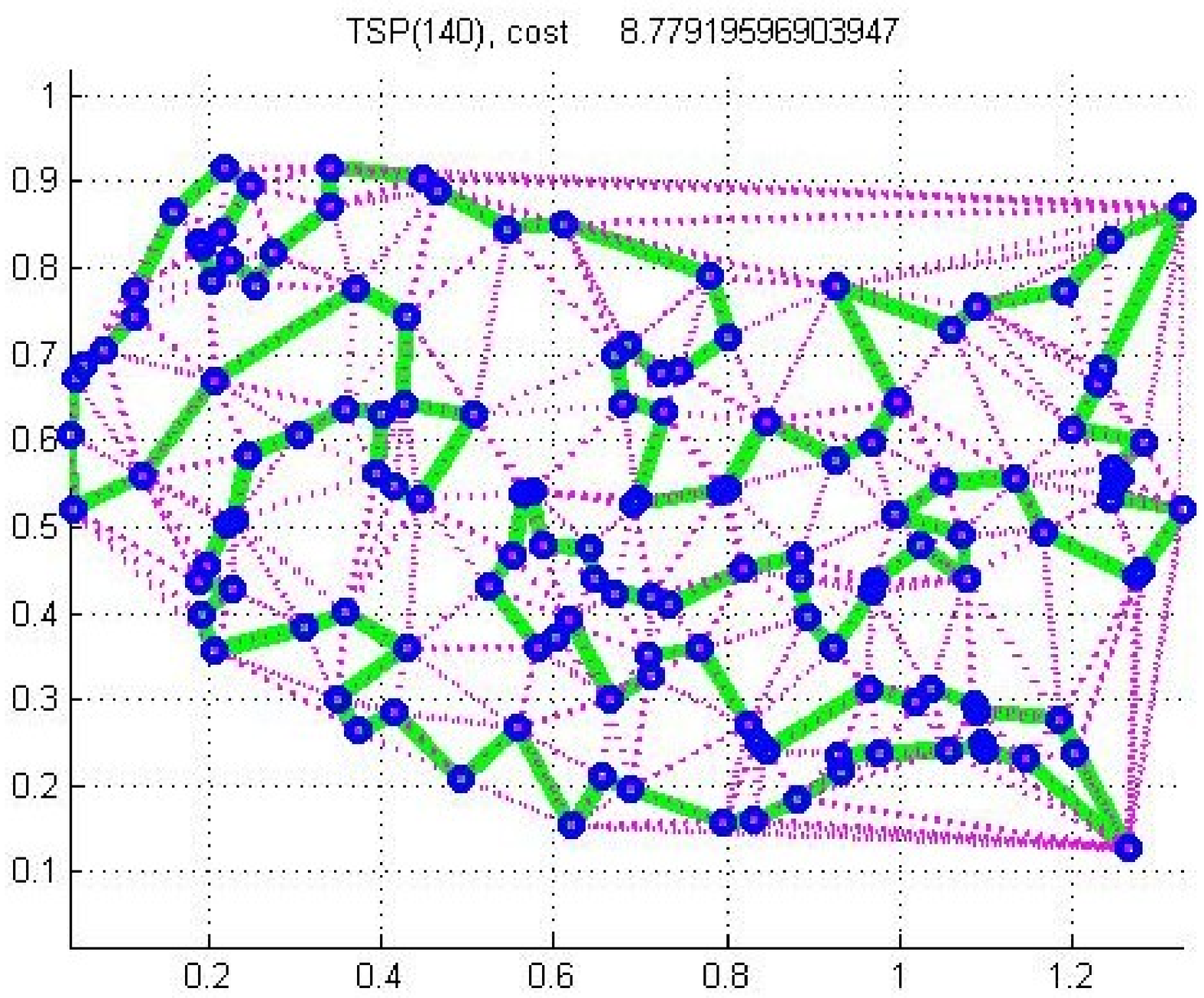,
height=60mm}
\psfig{figure=\IMAGESPATH/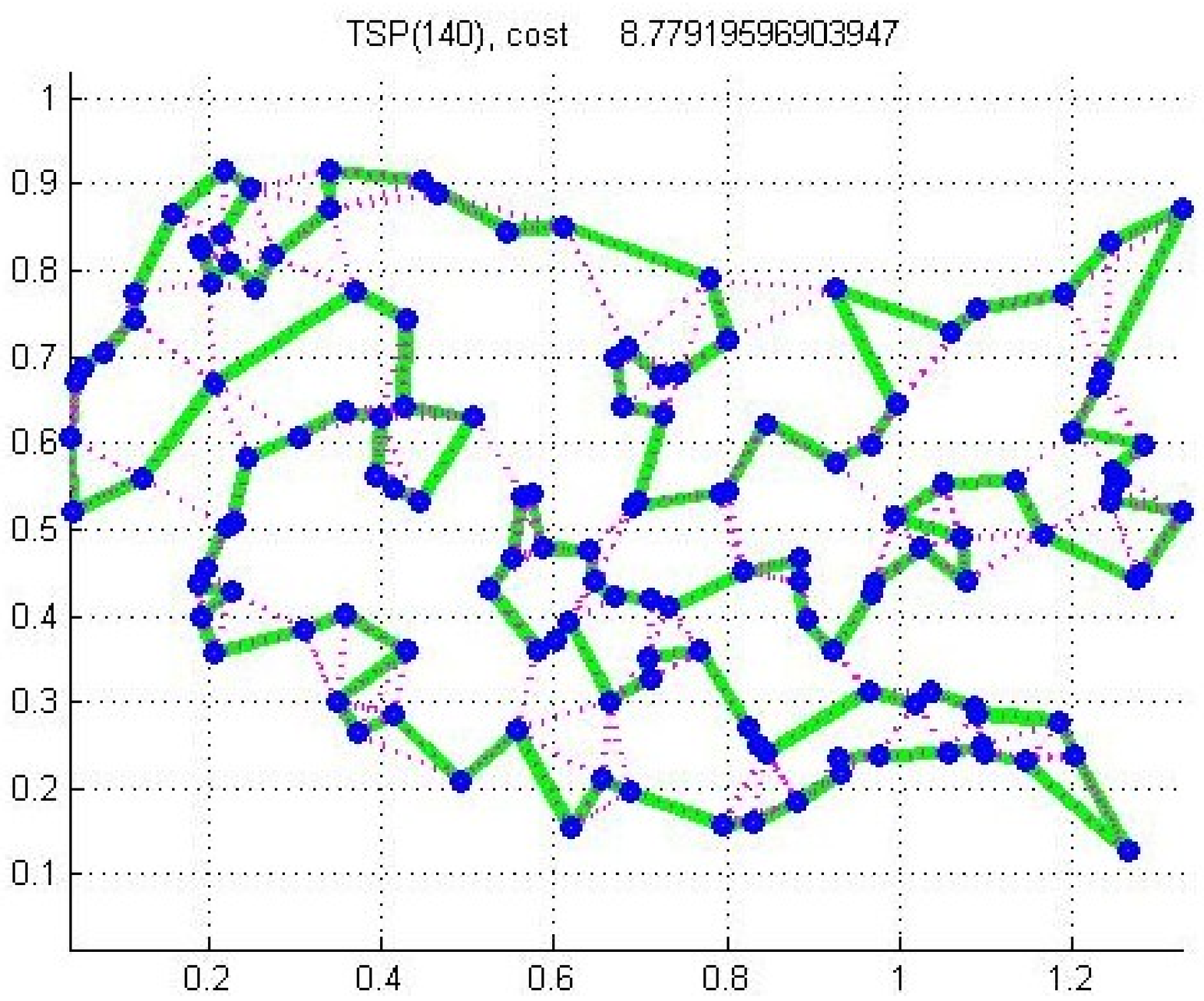,
height=60mm}}
 \caption{Triangulation (magenta dotted lines) complete and partial for
 the example 2D Euclidian TSP$_{140}$,
 the optimal cycle is depicted by the green lines,
 and the vertices are depicted the blue dots.}~\label{fig:TSP_140_Triangulation}
\end{figure}

\begin{rem}
The previous proposition states that a growing strategy vertex by
vertex can be used not to solve but to verify efficiently the 2D
Euclidian TSP$_n$'s solution.  Proposition~\ref{Prop:GAP_vs TSP} shows
that for the TSP$_n$, it is possible to stop computing paths in the
same branch when a reference value is reached, but the idea of the
reduction of the research space, also does not consider such paths,
therefore the algorithm~\ref{alg:VerificationSol} does not need to
include or use this property of the TSP$_n$, i.e. it is not necessary
an ``if얎 inside of cycles generator when $c(p_2) \geq c^*$, $c(y^*) =
c^*$ where $p_1$ is a path of TSP$_n$. The
algorithm~\ref{alg:VerificationSol} is not such algorithm as it is
depicted here. For the 2D Euclidian TSP$_n$ examples of this paper, I
did manually ad-hoc initial solution with its enumeration of the
vertices (a greedy manual process), after this, a reorder of the cost
matrix to build an equivalent GAP$_n$ (see
proposition~\ref{prop:enumeraCycles_descending_n_first}), and with the
algorithm~\ref{alg:GreedyAlgorithm_initialCycle} an initial cycle was
computed, finally,
 after that to verify and compute the
solution of the 2D Euclidian TSP$_n$ examples the
algorithm~\ref{alg:VerificationSol} was applied considering neighbors
with $\eps_{r_i} \in [-0.9, 0.6]$. It is not be confused that the
manually reduction described above missed or assumed the global optimal
cycle of examples of 2D Euclidian TSP$_n$, it is obvious that the
global optimal solution is a local optimal solution as necessary
condition. Therefore, the previous proposition shows that the reduced
search space allows to verify the locally of the putative solution, and
with a carefully reduction of the alternatives based in the 2D
Geometry, the global solution remains inside of the alternatives of
each putative cycle founded. Figures~\ref{fig:TSP_140_Triangulation},
and~\ref{fig:TSP_37_triangulation} depict examples of appropriate
complete and partial triangulation that they keep inside the solutions.

The key to build a efficient algorithm for varying the solution is the
triangle inequality property and the monotony of a finite norm in 2D
that allows to have locally vertices' vicinity and to keep a
triangulation, and in 2D a few paths closed from where it is possible
to obtain an appropriate numeration of vertices. The triangle
inequality helps to walk by short edges avoiding large diagonal (see
proposition~\ref{prop:trapecio_Triangle_inequality}), and helps to
define $\eps_{r_i}$ to find an appropriate local vertex's vicinities
(see fig.\ref{fig:TSP_37_triangulation}). The parameter $\eps_{r_i}$
correspond to the vertex $i$. The next algorithm estimates the
alternatives, and the parameters $\eps_{r_i}$ by selecting at least two
alternatives.

\begin{algorithm}~\label{alg:Estimate_alternatives eps_r}
\textbf{Input:} GAP$_n$, $y=(n,1,2,\ldots,n-1)$ the vertices of the putative cycle
solution, and $c(\cdot,\cdot)$ cost matrix.
\textbf{Output:} $a \in\Natural^{n \times (n-1)}$ ($v$'s alternatives, i.e., vertex related with $v$
by the $c(v$,vertex$)$ closed to $c(v,y(v))$),
$\eps_r\in\Real^n$ (array of the estimations of $\eps_{r_v}$).
\begin{enumerate}
    \item ~ \textbf{for} v:=1 \textbf{to} n \{
    \item~\hspace{0.5cm} $\eps_r(v)$ $:=$ $-1.0;$
    \item~\hspace{0.5cm} \textbf{while} (True) \{
    \item~\hspace{1.0cm} $c_{vk} := c(v,y(v))(1.0 + \eps_r(v));$
    \item~\hspace{1.0cm} $t:=0;$ $t\in \Natural^{(n-1)}$ (array of vertices)
    \item~\hspace{1.0cm} $t=\hbox{find}(c(k,:) \leq c_{vk}) \cup {y(v)};$
    (vertices of the $v$'s vicinity)
    \item~\hspace{1.0cm} $n_t$ $:=$ length($t$);
    \item~\hspace{1.0cm} \textbf{if} $n_t$ $>$ 1 \textbf{then}
    \item~\hspace{1.5cm} break;
    \item~\hspace{1.0cm} \textbf{if} $n_t$ $>$ $(n-1)$ \textbf{then}
    \item~\hspace{1.5cm}  $\eps_r(v)$ $:=$ $\eps_r(v)$ - $0.01;$
    \item~\hspace{1.0cm} \textbf{else}
    \item~\hspace{1.5cm}  $\eps_r(v)$ $:=$ $\eps_r(v)$ + $0.01;$
    \item~\hspace{0.5cm} \} (end while)
    \item~\hspace{0.5cm} $a(v,:)$ $:=$ $t;$ (store $v$'s alternatives)
    \item~ \} (end for)
\end{enumerate}
\end{algorithm}

The following
proposition shows that for any quadrilateral, its perimeter is the
lowest cycle'length.

\begin{figure}
\centerline{ \psfig{figure=\IMAGESPATH/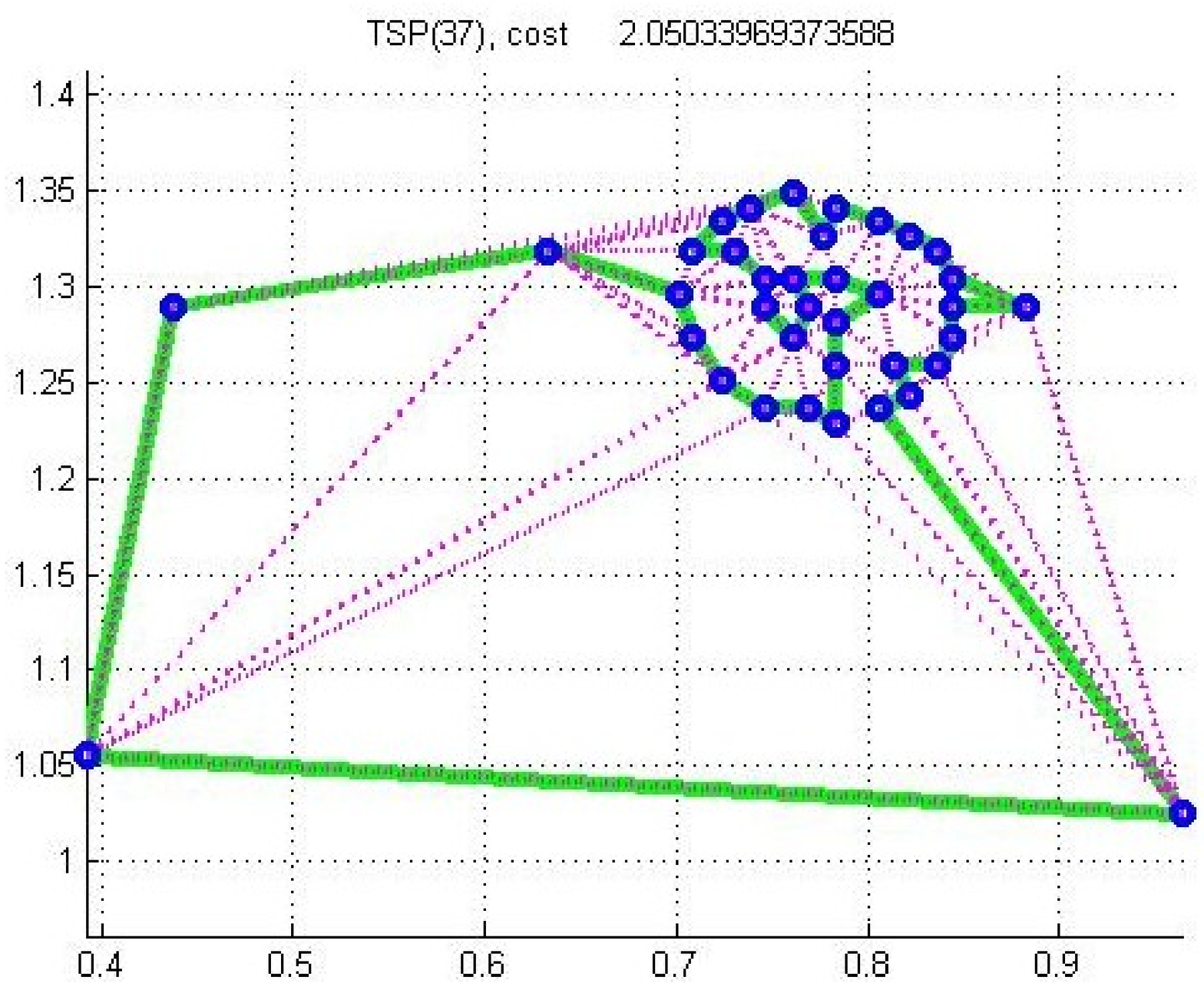,
height=60mm} \psfig{figure=\IMAGESPATH/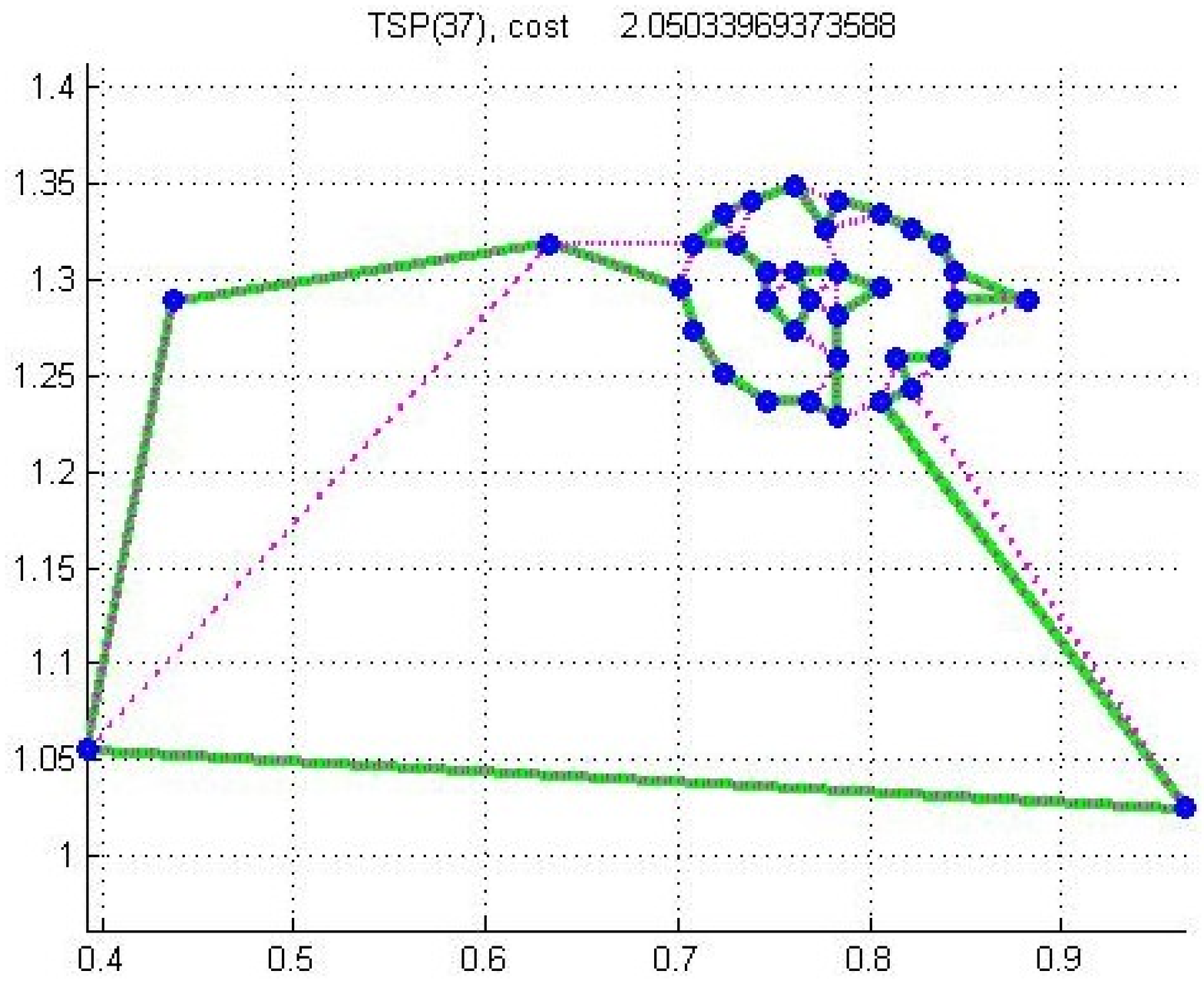,
height=60mm}}
 \caption{
 Triangulation (magenta dotted lines) complete and partial for
 the example 2D Euclidian TSP$_{37}$,
 the solution is depicted by the green lines,
 and the vertices are depicted by the blue dots.}~\label{fig:TSP_37_triangulation}
\end{figure}

\begin{figure}
\centerline{ \psfig{figure=\IMAGESPATH/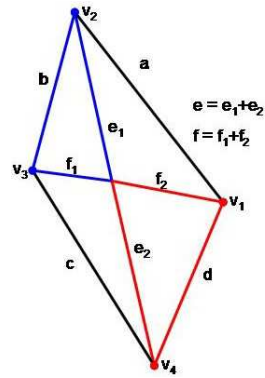, height=50mm}}
 \caption{Arbitrary  quadrilateral. }~\label{fig:quadrilateral}
\end{figure}

\begin{prop}~\label{prop:trapecio_Triangle_inequality}
Given four points in 2D as vertices, its quadrilateral's perimeter is lower
than the cycle's Euclidian length using the diagonals and the opposites sides.
\begin{proof}
Figure~\ref{fig:quadrilateral} depicts an arbitrary quadrilateral with
vertices $v_1$, $v_2$, $v_3$, and $v_4$, and edges' Euclidian length
$a$, $b$, $c$, and $d$. Its diagonals' length are $e$, and $f$. Without
loss of generality the opposites sides are $a$, and $c$. The next
relations are valid by triangle inequality for the blue triangle, and
the red triangle:
\begin{eqnarray*}
  b & \leq & e_1 + f_1 \\
  d & \leq & e_2 + f_2.
\end{eqnarray*}

Then, these relations and  the quadrilateral's perimeter comply:
$$ a+b+c+d \leq a + e_1 + f_1 + c + e_2 + f_2 = a+e+c+f
$$

\end{proof}
\end{prop}

In this
case the verification of the solution is localely around of the
oscillating cycles closed to the frontier, as the a vertex is far away of a local vicinity
the monotony of the distance allows that such vertices can be ignoring
because they include pair of vertices with higher cost. Finally
the given cycles by the initial enumeration and the cycles
computed do not have crossing edges, this means that all edges of
the solution does not cross the edges of an appropriate triangulation
(see figures~\ref{fig:TSP_37_triangulation}, and~\ref{fig:TSP_140_Triangulation}).
This is the case for Euclidian length, $d(x,y) = \sqrt{x^2+y^2}$ but
$d_{\inf}(x,y)=\max(|x|, |y| )$ allows crossing edges because
diagonal and sides have the same length in a quadrilateral.
The symmetry of
the matrix cost is not used in any way in the process of
verification. Other way to see that the 2D Euclidian TSP$_n$ can
be solved in polynomial time is the property that its sorted
$\mathcal{M}_{n+1}$ has a left frontier (this mean very few
alternatives around of the solution) and there is always a
vertices' numeration which it takes in consideration a closed
vertex with a consecutive number, so the cycle's numeration is
consecutive and correspond to the first cycle in the enumeration
given by the proposition~\ref{Prop:GAP_basic_properties}. Also,
its integer formulation can be also solve in polynomial time, and
pivot strategies get advantage of the geometric properties
(the variables of a pivot for TSP$_n$ are limited,
see proposition~\ref{Prop:GAP_vs TSP}),
which
are reflected on the structure of the sorted $\mathcal{M}_{n+1}$
and its frontier. But in 3D or higher dimensions the initial
enumeration is not easy to do, and could be many paths around a
new vertex, even so, in numerical experiments of  $m$D Euclidian
TSP$_n$ ($m \geq 3$), its the sorted $\mathcal{M}_{n+1}$ has a
left frontier lightly shifted to the right. The reduction of the alternatives in the 2D
Euclidian is limited by the number of neighbors that a vertex can
have in a plane to keep a triangulation that does not change the
global solution but reduce the alternatives to preserve the
solution in a local or reduced research space. Other property in
the solutions of 2D Euclidian TSP$_n$ is that the cycles have not
cross over so the new vertex has limited way to be connected. In
3D the Euclidian TSP$_n$ has a growing alternatives quite
different. A vertex could has from one to 12 neighbors and many
path around it but a cheap triangulation opens the possibility to
 formulate an efficient algorithm for solving it.
\end{rem}

On the other hand, the integer programming version IPGAP$_n$
depends on pivot strategies to explore the feasible points
(see~\cite{ siam:Kolda2003})
$x_{ij}$. For an arbitrary cost matrix without a pattern ,the
exploration even by oracles' strategies are not suitable because
the set of values of the edges' cost has not a relation or
property. On the section~\ref{sc:integer} was noted that at least
two vertices are need to change for three $x_{ij}$ by three
$x_{ij}$ for two feasible points. A pivot strategy of using two
vertices is not sufficient to explore all feasible points, a cycle
could by different of other cycle for more than two vertices. The
cycles obtained over a given cycle of GAP$_n$ by using an strategy
of two vertices is $2!(n-2)$, $n\geq 3$; for 3 vertices the cycles
are $3!(n-3)$ $n\geq 4$ but the previous ones are cover. For k
vertices the cycles are $k!(n-k)$ $n \geq k+1$ over a given cycle.
By the proposition~\ref{Prop:GAP_vs TSP} to limit the number of
variables on a pivot's strategy could cause to miss the
solution (the cost function of
GAP$_n$ is oscillating). The limits is an strategy of using $n-1$ active variables but this means
to explore all the $(n-1)!$ cycles of the equivalent GAP$_n$.

This point could be debatable but the complexity of matrix cost of a
given GAP$_n$ can grows. This can be done by assigning the edges' cost
as a random field or white noise with values in $[0,1]$ or by mixing
vertices at different but similar distances (distance as edge's cost).
Then without properties it is not possible to reduce the research
space, and properties can not be inherited. For example, take an 2D
Euclidian TSP$_n$ and add $m$ arbitrary vertices and their costs, it
loses its triangulation, i.e., the new $m$ vertices and the $n$
original vertices are not longer related by the Euclidian distance but
arbitrary values let say in the interval $[-v, v]$, $ v > 0$, the
resulting problem is a GAP$_{m+n}$ with $ m \gg n$ and without the
properties of an Euclidian Space. Also, a GAP$_k$ with cost matrix with
values on $[-\eps/k, \eps/k]$ has cycles' cost bounded by $[-\eps,
\eps]$ and ad-hoc probabilistic algorithm could give a putative
solution in polynomial time but if new $m$ vertices and their arbitrary
costs are added, the solution of the GAP$_k$ and properties what helps to
solve it, they do not contribute and be out of the solution in many cases
for arbitrary and large GAP$_{k+m}$.

\begin{prop}
An arbitrary and large GAP$_n$ of the NP-Class has not an
polynomial algorithm for checking their solution.
\begin{proof}
If such algorithm exist for any problem size $m$. Then it can
verify the solution of any problem GAP$_n$ with $n \gg m$. Let the algorithm to
estimate the solution of an arbitrary GAP$_n$, constructed by an
arbitrary combination of vertices with arbitrary edges' cost. This
means that the cost matrix has not properties or relations between
the edges' cost of the vertices of GAP$_n$. Now for the putative
solution, reorder and renumbering the vertices and cost matrix
such that the first  cycle $n,n-1,\ldots,2,1,n$ is the solution.
Compute sorted $\mathcal{M}$ and looks the images of the sorted
cost matrix, and their vertices of $\mathcal{M}$. If the there is
a pattern then the matrix is not arbitrary, which is not the case
by the construction of the cost matrix of GAP$_n$. If the frontier
is on almost over the left size then the values of the cost matrix
by row are related by an order, which is not the case by the
construction of GAP$_n$. Therefore, most of vertices of the
frontier are not close to the left side. The lack of properties of
the given cost matrix does not allow to use other algorithm but
the depicted by the Fig.~\ref{fig:Generator_test_cycles}. But this
means that even in an optimism case for at least tree alternatives
for each vertex, the complexity of the loop to verify the solution
is $\bigO(3^n).$ There is not exist such efficient algorithm.
\end{proof}
\end{prop}

The proposition shows that an algorithm could claim and maybe
solve a NP problem but the verification without properties over an
a worst case with arbitrary data can not be demonstrated in
polynomial time, and if the research space could be reduced then
input problem is not a arbitrary worst case with many oscillating cycles.

\section*{Conclusions and future work}~\label{sc:conclusions and future work}

Here the concept of a tube and the parallel solution were drafted as an
example of a characteristic that in order to estimate the complexity
does not relates to the complexity of the input data's size. This means
that there are easy to solve instances of TSP$_n$ and GAP$_n$. The
images of the frontier depicts and corroborates the existence of this
tubes for the examples of 2D Euclidian TSP$_n$ but there are more work
to do. The reducibility of 2D Euclidian TSP$_n$ allows to verify its
solution in polynomial time. It is consequence of the 2D geometric
restriction, the monotony of a distance and their geometric properties.
For LJP$_n$ the reducibility allows to define the lattice CB but
arbitrary and large optimal clusters can not be efficiently estimated.
I suspect and has not proved that an algorithm to solve an arbitrary
$m$D
 Euclidian TSP$_n$ ($m > m_0$) has exponential complexity starting with some $m_0>4$, but
 if the vertices can be numerated, and the initial cycle has a small cost
 then the complexity for verifying or updated the
 initial putative cycle could be limited to polynomial time by geometric properties.
 The key is to have an appropriate way to numerate,
and also, to reduce the alternatives by the geometric properties. It
could be possible to find a reduced research space using an appropriate
triangulation so that the global solution does not escape. But
Triangulation algorithms has $\bigO\left(n^{\frac{m-1}{p}}\right)$,
$\forall p \in [2, m-1]$ (see~\cite{ SODA:Amenta07}). Therefore, it is
highly possible that it exists an algorithm to solve arbitrary and
large $m$D Euclidian TSP$_n$ with large polynomial time until some
finite $m>0$. There is not an efficient algorithm for solving 2D
Euclidian TSP$_n$  but it is highly possible to  build it. However, for
arbitrary and large GAP$_n$, it does not exist an efficient algorithm
for solving it.

Moreover, even with the reduction of the research space to the
lattice CB, it was expected that LJ problem or the Searching of
the Optimal Geometrical Structures of clusters of $n$ particles
remains out of the existence of an efficient algorithm but now it
is proved that it has exponential complexity. On the other hand,
the putative optimal cluster that I tested in the appropriate
reduced lattice CB  from 2 to 55 particles are now considered
global optimal clusters for the LJ potential. There are more work
to verify the long list of putative optimal clusters from 56 to
1501 particles given by Shao or The Cambridge Cluster Database
(CCD), i.e., to shift them from putative to global optimal clusters.

The line to divide when is possible to verify in polynomial time
the solution is given by the frontier압 position on the sorted
matrix $\mathcal{M}$. This brings the solution of Conjecture of
the P-Class and the NP-Class, the later has not polynomial
complexity in the general or worst case scenario (large and
arbitrary NP problems) but exponential, therefore, the Classes P
and NP are different except for trivial cases when it is possible to verify the
solution by the reducibility. It took a while to understand and
accept that they lack of properties do not allow to design an
algorithm able to tackle the growing of the complexity in term of
the input data's size. Algorithms are based on properties or data
relations given or constructed by organizing the data on
appropriated data structures as the algorithm progress. Here the
problems of class NP are not only a worst cases but without
reducibility, arbitrary and large ones.

The result shows that does not exist a general property that allows to
solve in polynomial time all NP problems, therefore is important to
continuing developing specific and ad-hoc algorithms for each type of
the NP problems. Also, it showed that is important in the design of
these algorithms to consider the problem's properties and
characteristics as insights to save operations. The polynomial
reduction procedure states that is possible to solve 2D Euclidian
TSP$_n$ by changing the cost matrix to an structure based in few
alternatives, i.e., as a rectangular matrix with few columns to
consider or an appropriate triangular sub-graph of its graph $G_n$.

The local or global dilemma, could be addressing as in here by defining
an appropriate search space where a local not heavy time consuming
algorithm could be sufficient for many scientific, technological, and
engineering issues. In fact, the landscape of the research space of the
NP Problems allows stationary properties from where is quite difficult
to escape for improving the putative solution. The evolution and the
living beings are the kind of solution of problems of this class. There
are many examples where the solution is know but we can not justify or
demonstrate and reproduce such results. Too many combinations,
relations, variables, restrictions, and process are perfectly  working
every day in a biological machine but we are not yet able to understand
them. For the NP problems the landscape of the research space could
have one or many (exponential) similar results but the solution could
be slightly different. We are yet able to find out these slightly but
important solutions because our computers technology  has time
limitations. Practical limitation are the main argument for an early
stop. However, besides that is not worth to expend exponential time for
look for the solution, it is possible that this slightly difference
could be the answer for the life mechanism or other complex biological
process. The Lennard-Jones problem is an example where the meaning of
practical purposes for limiting the time for determining the optimal
geometry for cluster of $n$ particles is debatable. The prediction of
real cluster structures and the possibility to help to understand the
cooling process of real cluster for this simple NP problem keeps many
researchers busy and greedier for new results.

For the future, new computing technology and models are needed, it
will be very interesting to define quantum computing models to
look for establishing novel possible efficient perspectives to
solve the problems of the NP-Class and exponential ones.

Finally, this novel formulation solves the Noted Conjecture of the
NP-Class and brings a promising theoretical perspective to address
the construction of efficient algorithm for arbitrary problems.

\section*{Acknowledgement}

This work is dedicated to the dreamers of a new era of peace,
understanding, sharing, and welfare for all beings.

\bibliographystyle{abbrv}
\bibliography{\BIBPATH/NPComplexity_a12}

\end{document}